\documentclass[%
twocolumn,
 amsmath,amssymb,
 pra,
]{revtex4-2}

\usepackage{graphicx}
\usepackage{dcolumn}
\usepackage{bm}

\usepackage[utf8]{inputenc}

\usepackage{xcolor}

\newcommand{\avZ}[1]{\ensuremath{\langle #1\rangle_Z}}	
\newcommand{\erf}[1]{\ensuremath{\mathrm{erf}\left( #1\right)}}	
	
\def\mr{\mathrm}
\def\mb{\mathbf}
\def\bs{\boldsymbol}

\def\Tr{\mathrm{Tr}}

\newcommand{\Ns}{\ensuremath{N_{\rm S}}} 
\newcommand{\Jw}{\ensuremath{J}} 
\newcommand{\Nt}{\ensuremath{M}} 

\usepackage{mathtools}
\DeclarePairedDelimiter\floor{\lfloor}{\rfloor}

\begin{document}

\title{Signal-to-noise-ratio and SNR-max detection statistics in template bank searches for exotic physics transients with networks of quantum sensors}

\author{T. Daykin}\email[]{tdaykin@unr.edu}
 	\affiliation{Department of Physics, University of Nevada, Reno, 89557, USA}
 \author{C. Ellis}
 	\affiliation{Department of Physics, University of Nevada, Reno, 89557, USA}
\author{A. Derevianko}\email[]{andrei@unr.edu}
	\affiliation{Department of Physics, University of Nevada, Reno, 89557, USA}

\date{\today}

\begin{abstract}
Quantum sensor networks such as the existing networks of atomic clocks and magnetometers
offer intriguing capabilities in searches for transient signals such as dark matter or fields sourced by powerful astrophysical events. The common matched-filter technique relies on signal-to-noise (SNR) detection statistic to probe such transients. For macroscopic dark matter objects, the network would register a sweeping signal as the object propagates through at galactic velocities. A potential event is registered when the SNR from a specific template exceeds a threshold set by a desired false positive rate. Generically, to span the continuous parameter space for the network-exotic-physics encounter, one has to deal with multiple templates. In such template bank searches, the natural generalization of the SNR statistic is the SNR-max statistic, defined as the 
maximum of the absolute values of SNRs determined from individual template matching. While  individual SNR realizations are Gaussian distributed, SNR-max probability distribution is non-Gaussian. Moreover,  as the individual template-bank SNRs are computed using the same network data streams, SNRs become correlated between the templates. Cross-template correlations have a sizable effect on the SNR-max probability and cumulative distributions, and on the threshold SNR-max values. Computing threshold SNR-max values for large template banks is computationally prohibitive and  we develop analytic  approaches to computing properties of SNR-max statistic. This is done for cases when the template bank is nearly orthogonal (small cross-template correlations) and for banks with cross-template correlation coefficients distribution ``squeezed'' about the most probable cross-template correlation value. Since the cross-template correlation coefficients quantify the similarity of templates, increasing  correlations tend to decrease SNR-max thresholds for specific values of false positive rates. Increasing the number of templates in the bank increases the SNR-max thresholds.
Our derivations are carried out for networks that may exhibit colored noise and cross-node correlations. Specific applications are illustrated with a dark matter search with atomic clocks onboard satellites of Global Positioning System (GPS) and with a ``toy'' planar network with cyclic rotational symmetry.
\end{abstract}

\maketitle

\section{\label{sec:level1}Introduction}

As precision quantum sensors~\cite{Degen2017-RMP-quantum-sensing} become ubiquitous, it is natural to combine them into networks.
In this paper, we focus on time-domain searches for exotic physics signals with networks of quantum sensors. Time-domain analysis is a natural fit for transient non-oscillating signals. We refer the reader to  frequency-domain analysis~\cite{Derevianko2016a} relevant to searches  for oscillating dark matter (DM) signals with networks or to combined time-frequency domain analyses~\cite{Romano2017,AndVraCre01,dailey2020ELF.Concept} for detecting oscillating transients such as bursts of gravitational waves~\cite{Romano2017,AndVraCre01} or low-mass exotic field (ELFs) emitted by powerful astrophysical events~\cite{dailey2020ELF.Concept}.

The spatial extent of networks can range from microscopic (atomic ensemble scales) to continental or trans-continental scales.  Specific examples include a constellation of microwave atomic clocks on-board GPS satellites~\cite{Roberts2017}, network of gravitational wave detectors~\cite{Canton2017,Sathyaprakash1999}, trans-European network of high-performance optical clocks~\cite{Roberts2019-DM.EuropeanClockNetwork}, and the global network of atomic magnetometers (GNOME)~\cite{GNOME2021search}. The nodes of the networks can be connected by physical links (e.g., optical fiber or microwave/laser links). Alternatively, synchronized cross-node comparison can be accomplished by a simple time-stamping of sensor data relying on world-wide broadcast of GPS time. The network can be composed of identical sensors (homogeneous network) or different
sensors (heterogeneous network). The position of the nodes can be stationary or evolving in time (like in the case of orbiting GPS satellites). Finally, individual sensors can use entangled  or spin-squeezed ensembles that operate below the quantum projection noise limit, see, e.g.~\cite{BlaWin08,WeiBelDer10,Komar2016}. The formalism developed in this paper  applies to all these cases and our results hold for any network with stationary white or colored Gaussian noise and with and without cross-node correlations.

This paper grew from our practical experience with the GPS.DM search~\cite{Roberts2017} for DM-induced transients using GPS network of atomic clocks. Multiple galactic scale astrophysical observations indicate that DM comprises $85\%$ of all matter in the universe, with only $15\%$ left to ordinary matter~\cite{Bertone2005}. These galactic scale observations  only characterize the gravitational interaction of ordinary matter  and dark matter. Little is known of the microscopic composition of DM or of its non-gravitational interactions with standard model particles and fields. The primary target for current DM searches are weakly interacting massive particles (WIMPs). With WIMP searches so far failing to provide convincing evidence,  alternative DM candidates are being considered~\cite{Bertone2005}. Self-interacting ultralight fields is one such alternative, where the DM candidate can take a form of a macroscopic object that is coherent over large scales~\cite{Vilenkin:1984ib,Coleman:1985ki,KusSte01,DerPos14}. These macroscopic DM objects is the target of the GPS.DM search.

In the context of DM models, ultralight fields are characterized by macroscopic mode occupation numbers and can thus be described as classical fields. 
The effect of exotic physics on quantum sensors depends on specific interactions~\cite{DerPos14} (portals) between exotic fields and standard model particles and fields. Particle detectors rely on energy-deposition. By contrast, ultralight fields, through their portal interactions, {\em gentle modulate}  atomic (qubit) energy levels. Detecting such gentle modulations is the domain of precision quantum sensors.
For example, for atomic clocks~\cite{DerPos14,ArvHuaTil15}, atom interferometers~\cite{GraKapMar2016,GeraciDerevianko2016-DM.AI,Carney_2021,FigueroaNataniel2021Dmsu,McNallyReesL2020Cdwd,YangWanpeng2016DTDD} %
, and optical cavities~\cite{Cavity.DM.2018,Carney_2021,RajendranSurjeet2017Amfd}
, DM objects can drive variation of  fundamental constants of nature.  The change in fundamental constants in turn affects atomic or cavity frequencies causing an atomic clock to speed up or slow down when the DM object overlaps with a sensor. 
Alternatively, gradients of objects composed of axion-like fields act as fictitious 
magnetic fields and can be detected with another class of quantum sensors, magnetometers~\cite{Pospelov:2012mt,Masia-RoigHector2020Amfd} 
In both examples, as the DM object sweeps through, the induced frequency shifts or fictitious 
magnetic fields would appear as a transient perturbation sweeping through the network. A geographically-distributed network can resolve the directionality and speed of the transient signals. Further, this provides vetoing mechanism  as the network sweep speed and directionality must be consistent with standard halo model (SHM) priors. Currently, there are several on-going DM searches that follow these ideas~\cite{Roberts2017,Wcislo-clock-network-2018,GNOME2021search,FigueroaNataniel2021Dmsu,Masia-RoigHector2020Amfd,YangWanpeng2016DTDD,RajendranSurjeet2017Amfd,McNallyReesL2020Cdwd}

Our GPS.DM collaboration~\cite{Roberts2017,Panelli2020}, focuses on searching for DM transient signatures by utilizing the network of atomic clocks aboard the GPS satellites. The GPS data has an inherent advantage over building a new network of quantum sensors as there is over two decades of publicly available archival data~\cite{JPLwebpage}, ready for data mining. DM signatures would consist of a correlated propagation of atomic clock frequency perturbations through the GPS network at galactic scale velocities ($\sim 300 \, \mr{km/s}$). Previously, our GPS.DM group has performed analysis of the archival GPS data in search for 2D DM walls~\cite{Roberts2017}. Although no DM signatures were found, prior astrophysical limits on certain DM couplings to atoms were improved by several orders of magnitude. Ref.~\cite{Roberts2018a} developed the application of the more sophisticated Bayesian search techniques which may extend the DM discovery reach further by several orders of magnitude in both the sensitivity and size/geometry of the DM objects. The main focus of Ref.~\cite{Panelli2020} was on the performance of an alternative frequentist approach, the matched filter technique (MFT). That work obtained several analytical results for an idealized network of white-noise sensors, including cross-node correlation. Here we extend the results of Ref.~\cite{Panelli2020} to template banks.

The MFT is an ubiquitous technique, utilized for example by the Laser Interferometer Gravitational Wave Observatory (LIGO) in their gravitational wave detection~\cite{Canton2017,Sathyaprakash1999}. Also, MFT has applications in astrophysics~\cite{Lenon2021,Ofek2018,Feng2017,Ruffio2017,Lenon2021}, geophysics~\cite{https://doi.org/10.1029/94JB00498}, and searches for exotic physics~\cite{Pustelny2013,Afach2018,Simeone2009,Sachdev2019,Owen1996n}. In applications, a large template bank determined by prior information on the expected signal, is used to match data streams. Quantitatively, the degree of the match is characterized by  a signal-to-noise ratio (SNR). In Ref.~\cite{Panelli2020}, the SNR for individual templates was shown to be a Gaussian normal random variable. Further maximization of the template specific SNR variables (ignoring dependence on sign), was performed over the template bank.  In this process, a Gaussian distribution was considered instead of the distribution for the maximum of the absolute valued SNRs (SNR-max). As we discuss in this paper, the SNR-max distribution is no longer  a Gaussian distribution, but a skewed (non-Gaussian) distribution. Moreover, there exist correlations in the SNR statistics computed for different templates in the bank, which alter the uncorrelated SNR-max distribution. Thresholds for specific false positive rates determined in Ref.~\cite{Panelli2020} overestimate the thresholds due to the effect of correlated templates in a template bank. One of the goals of this paper is to determine the thresholds correctly.

The structure of this paper is as follows. Sec.~\ref{sec:SNRprobsetup} reviews the SNR and SNR-max statistics in the context of detecting transients and introduces notation used in the rest of the paper. In Sec.~\ref{sec:signaltempgen}, we discuss
various aspects of generation of template banks in the network searches for transients. This is illustrated with a pedagogical toy problem of a highly-symmetric planar network and with a more practical GPS.DM search. Strategies for generating optimal template banks are discussed in Sec.~\ref{Sec:StrategiesBankChoice}.
Sec.~\ref{sec:snrcorr} focuses on correlations between SNRs for different templates and introduces template bank covariance matrix.  Computations of the template bank covariance matrix are illustrated with the toy problem and with the GPS.DM search. Sec.~\ref{Sec:MAXSNR} derives the SNR-max probability distribution (PDF) and cumulative distribution (CDF) distribution for several practical yet analytically treatable cases:
for a fully-orthogonal template bank, a nearly orthogonal template bank, and for  template bank "squeezed" about the mean value of covariance matrix elements. Sec.~\ref{sec:thresholds} discusses the effect on thresholds when the template bank is fully-orthogonal, a special case of $\Nt=2$ template bank, a nearly-orthogonal bank, and a ``squeezed'' bank. Lastly, Sec.~\ref{sec:conclusion}  draws the conclusions and 
addresses the utility of  networks with entanglement shared by geographically-separated nodes. The paper contains several appendices where we present detailed derivations for some of the paper results.

\section{SNR and SNR-max statistics}
\label{sec:SNRprobsetup}
 In the previous work~\cite{Panelli2020} by our group, we explored the properties of the SNR statistic in the context of networks of quantum sensors.  In this section we set up the problem, review relevant results from Ref.~\cite{Panelli2020}, and then introduce the SNR-max detection statistic. 
 
 Throughout this paper, we assume that the network data streams are stationary and Gaussian-distributed. The sensor noise can be colored and the noise can be correlated across the network. For an atomic clock network, the cross-network correlation comes from referencing every clock phase to that of  ``reference'' clock common to all clocks~\cite{Roberts2017}. Notice our assumption of the noise stationarity; for clocks this implies working with frequency data instead of phase (time or time bias as recorded by the clock) data. 
 
Network data streams $\mb{d}$ are comprised of sensor measurements $d_i^a$, where the subscript $i$ refers to the time stamp (time $t_i$)  and superscript $a$ identifies a specific sensor with the total number of sensors $\Ns$. The time grid is assumed to be uniform with a step size $\Delta_t$. We will refer to a time interval $[t_i, t_{i+1})$ as an epoch $i$.

 In the most basic search for transient signals,  network data streams $\mb{d}$ are partitioned into data windows of fixed length $\Jw$ and the data inside each window are matched to a bank of  $\Nt$ templates. We enumerate windows with an index $w$ and refer to the portion of the data stream within a specific data window as $\mb{d}^w$. The template bank contains multiple signal templates, $\mb{s}_k$, $k=\overline{1,\Nt}$. The templates must reflect  underlying physics of the sought signals. The length of each template in time domain is equal to that of the data window $\Jw$. For a network, each template is represented by an $\Ns \times \Jw$ matrix.

 As a concrete example of physics-motivated generation of a signal template bank, in Sec.~\ref{sec:signaltempgen} we discuss the bank used in the GPS.DM search. Briefly, in the GPS.DM search we probe data streams for sweeps by  walls of DM bubbles, with the assumption that the radius of the DM bubble is much larger than the network spatial extent. Then the network is swept by planar walls. These hypothesized walls are to sweep through the network with velocities within a range prescribed by the SHM~\cite{Bovy2012}. Each template is distinguished by the specific values of a DM wall incoming velocity and orientation. We typically use $\Nt=1024$ templates in the bank to ensure a sufficiently dense coverage of the sweep parameters.  There are nominally $\Ns = 32$ atomic clocks on-board GPS satellites populating orbits on a spherical shell of diameter $D \sim 5 \times 10^4\, \mathrm{km}$. The atomic clock time readings (biases) are sampled every $\Delta_t= 30 \, \mathrm{s}$ and the length of data window $\Jw=61$ is determined by the ratio of the maximum network sweep time to the sampling time,  
 $(D/v_\mr{min})/\Delta_t$, where the minimum speed $v_\mr{min} \approx 25 \, \mathrm{km/s}$. The minimum value of the speed is determined by the requirement that the spatial network configuration does not change substantially over the duration of the sweep, since the orbital velocity of the satellites is $\sim 5\,\mr{km/s}$.
 
The sought physics signal can be parameterized as $h \mb{s}_{k}$, with $h$ being the amplitude or the strength of the signal. $h$ is to be estimated from the match to the template. 
 Typically, $h$ encodes the coupling of the sensor to the exotic physics.
 Formally, the signal strength estimator
for a specific window $w$ and a template $k$  reads~\cite{Panelli2020}
\begin{equation}
  \hat{h}^{w}_k=\frac{\left(  \mathbf{d}^{w}\right)^{T}\mathbf{E}^{-1}%
\mathbf{s}_k}{\mathbf{s}_k^{T}\mathbf{E}^{-1}\mathbf{s}_{k}}\,,  \label{Eq:hhat}
\end{equation}
where $\mb{E}^{-1}$
is the inverse of the network noise covariance matrix $\mb{E}$, discussed in Sec.~\ref{Sec:NoiseCovMatrix}.

Further, the standard deviation of the estimated signal amplitude, Eq.~(\ref{Eq:hhat}), is~\cite{Panelli2020}
\begin{equation}
\sigma_{h,k}=\frac{1}{\sqrt{\mathbf{s}_k^{T}\mathbf{E}^{-1}\mathbf{s}_k}}\,.
\label{Eq:sigmah}
\end{equation}
This uncertainty does not depend on the measurement data but it does depend on a specific template.
Finally, the SNR, $\rho^w_k$, in window $w$ for a template $k$, is simply the signal strength estimator (\ref{Eq:hhat}) divided by the standard deviation of the signal strength (\ref{Eq:sigmah}),%
\begin{equation}
\rho^{w}_k\equiv\frac{\hat{h}^{w}_k}{\sigma_{h,k}} = \frac{ \mathbf{s}_k ^{T}\mathbf{E}^{-1}\mathbf{d}^{w}
}{\sqrt{\mathbf{s}_k^{T}\mathbf{E}^{-1}\mathbf{s}_k}} .
\label{Eq:snr}
\end{equation}

As shown in Ref.~\cite{Panelli2020}, in the absence of a signal (i.e. when $\bm{d}^w \equiv \bm{n}^w$, noise of sensors), the SNR statistic is Gaussian-distributed with zero mean and variance of 1,
\begin{equation}
\rho^{w}_k \sim \mathcal{N}(0,1).
\label{Eq:snrdist}
\end{equation}
Here $\mathcal{N}(\mu,\sigma)$ is the normal distribution with mean $\mu$ and variance $\sigma^2$.
This can be seen by examining the SNR definition (\ref{Eq:snr}). The denominator, $\sqrt{\mathbf{s}_k^{T}\mathbf{E}^{-1}\mathbf{s}_k}$, is a constant as the template and the network noise covariance matrix are fixed, while the numerator $\mathbf{s}_k ^{T}\mathbf{E}^{-1}\mathbf{d}^{w}$ is a linear combination of Gaussian random variables. 
Thereby, the SNR is also a Gaussian random variable with mean $\left\langle
\rho^w\right\rangle \propto \langle
\mb{n}^w \rangle = 0$.  The SNR variance is $\left\langle
(\rho^w_k)^{2}\right\rangle = \mathrm{Var}((\mb{n}^w)^{T}\mb{E}^{-1}\mb{s_k})/\mb{s}_k^{T}\mb{E}^{-1}\mb{s}_k$.  Ref.~\cite{Panelli2020} (Appendix B) shows that $\mathrm{Var}((\mb{n}^w)^{T}\mb{E}^{-1}\mb{s}_k) = (\mb{s}_k)^{T}\mb{E}^{-1}\mb{s}_k$, which results in an SNR variance  $\left\langle
(\rho^w_k)^{2}\right\rangle = 1$.

Eq.~(\ref{Eq:snrdist}) means that given multiple realizations of the signal-free data streams, the values of SNR $\rho_k^w$ for a fixed template $\mb{s}_k$ are distributed as
\begin{equation}
    p_{\mr{SNR} }( \rho_k^w ) = \frac{1}{\sqrt{2\pi}} e^{-(\rho_k^w)^2/2} \,.
    \label{Eq:snrdistExplicit}
\end{equation}
The above definitions and properties of the SNR statistic hold for a general case of colored noise and multi-sensor networks.
If the subsequent windows overlap, then the SNR values for the two  windows are correlated as they are based on partially overlapping data. Even then the SNR probability distribution in individual windows still follows Eq.~(\ref{Eq:snrdistExplicit}). This can be proven by marginalizing joint probability distribution (multi-variate normal distribution) over SNRs in all other overlapping windows.

So far we reviewed the properties of the SNR statistic for a single template.
Practical applications involve  banks containing multiple templates. 
For a given  window $w$, all $\Nt$ individual template SNRs are compared and the maximum of the {\em absolute} values of the individual template SNRs is used as a detection statistic, 
\begin{equation}
 z^w \equiv \max( \vert \rho _{1}^w  \vert  ,.., \vert \rho _{\Nt}^w \vert  ) \,. \label{Eq:SNRmaxDefinition}
\end{equation}
Following the literature~\cite{Owen1996,Babak2008new,Apostolatos1995}, we refer to this quantity as the {\em SNR-max statistic}. To simplify notation, from this point forward we drop the window superscript $w$, so that the above definition reads  $z \equiv \max( \vert \rho _{1}  \vert  ,.., \vert \rho _{\Nt} \vert  )$.

Matching the same data stream to multiple templates
induces cross-template correlations in the SNR statistics, $\langle  \rho_{i} \rho_{j} \rangle \ne 0$. In addition, comparing SNRs and taking their maximum per Eq.~(\ref{Eq:SNRmaxDefinition}) suppresses smaller values of SNR. 
This suppression and  cross-template correlations substantially modify the single template SNR distribution (\ref{Eq:snrdistExplicit}). It is the focus of our paper to quantitatively characterize  the  SNR-max probability and cumulative distributions. Specifically, we are interested in a practically important threshold value for false positives. We remind the reader that the detection can be claimed when the value of chosen statistic exceeds the detection threshold~\cite{Allen2021}. The threshold values for  SNR-max are required to ensure (at a given confidence level) that the observed SNR-max values exceeding this threshold are due to probed physics and not due to randomness.  Our previous analysis~\cite{Panelli2020} did not take into account that the detection statistics may be correlated by templates or that our statistic is no longer Gaussian-distributed. 
Here we present a more sophisticated analysis to improve upon the treatment of Ref.~\cite{Panelli2020}. Before proceeding to our analysis, below we review relevant properties of the network noise covariance matrix $\mb{E}$, entering the SNR definition~(\ref{Eq:snr}).

\subsubsection{Network noise covariance matrix $\mb{E}$}
\label{Sec:NoiseCovMatrix}
Network noise covariance matrix $\mb{E}$ is defined by matrix elements 
\begin{equation}
E_{(al)(bm)} = \left\langle n_l^a n_m^b\right\rangle \,.
 \label{Eq:Ecov}
\end{equation}
This definition involves averaging over realizations of signal-free data $n_l^a$, i.e. intrinsic noise of the sensors. Without loss of generality, we assume that the noise has zero mean, $\langle n_l^a \rangle =0$.
In our index convention, the letters at the beginning of alphabet $a,b,\ldots$ enumerate the sensors (range $\overline{1,\Ns}$) , while letters in the middle of alphabet $l,m,\ldots$ --- the epochs (index range $\overline{1,\Jw}$ in a data window). We also introduced a compound index notation $(al)$, which organizes sensor-epoch data points into a vector, i.e. we unroll a $\Ns \times \Jw$ matrix into a 1D array.  For example, 
for the first window, $\mathbf{d}^{w=1} = (d_1^1, d_2^1, \ldots d_\Jw^1,  d_1^2, d_2^2, \ldots d_\Jw^2,\ldots d_1^{\Ns}, d_2^{\Ns}, \ldots d_\Jw^{\Ns})^T$. We will use Greek letters for such compound indexes, e.g. Eq.~(\ref{Eq:Ecov}) can be rewritten as $E_{\alpha \beta} = \left\langle n_\alpha n_\beta\right\rangle$.
We remind the reader that the covariance matrix is symmetric and positive semi-definite.

As an illustration, for a network of white-noise sensors without a common reference sensor, the noise covariance matrix  is diagonal, 
\begin{equation}
    E_{(al)(bm)} = \sigma^{2}_{a}\delta_{ab}\delta_{lm} \, ,
    \label{Eq:whitenetEcovnoref}
\end{equation}
with $\sigma_a^2$ being the noise variance for sensor $a$. The matrix elements of its inverse are
\begin{equation}
    \left(E^{-1}\right)_{(al)(bm)}=\frac{1}{\sigma_a^{2}} \delta_{lm}\delta_{a b}\, ,
    \label{Eq:whitenetEinvcovNoRef}
\end{equation}
i.e., $\mathbf{E}^{-1}$ is a diagonal matrix.

The same network but referenced to a common sensor (variance $\sigma_{R}^{2}$) has the noise covariance matrix
\begin{equation}
E_{(al)(bm)}  = (\sigma^{2}_{a}\delta_{ab} + \sigma_{R}^{2})\delta_{lm}\,.
\label{Eq:whitenetEcov}
\end{equation}
Compared to the no-reference sensor case, Eq.~(\ref{Eq:whitenetEcovnoref}), the reference sensor introduces off-diagonal matrix elements.
When all the sensors, except for the reference sensor, are assumed to have the same variance $\sigma^2$,
 the inverse of the covariance matrix $\mb{E}^{-1}$ can be found analytically~\cite{Panelli2020} 
\begin{equation}
    \left(E^{-1}\right)_{(al)(bm)}=\frac{1}{\sigma^{2}} \delta_{lm}\left(\delta_{a b}-\frac{1}{\Ns} \frac{\xi}{1+\xi}\right) \, ,
    \label{Eq:whitenetEinvcov}
\end{equation}
where the parameter
\begin{equation}
\xi = \Ns \sigma_{R}^{2}/\sigma^2 \label{Eq:xi-parameter}
\end{equation}
has the meaning of the square of the ratio of network-averaged sensitivity ($\propto \sigma/\sqrt{\Ns}$) to the reference sensor sensitivity ($\propto \sigma_R$). We will refer to $\xi$ as the relative reference-sensor sensitivity parameter.

It is instructive to consider a limiting case of a single sensor ($\Ns=1$) data stream with white noise of zero mean  and variance $\sigma^{2}$. In this case, the inverse of the covariance matrix is simply
$
\mathbf{E}^{-1}=\sigma^{-2}\mathbf{I},%
$
where $\mb{I}$ is the $\Jw\times \Jw$ identity matrix. Further, Eqs.~(\ref{Eq:hhat}), (\ref{Eq:sigmah}), and (\ref{Eq:snr}) 
reduce to
$
\hat{h}^{w}_k   =\left(  \mathbf{d}^{w}\cdot\mathbf{s}_k\right)  /{\left(
\mathbf{s}_k\cdot\mathbf{s}_k\right)  }$, 
$\sigma_{h}   =\sigma/{\sqrt{\left(  \mathbf{s}_k\cdot\mathbf{s}_k\right)
}}$, and 
$\rho^{w}_k   =\left(  \mathbf{d}^{w}\cdot\mathbf{s}_k\right)/%
{(\sigma \sqrt{\left(  \mathbf{s}_k\cdot\mathbf{s}_k\right)  })},%
$
where we used the conventional definition of the dot product of two vectors.

\section{Generation of template banks in the searches for transients}
\label{sec:signaltempgen}
Having defined the SNR and SNR-max statistics, now we formalize the procedure for template bank generation.
Template banks must reflect  signals prescribed by the underlying physics to be probed. While the signal templates are application-specific, some generic requirements and challenges can be illustrated with specific examples. 
First, in Sec.~\ref{Sec:ToyTemplateBankGeneration} we consider a ``toy'' 2D problem of a network search for transient signals induced by sweeps of randomly oriented lines  through a circular network of high-degree symmetry. This toy problem introduces a number of relevant issues specific for networks, such as the angular and speed resolutions.  
After understanding the simplified problem, in Sec.~\ref{Sec:signaltempgen:GPSDM}, we discuss a more practical case of  the GPS.DM search~\cite{Roberts2017,Panelli2020,Derevianko2014,Derevianko2018}.

\subsection{Toy problem: Circular topology network with $C_{\Ns}$ rotational symmetry}
\label{Sec:ToyTemplateBankGeneration}

\begin{figure}[h!]
	\begin{center}
		\includegraphics[width=0.42\textwidth]{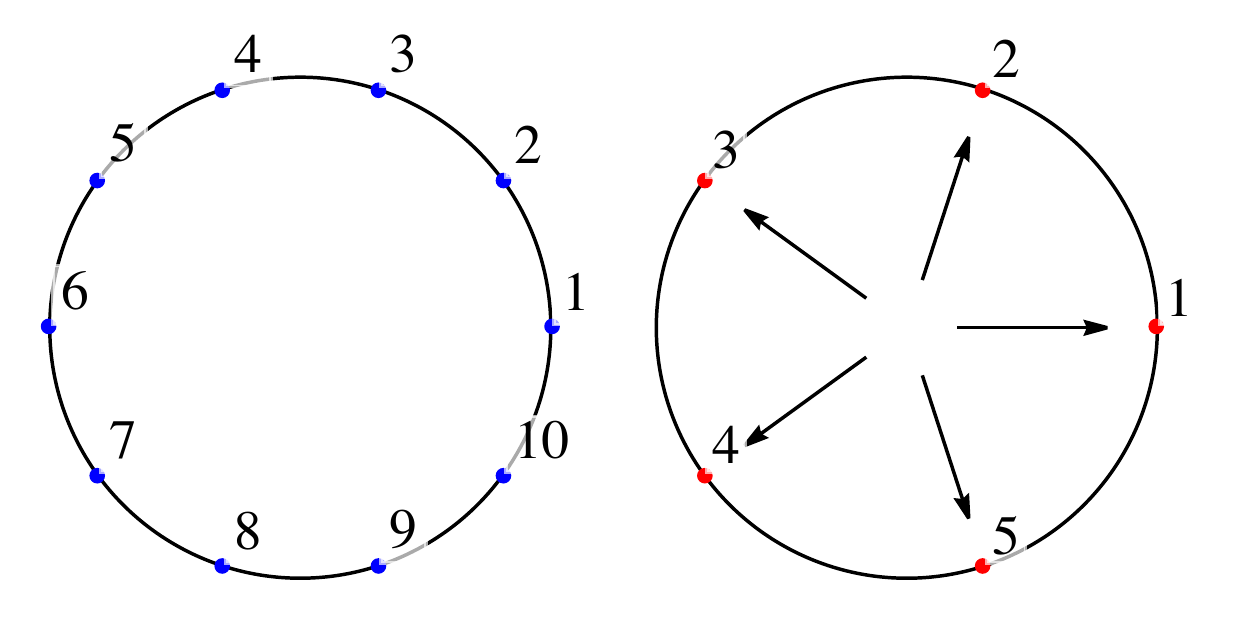}
	\end{center}
	\caption{(Color online) (Left panel) A planar network of $\Ns=10$ sensors with $C_{10}$ symmetry. (Right panel) Graphical representation of a bank of $\Nt=5$ templates. Arrows indicate direction of the line sweep $\hat{u}_i$. The arrows $\hat{u}_i$ are normal to the sweeping line. For example, template 1 corresponds to a vertically-oriented line sweeping from the left.}
	\label{Fig:tempplanarnet}
\end{figure}

As an illustration of generating template banks and their properties, we consider a planar network with a circular topology where identical point-like sensors are distributed evenly on a circle of radius $R$, see Fig.~\ref{Fig:tempplanarnet}.


Sensors' polar angles are given by 
\begin{equation}
    \phi_a = \frac{2\pi}{\Ns}(a-1),
    \label{Eq:azimuthangles}
\end{equation}
so that the sensor coordinates are
\begin{equation}
    \bm{r}_a = R(\cos\phi_a, \sin\phi_a )^T = R \hat{n}_a ,
    \label{Eq:sensorcoords}
\end{equation}
where we used the  coordinate system with the origin at the circle center and
$\hat{x}$ pointing towards sensor 1. We also introduced a unit vector $\hat{n}_a$ pointing towards sensor $a$. For example, for the network of $N_s=10$ sensors, shown in Fig.~\ref{Fig:tempplanarnet}, $\phi_1 = 0$ and $\phi_2 = \pi/5$.
This arrangement of sensors has the $C_{\Ns}$ cyclic group symmetry, as rotating the network by multiples of the angle $2\pi/\Ns$ about the circle center maps the network onto itself. In this toy problem, we do not include effects of the reference sensor.

We note that this 2D network topology is of relevance to the configuration of satellites in the GPS constellation. The satellites in the GPS network are arranged into six equally-spaced orbital planes with an inclination of $55^{\circ}$, and each nearly-circular orbit is populated with 4-5 satellites.
Because the satellites are not evenly spaced apart within each orbit however, the $C_{\Ns}$ symmetry is broken. Still, understanding our toy network provides a number of useful insights applicable to general network topology.

With the toy network topology specified, now we generate a template bank for sweeps of this 2D network by ``lines''. These lines can be thought as 2D cross-sections of thin DM walls. The lines sweep through the network, leaving a geographically distributed imprint on the sensor signals. The sensor is considered to be affected in a given epoch $l$, when the line crosses the sensor in the time interval $[t_l,t_l + \Delta_t)$. 

In our 2D example, the velocities $\bs{v}$ of the lines lie in the network plane. However, the component of the velocity parallel to the line leads to an unobservable translation of the line along itself. Different templates are then uniquely specified by the direction and value of velocity component $\bs{v}_{\perp} = v_{\perp} \hat{u}$, where $\hat{u}$ is the unit vector perpendicular to the line and pointing in the direction of the sweep.
We will use polar angle $\theta$ in the network reference frame to characterize incidence direction $\hat{u}$. For example, template bank in Fig.~\ref{Fig:tempplanarnet}
samples $M=5$ directions with $\theta_k = (2 \pi/M) \, (k-1)$, $k=\overline{1,\Nt}$. 
For simplicity, we assume that all the directions $\hat{u}$ are equally probable. The bank should provide an adequate sampling of both incidence directions $\hat{u}$ and speeds $v_{\perp}$.

Templates can be generated based on the following argument.
Suppose $t_O$ and $t_a$ are the moments of time when the line reaches the network center and the sensor $a$, respectively. The difference between these two times is 
\begin{equation}
    t_a - t_O = (\hat{u} \cdot \hat{n}_a ) \frac{R}{v_{\perp}}.
    \label{Eq:timedifference}
\end{equation}
Since our time sampling is finite, the sensor signal time series is simply $s_l^a = \delta_{l,l_a}$, where the epoch $l_a$ is given by
\begin{equation}
    l_a = \floor*{(\hat{u} \cdot \hat{n}_a) \frac{R}{v_{\perp}\Delta_t}} + l_O.
    \label{Eq:epochsata}
\end{equation}
Here $\floor{x}$ is the floor function and $l_O = (\Jw +1)/2$ is the center of the data window.


\begin{figure}[h!]
	\begin{center}
		\includegraphics[width=0.42\textwidth]{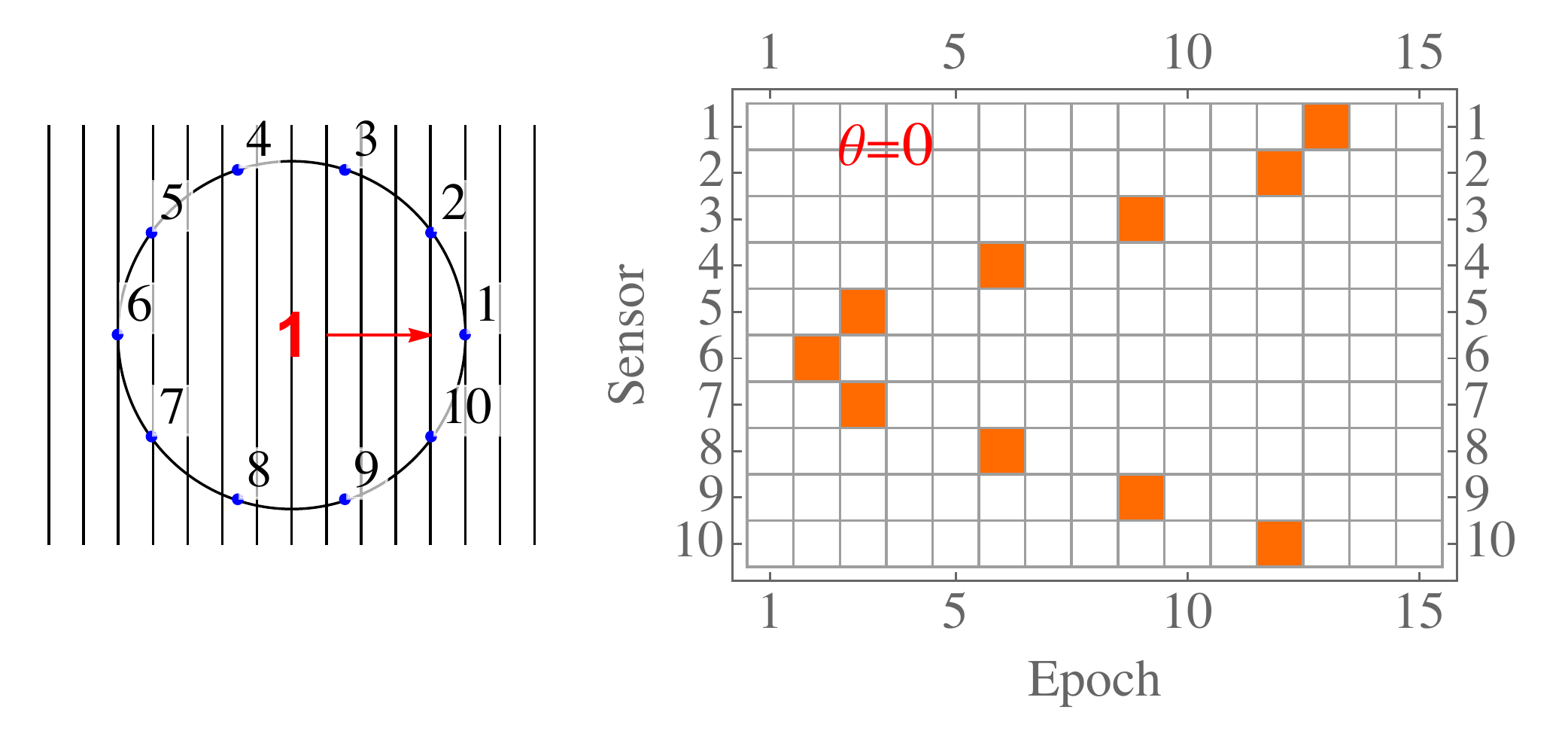}
	\end{center}
	\caption{(Color online) Construction of the template corresponding to a vertically-oriented line sweeping from the left (polar angle $\theta=0$). Sensor 6 is affected first and sensor 1 --- last. Here we used $R/(v_{\perp}\Delta_t) = 5.01$ and the number of epochs $\Jw=15$. The center of the network is crossed in the $l_O=7$ epoch. }
	\label{Fig:tempplanarnetslice}
\end{figure}


Fig.~\ref{Fig:tempplanarnetslice} presents a construction of a template for a vertically-oriented line sweeping our network from the left. This is template 1 from the template bank shown in Fig.~\ref{Fig:tempplanarnet}. The left panel shows slicing of the network
by the sweeping line with the spatial slice width $v_\perp \Delta_t$. Slice boundaries correspond to the sweeping line snapshot at the beginning and the end of a given  epoch, e.g. the left-most slice is epoch 1.

Since $-1 \le \hat{u} \cdot \hat{n}_a \le 1$ in Eq.~(\ref{Eq:epochsata}), the constructed template spreads over 
\begin{equation} 
  \Delta l = 2R/(v_{\perp}\Delta_t) \label{Eq:TemplateSpread} 
\end{equation}
epochs. Clearly, the number of epochs in the window $\Jw$ should be greater than $\Delta l$, otherwise a useful portion of the signal is cut out. For fixed network geometry and sampling time $\Delta_t$, $\Delta l$ increases indefinitely as $v_{\perp}$ approaches small values. Practical considerations, however, introduce a cutoff at small $v_{\perp}$. For example, GPS satellites positions change with time due to their orbital motion; this limits $v_{\perp}$ from below, see Sec.~\ref{sec:SNRprobsetup}.

If $v_{\perp} > (v_{\perp})_{\mathrm{max}} = 2R/\Delta_t$, all the sensors are affected in the same epoch, so the template width $\Delta l = 1$. $(v_{\perp})_{\mathrm{max}}$ is the speed at which the line sweeps through the entire spatial extent (diameter = $2R$) over the duration of the sampling interval $\Delta_t$. In this case, the network losses directional and velocity resolutions. That is for any incident line orientation and any $v_{\perp} > (v_{\perp})_{\mathrm{max}}$, all the templates are identical, collapsing into a vertical single-epoch column at $l_a = l_O$ in Fig.~\ref{Fig:tempplanarnetslice}. Underlying physics may constraint $(v_{\perp})_{\mathrm{max}}$ to lower values, see, e.g., discussion of the GPS.DM search in Sec.~\ref{Sec:signaltempgen:GPSDM}, where $(v_{\perp})_{\mathrm{max}}$ is determined by the DM velocity distribution cutoff.

As $v_{\perp}$ decreases, the template width $\Delta l$ increases and different sensors begin to be perturbed at different epochs. On average, we expect ${\Ns}/{\Delta l}$ sensors to be affected at the same epoch. We refer to such groups of sensors as being degenerate and introduce the average degeneracy factor
\begin{equation}
    \bar{g} = \Ns \frac{v_{\perp} \Delta_t}{2R}.
    \label{Eq:gbar}
\end{equation}
This degeneracy factor is a rule of thumb.
For example, for the bank and network parameters in Fig.~\ref{Fig:tempplanarnetslice}, from Eq.~(\ref{Eq:gbar}) we expect $\bar{g} = 1$, while we observe mostly two-fold degeneracies in the resulting template shown in Fig.~\ref{Fig:tempplanarnetslice}. This is a consequence of the line propagation direction $\hat{u}$ being aligned with one of the network axes of symmetry.
Generically, in the limit of small degeneracy factors, $\bar{g} \le 1$, we expect all the sensors to be perturbed at different epochs.

To generate the template bank we need to sample continuous parameter space of directions $\hat{u}$ and normal speeds $v_\perp$. This leads to the question of angular and velocity resolutions of our network. 

We start with the angular resolution. Consider two sweeping lines which have 
the same $v_\perp$ but differ by their angles of incidence: $\theta$ and $\theta'= \theta + \Delta_\theta$. Because both the number of sensors and time sampling are finite, for sufficiently small relative tilt angles $\Delta_\theta$, sweeps by both lines 
produce the same template. As we increase the tilt, there is some critical value 
of $\Delta_\theta$ when the two sweeps start mapping onto two distinct templates.
Then the two incident directions can be resolved in the network search.
The critical value of the tilt determines the network angular resolution.

As an illustration, consider the template construction in Fig.~\ref{Fig:tempplanarnetslice} for a line sweeping from the left, $\theta =0$. 
Templates for tilted sweeps can be generated by rotating the network sensors counter-clockwise by the angle $\Delta_\theta$ about the circle center. Since the sensors 5 and 7 are near the edge of epoch 3 slice (and sensors 2 and 10 are near the edge of epoch 10), even small tilts $\Delta_\theta>0$ move sensor 7 into epoch 4 and sensor 2 into epoch 11. However, sensors being at the slice edge is an exceptional situation. Generically, the sensors are more likely to be found in the middle of epoch slices. Then the angular resolution 
is determined by the maximally separated sensors within the same slice, i.e. sensors swept within the central epoch. This leads to the worst-case-scenario angular resolution estimate,
\begin{equation}
    \Delta_\theta \approx \sin^{-1} \left( \frac{ v_\perp \Delta_t}{2R} \right) \,. 
    \label{Eq:AngularResolution}
\end{equation}
Recalling the relation~(\ref{Eq:TemplateSpread}) for the template width $\Delta l$, we observe that 
$ \sin \Delta_\theta = 1/\Delta l$. When the template collapses into a single column due to high values of normal velocity, $\Delta l =1$, and the angular resolution is lost  ($\Delta_\theta \rightarrow \pi/2 $) in agreement with our previous qualitative statements. In the opposite limit of $\Delta l \gg 1$,  $\Delta_\theta \approx 1/\Delta l$. For the template of Fig.~\ref{Fig:tempplanarnetslice}, $\Delta_\theta \approx 1/12 \, \mathrm{radians}$. Full, resolution-limited, coverage of the  incidence angles by the bank then requires $\Nt= 2 \pi/\Delta_\theta \approx 75 $ templates.  Having degenerate (affected in the same epoch, but spatially separated) sensors improves the angular resolution. It is worth emphasizing that the derived angular resolution is a pure consequence of the line-network encounter geometry and it does not account for the intrinsic sensor noise, see Refs.~\cite{Roberts2018a,Panelli2020}, which will worsen the angular resolution~(\ref{Eq:AngularResolution}).

Angular resolution of a geographically-distributed network has been considered in several papers. For example, 
in Ref.~\cite{dailey2020ELF.Concept}, there is a brief discussion on the angular resolution for bursts of exotic fields propagating through a network of quantum sensors. Their discussion suggests that the angular resolution is roughly the ratio of the temporal resolution $\Delta_t$ to the field burst propagation time through the sensor network ($2R/v_\perp$), in agreement with 
our formula (\ref{Eq:AngularResolution}) in the limit of high sampling rates, $\Delta_t \ll 2R/v_\perp$. We also refer the reader to papers~\cite{Monnier2003,Romano2017} discussing ``Rayleigh criterion'' angular resolution for a network of interferometers.

Now we find the resolution in normal velocities $v_\perp$. We apply the same logic of finding the minimum increment $\Delta_v$ so that the sweeps with two velocities $v_\perp$ and $v_\perp + \Delta_v $ map into two distinct templates. 
Since the number of sensors and time sampling is finite, vanishingly small variations in the normal velocity would produce identical templates. A distinct new template is generated when the normal velocity is increased by a critical value $\Delta v$,
\begin{equation}
    \Delta_v \approx \frac{v_{\perp}^2 \Delta_t}{2R}.
    \label{Eq:VelocityResolution}
\end{equation}
This velocity resolution improves for slower sweeps, as the template spreads over more epochs, see Eq.~(\ref{Eq:TemplateSpread}).

Most of our observations are transferred to the GPS constellation in the following section.
We will return to this toy problem in the context of computing template bank covariance matrices in Sec.~\ref{Sec:snrcorr:ToyProblem}.

\subsection{GPS.DM search}
\label{Sec:signaltempgen:GPSDM}

Signal templates in the GPS.DM search are determined by the particular DM model used (monopoles, strings, domain walls, etc...) and the parameters associated with the event. For concreteness, we focus on DM wall events, which are observed as a sweep through the geographically-distributed network, with a given sensor perturbed when the wall overlaps with the sensor. The relevant parameters for a particular event are the speed, incident direction, time of the event, and the thickness of the  wall. As in the toy problem, only the normal component of the sweep velocity matters, however the velocity vector now spans 3D. This parameter space is continuous and it is necessary to discretize it to keep the number of templates finite. We discuss the discretization strategies  in Sec.~\ref{Sec:StrategiesBankChoice}.

\begin{figure}[h!]
	\begin{center}
		\includegraphics[width=0.4\textwidth]{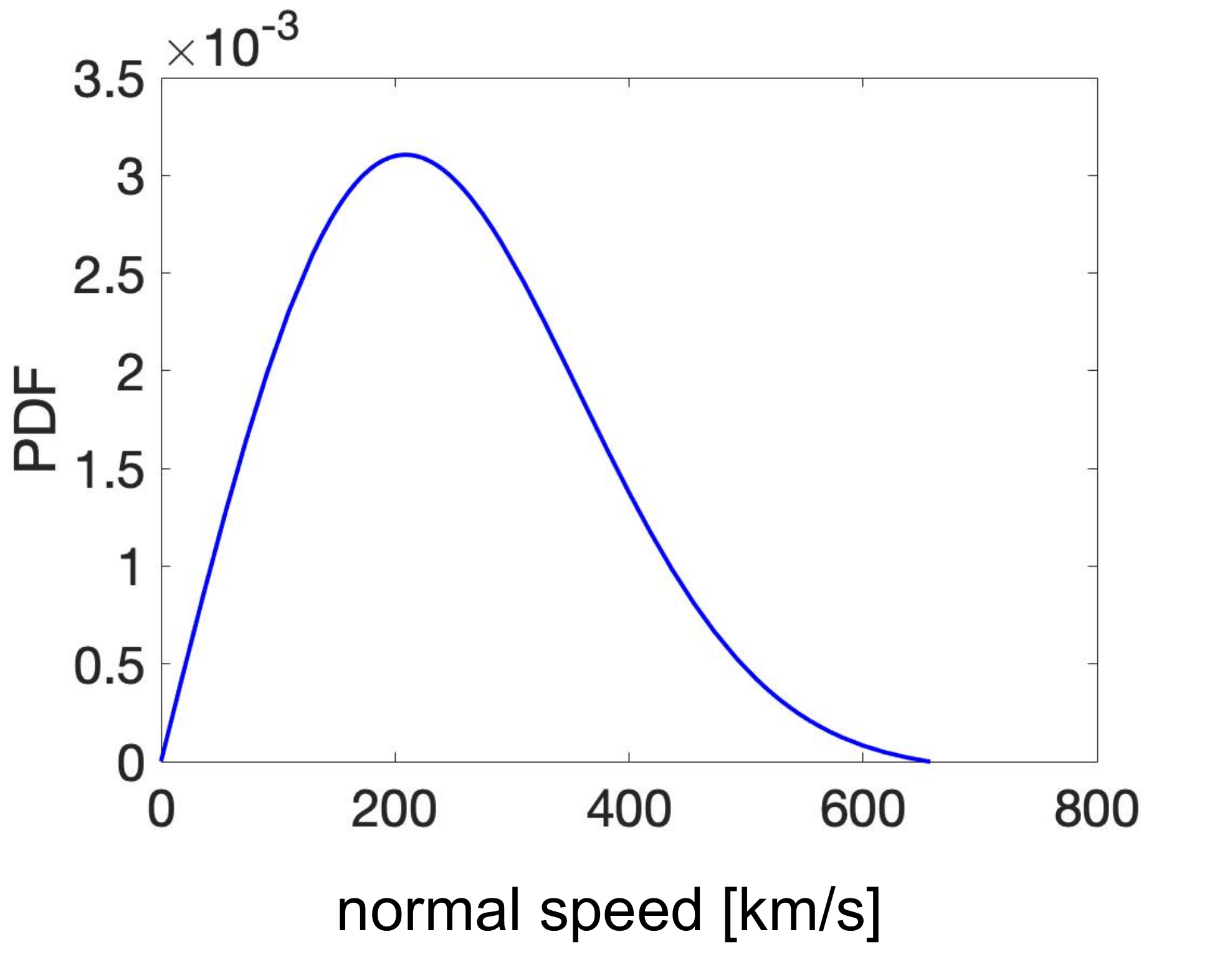}
	\end{center}
	\caption{ (Color online)
	Normal speed $v_\perp$ probability distribution~\cite{Roberts2017} for galactically bound DM wall objects that cross the network. The distribution has a maximum at $v_{\perp} = 209 \, \mathrm{km/s}$ and a cut-off at $v_\perp = 760 \, \mr{km/s}$. }
	\label{Fig:normalvelocitydist}
\end{figure}

The standard halo model (SHM) dictates the necessary parameter prior distributions. For DM objects the velocity distribution in the halo frame is quasi-Maxwellian and isotropic with dispersion $v \approx 300 \, \mathrm{km/s}$~\cite{Bovy2012} and with a cutoff set by the escape velocity of $v_\mr{esc}  \approx 550 \, \mathrm{km/s}$.
Additionally, the solar system travels through the halo at galactic velocities of $v_g \approx 200 \, \mr{km/s}$ toward the Cygnus constellation,
implying that over $90\%$ of DM sweeps come from the forward facing hemisphere~\cite{Roberts2017}. In the GPS.DM search, we are interested in the normal velocity of the DM wall in the Earth Centered Inertial (ECI) frame. In Fig.~\ref{Fig:normalvelocitydist} we reproduce the normal velocity distribution for DM walls bound by the galaxy in the ECI frame~\cite{Roberts2017}. The maximum normal velocity, $v_{\perp}$, in this frame is the addition of the DM escape velocity and the velocity of the solar system, $v_{\perp} < v_\mr{esc} + v_g \approx 770 \, \mr{km/s}$. As previously stated, the minimum normal speed that can be resolved must be larger than the velocities of the GPS satellites $v_a \approx 5 \, \mr{km/s}$. This comes from the requirement that we can fix the positions of orbiting satellites for a given time window. Similar to the toy problem of Sec.~\ref{Sec:ToyTemplateBankGeneration}, the event parameters determine in which order the sensors are swept and the overall sweep time. These characteristics  distinguish the templates within the signal template bank.

The GPS data-streams from satellite clocks are given as time measurement biases with with respect to a fixed terrestrial reference sensor (clock) $R$. DM-induced sensor bias (phase) of a given sensor $a$ is proportional to an integral of the DM-induced frequency shift ~\cite{DerPos14}. Since the clock biases are a subject to a random-walk noise, we whiten the data by differencing the data streams. 

Both satellite and reference clocks are affected by the DM sweeps. In general, sensitivity of the clock to the variation of fundamental constants depends on the clock atom. The GPS 
constellation employs mostly Rb and a few Cs clocks. A precision reference clock can be either an H-maser or a Rb clock (see Ref.~\cite{Roberts2017} for details) and one can choose among several terrestrial clocks to serve as the reference clock. To simplify our consideration, we assume that we are only dealing with a network of Rb clocks (including the reference clock). Then the DM object affects each clock with the same signal amplitude $h$. This is the limiting case of a homogeneous network. To bring out
the essentials, we further focus on ``thin'' walls characterized by an interaction time with each sensor (clock) being much shorter than the sampling time interval $\Delta_t$. This reduces  the perturbation of atomic frequencies of each clock to Kronecker deltas.

The terrestrial reference clock is centered in the network and we fix its position in the center of our window of length $\Jw$. Indeed, 
the reference clock is located on the Earth, while the satellite clocks are orbiting approximately 20 thousand kilometers above the Earth. $\Jw$ is chosen to be an odd number, so that the reference clock is perturbed in the epoch $l_R = (\Jw+1)/2$. 
Then the sought signal in the differenced data stream 
reads 
\begin{equation}
     s_{l}^{a} = \delta_{l,l_a} + (-1)\delta_{l,l_R} \, ,
     \label{Eq:signaltemplate}
\end{equation}
where $l$ enumerates epoch in the data window and  $l_a$ and $l_R$ are the epochs which the satellite clock and reference clock interact with the DM object, respectively. 

In general, the order in which the sensors are swept by the DM object depends on the geometry of the network. In our setup,  no matter the directionality of the DM object some of the  satellite sensors are affected prior to the reference clock. However, for a terrestrial network, or when the reference clock role is assigned to one of the satellite clocks, one can not ascertain which sensors (reference or network) would be affected first in the window. For this reason, we always set the reference sensor (on Earth or a satellite) to be affected in the center on the window, thus mitigating any possibility of missing information from a network sensor that is affected prior or after the reference sensor.   

\begin{figure}[h]
	    \begin{center}
		    \includegraphics[width=0.4\textwidth]{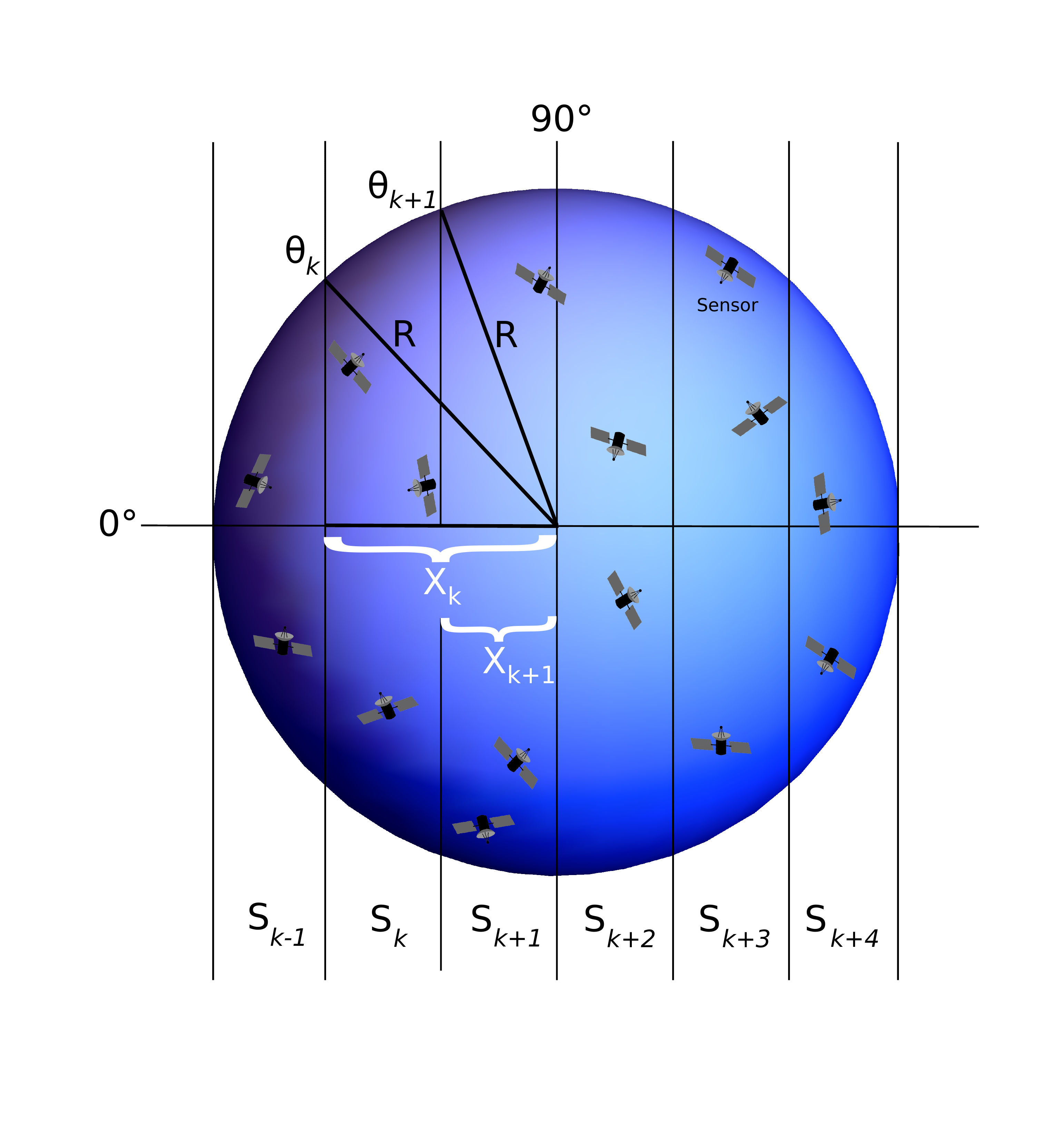}
	    \end{center}
	\caption{(Color online) A spherical shell network sliced by a thin wall propagating from the left. The wall is normal to the page. The shell is segmented into $\Delta l$ slices and each slice segment of the sphere $S_{k}$ is defined by the polar angles $\theta_{k}$ and $\theta_{k+1}$ spanning its width $L= v_\perp \Delta_t$. Angles are  $\theta_k = \cos^{-1}(X_k/R)$ with $X_{k} = \frac{2R}{\Delta l} (\frac{\Delta l}{2} -k)$. Reference clock is not shown.}
	\label{Fig:constellation}
    \end{figure}

The GPS.DM template construction can be visualized  similar to the toy problem. Fig.~\ref{Fig:constellation} shows a spherical shell of diameter $D=2R$ with sensors  distributed on the surface. A planar DM object passing through the network with normal velocity $v_{\perp}$ and time resolution $\Delta_{t}$ segments the sphere into $\Delta l = D/(v_\perp \Delta_t)$ slices. As in  
our toy problem, $\Delta l$ has the meaning of template spread. Each segment has a spatial width  $L = v_{\perp} \, \Delta_t$. The normal velocity of thin walls determines the number of equally spaced slices that cut the spherical shell. 

\begin{figure}[h!]
	\begin{center}
		\includegraphics[width=0.42\textwidth]{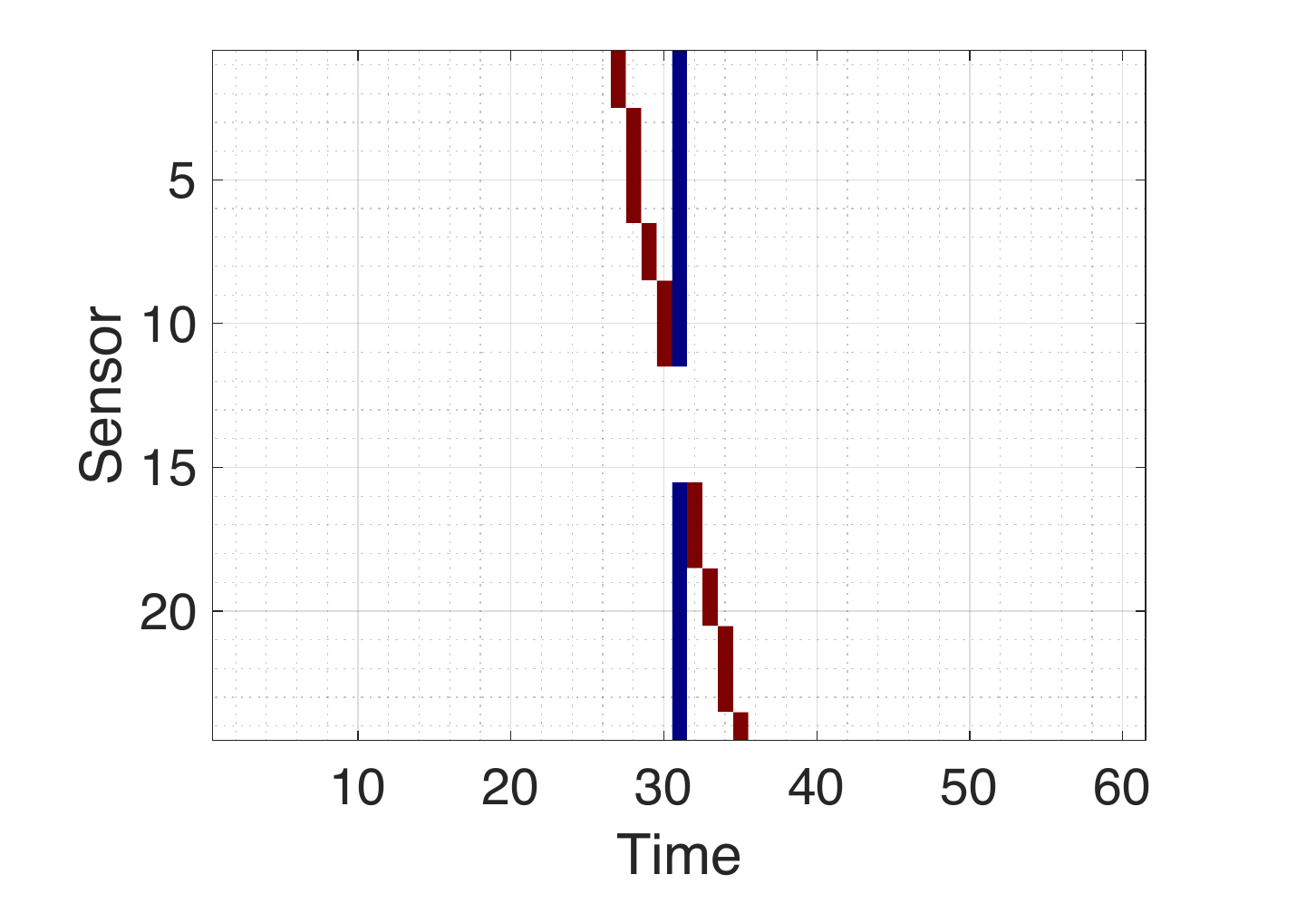}
	\end{center}
	\caption{(Color online) Template for a thin wall signal sweeping through the GPS constellation
	with the  normal speed of  $v_\perp=185 \, \mathrm{km/s} $. This network on January 3, 2012 has 
	$\Ns=25$  satellite clocks  and the reference clock. Data window size is $\Jw=61$.  When the wall interacts with the satellite clock, a positive spike of magnitude 1 (red tile) is observed. When the wall interacts with the reference clock, there is a   spike of opposite sign  (blue tile) at the epoch 31.}
	\label{Fig:templateexample}
\end{figure}

Fig.~\ref{Fig:templateexample} shows a signal template for a wall sweep with normal speed $v_{\perp}=185 \, \mathrm{km/s}$. The network has $\Ns = 25$  sensors and data window contains $\Jw=61$ epochs. The sensors are numbered in such a way that the sensor $a=1$ is swept first and the sensors are swept in increasing order. The slope of the blue line reflects the sweep speed. Notice, that once the sensors are numbered, the sweep from a different direction would result in a different pattern, with the straight red line converted into a dis-ordered sequence of red tiles, see, e.g., Fig.~\ref{Fig:tempplanarnetslice}. The blue vertical line would remain the same with some of the tiles (corresponding to satellites degenerate with the reference sensor) missing.


Our toy problem discussion (Sec.~\ref{Sec:ToyTemplateBankGeneration}) of template epoch spread, sensor degeneracy, and angular and normal speed resolutions applies to the GPS network search. The presence of the reference sensor requires refinement of  some of the considerations. For example, consider  the limiting case when  all the  satellite clocks are affected by the DM wall within the same $\Delta_t$ time interval (epoch) as the reference sensor. In this case, $l_a=l_R$ and the signal template~(\ref{Eq:signaltemplate})  collapses into a ``null'' template, where all elements $s_l^a$ of the sought DM signal stream are zero. This effectively  eliminates  detectable DM effects on the data stream. Even when only some of the satellites are degenerate with the reference sensor, part of the network sensitivity is effectively eliminated.  This effect can be seen in Fig.~\ref{Fig:templateexample}, where satellites 12 through 15 are degenerate with  the reference clock and lead to the lost (null) information from the satellites  and reference sensor (blank rows in Fig.~\ref{Fig:templateexample}). 

Another peculiarity of the GPS network is that the network nodes move due to the orbital motion of the satellites.  The spatial positions of the orbiting satellites is a part of the archival dataset~\cite{JPLwebpage} and this information is used in the GPS.DM search. In analytical estimates, however, a more tractable approach is to introduce a uniform occupancy approximation,  where the sensors  populate a sphere of radius $R$ surrounding Earth with a random (but uniform in probability) distribution of sensors on the sphere surface.  In this approximation, the probability for a sensor $a$ to reside on a spherical surface segment $S_{k}$  segment is simply $P(a \in S_{k}) =  \Omega_{k}/4\pi$, where solid angle $\Omega_{k}$ spans the segment $S_{k}$, see Fig.~\ref{Fig:constellation}. Explicit computations show that $ \Omega_{k}  = 4\pi/\Delta l$, 
so that $P(a \in S_{k}) = 1/\Delta l$, i.e. the probability is independent of the slice. The average number of sensors in a given slice is the total number of network sensor  $\Ns$ multiplied by the probability $P(a \in S_{k})$ for a sensor $a$ to reside on the segment $S_{k}$.
Thereby,  in this uniform occupancy approximation, the degeneracy factor introduced in Sec.~\ref{Sec:ToyTemplateBankGeneration}
 is the same as for the 2D toy problem,
\begin{equation}
    \bar{g} = \Ns \frac{v_{\perp} \Delta_t}{2R}.
    \label{Eq:gbar3D}
\end{equation}
As an illustration, template in Fig.~\ref{Fig:templateexample}
was generated for $\Ns=25$ sensor network and  sweep velocity $v_{\perp}=185 \, \mr{km/s}$ with realistic positions of GPS satellites. Based on Eqs.~(\ref{Eq:TemplateSpread},\ref{Eq:gbar3D}) we estimate a template spread $\Delta l \approx 9$ with a degeneracy of $\bar{g} \approx 3$ sensors per epoch. This is consistent with Fig.~\ref{Fig:templateexample}, which, depending on the epoch, exhibits between 1 to 4 degenerate sensors.  While not exact, the uniform occupancy is a reasonable approximation for the true GPS network geometry. 

The angular and speed resolutions remain the same as in the toy problem, Sec.~\ref{Sec:ToyTemplateBankGeneration}. The angular resolution  depends on just the normal speed of the DM wall, while the normal speed resolution is inversely proportional to the square of the normal speed,
\begin{align}
    \Delta_{\theta} & \approx \sin^{-1} \left( \frac{ v_{\perp} \Delta_t}{2R} \right) \,, \label{Eq:Dtheta} \\
    \Delta_v & \approx \frac{v_{\perp}^2 \Delta_t}{2R}.  \label{Eq:Dv}
\end{align}

Because the velocity distribution is peaked at $v_\perp^p \approx 209 \, \mr{km/s} $, we can introduce the most probable quantities, such as the  most probable template spread
\begin{equation}
      \overline{\Delta l} = 2R/(v_{\perp}^p\Delta_t)\,,  \label{Eq:AverageTemplateSpread}  
\end{equation}
which evaluates to $\overline{\Delta l}=8$ for GPS data.

\subsection{Strategies in choosing template banks}
\label{Sec:StrategiesBankChoice}
Choosing templates in the template bank is a trade-off between computational cost and the false-negative probability of missing the sought signal. Indeed, one would like to provide dense coverage of the transient parameter space to minimize false negatives, however increasing the number of templates $\Nt$ raises computational cost of the search. We address this question in this section and refine some of the statements further in the text.

There is a natural network resolution-limited choice of templates. 
For example, in the GPS.DM search for thin walls, Sec.~\ref{Sec:signaltempgen:GPSDM}, one may visualize the template parameter space (normal speed, polar and azimuthal angles of incidence for DM wall)  as points in 3D space. Network resolutions ($\Delta_v$ and $\Delta_\theta$, Eqs.~(\ref{Eq:Dv},\ref{Eq:Dtheta})) determine a small ``resolution volume'' enclosing each point in this continuous space. Points within the ``resolution volume'' lead to identical templates.
Thereby, the resolutions $\Delta_{\theta}$ and $\Delta_v$ effectively determine the grid
in this 3D parameter space. As an illustration, consider a thin wall sweeping the GPS constellation at the most probable normal speed $v_{\perp} = 209 \, \mathrm{km/s}$ (from Fig.~\ref{Fig:normalvelocitydist}). Then for $\Delta_t=30 \, \mathrm{s}$ and $2R \approx 5 \times 10^4 \, \mathrm{km}$, the normal speed resolution is $\Delta_v \approx 26 \, \mathrm{km/s}$ and an angular resolution in both polar and azimuthal angles of incidence is $\Delta_\theta \approx 7.2^{\circ}$. Complete resolution-limited coverage in normal speed coordinate requires 68 points  (with $ 25 \, \mr{km/s} \le v_\perp < 760 \, \mr{km/s}$; the grid is non-uniform because $\Delta_v$ depends on $v_{\perp}$). For a fixed value of speed $v_{\perp} = 209 \, \mathrm{km/s}$, 25 points for the polar angle and 50 points for the azimuthal angle are required for resolution-limited angular coverage. Covering the entire 3D cube in the template parameter space requires $\sim 2\times 10^4$ points. This is still computationally prohibitive in our experience.

Since the most important criteria in choosing the bank is reducing the probability of false negatives, the discussed resolution-limited bank construction can be further augmented by the importance sampling technique~\cite{alma991005880979706781}, where the parameter priors determine the density of sampling the template parameter space. For example, in the GPS.DM search, the normal velocity distribution, shown in Fig.~\ref{Fig:normalvelocitydist},
is a peaked distribution. Thus we need more points on the $v_\perp$ grid near the maximum of the distribution 
and less points in the tails to maintain the probability of false negatives uniform throughout the bank.
This can be accomplished by using a non-linear grid or using Monte-Carlo sampling. To optimize the resulting grid, we would further limit the spacing between the velocity grid points to be greater than the resolution $\Delta_v$.
The angular grids can be optimized with symmetry-adapted choice of template bank that takes advantage of the network symmetry, see Sec.~\ref{Sec:snrcorr:ToyProblem}.

 Ref.~\cite{CannonKipp2010Svda} addressed a similar question of bank optimization in the context of gravitational wave searches.
 They proposed to optimize the bank using the singular value decomposition technique. However, these authors dealt with a single sensor (not a network) and it remains to be seen if their technique can be generalized to the network search.

An issue related to the network resolution, is the degree of similarity between the templates. This degree of similarity will be quantified in the following Sec.~\ref{sec:snrcorr} with cross-template correlation coefficients. The generated bank can be optimized further based on this consideration. We also remind the reader that the SNR-max statistic is the quantitative measure governing the results of the search. The choice of template bank affects the SNR-max PDF, CDF, and threshold values for false positives. These characteristics of the SNR-max detection statistic is the focus of the remainder of this paper.

\section{Template bank covariance matrix}
\label{sec:snrcorr}
Consider a template bank comprised of $\Nt$ templates. 
Since the same data, $\mathbf{d}^{w}$, is used to compute the SNR~(\ref{Eq:snr}) $\rho_k$ for each template, $\mathbf{s}_{k}$, then in general,  different template SNRs are correlated, i.e.  cross-template correlation coefficients $\Sigma_{ij} \equiv  \langle  \rho_{i} \rho_{j} \rangle \ne 0$. 
Because each SNR is a Gaussian random variable with zero mean, see Sec.~\ref{sec:SNRprobsetup},
their joint PDF $f\left(  \boldsymbol{\rho}|\mb{\Sigma}\right)$ is given by the multivariate normal distribution uniquely  specified by the $\Nt$-dimensional template bank covariance matrix $\mathbf{\Sigma}$, 
\begin{equation}
f\left(  \boldsymbol{\rho}|\mb{\Sigma}\right)  =\frac{1}{\sqrt{\det(2\pi\boldsymbol{\Sigma}})}\exp\left(  -\frac{1}{2}%
\boldsymbol{\rho}^{T}\boldsymbol{\Sigma}^{-1}\boldsymbol{\rho}\right)\,.
\label{Eq:jointsnrpdf}
\end{equation}
Here  $\boldsymbol{\rho}=\left(  \rho_{1},...,\rho_{\Nt}\right)  ^{T}$  and $\boldsymbol{\Sigma}^{-1}$ is the inverse of the covariance matrix $\boldsymbol{\Sigma}$.

We compute  matrix elements of  $\mathbf{\Sigma}$ in Appendix~\ref{App:tempcorr}. The result reads
\begin{equation}
   \Sigma_{ij}  = \frac{ \mb{s}_{i}  ^{T}\mathbf{E}^{-1}\mb{s}_{j}}{\sqrt{ \left(   \mb{s}_{i}^{T}\mathbf{E}^{-1}\mb{s}_{i} \right)  \left(  \mb{s}_{j}^{T}\mathbf{E}^{-1}\mb{s}_{j} \right) }} \,,
   \label{Eq:tempcovelements}
\end{equation}
where the subscripts $i$ and $j$ identify templates. This expression  depends only on the templates and the noise covariance matrix $\mb{E}$. The diagonal matrix elements of $\mb{\Sigma}$ are equal to 1, in agreement with Ref.~\cite{Panelli2020}. The off-diagonal matrix element vanishes if the templates are orthogonal in the generalized sense, $ \mb{s}_{i}^{T}\mathbf{E}^{-1}\mb{s}_{j}=0$. Eq.~(\ref{Eq:tempcovelements}) is simply $(\mb{s}_i \cdot \mb{s}_j)$ divided by the generalized norms of $\mb{s}_i$ and  $\mb{s}_j$, where the generalized dot product is understood as $(\mb{a} \cdot \mb{b}) \equiv \mb{a}  ^{T}\mathbf{E}^{-1}\mb{b}$.

Qualitatively, the cross-template correlations \linebreak $\Sigma_{ij}$'s quantify the degree of similarity between two templates. If the two templates are completely dis-similar (fully orthogonal), $\Sigma_{ij}=0$. If the two templates are identical (up to an overall sign), $\Sigma_{ij}=\pm 1$. To refine statements of Sec.~\ref{Sec:StrategiesBankChoice}, it is desired to generate banks with the smallest possible values of cross-template correlations to avoid testing data streams against  nearly-identical templates. In this sense, the most optimal choice of the bank templates would be a fully-orthogonal bank. However, as we show in Secs.~\ref{Sec:snrcorr:ToyProblem} and \ref{Sec:snrcorr:GPSDM}, constructing a fully-orthogonal bank is not possible for at least some of the applications.

The result~(\ref{Eq:jointsnrpdf}) holds
for any network with stationary white or colored Gaussian noise and with and without cross-node correlations. It also holds for any template bank choice. Namely the properties of $\mb{\Sigma}$ determine the SNR-max statistic PDF, CDF, and threshold values. In the rest of this section, we illustrate computation and properties of the template bank covariance matrix with our toy problem of Sec.~\ref{Sec:ToyTemplateBankGeneration} and the GPS.DM templates of Sec.~\ref{Sec:signaltempgen:GPSDM}.

\subsection{Toy problem and symmetry-adapted bank construction}
\label{Sec:snrcorr:ToyProblem}

Now we return to the toy problem of searching for line sweeps through a circular 
topology network of Sec.~\ref{Sec:ToyTemplateBankGeneration}. In this problem, a template $\mb{s}_i$ is uniquely specified by $(\bs{v}_{\perp})_i$, i.e. by the incident direction $\hat{u}_i$ and the normal speed $(v_{\perp})_i > 0$ of the sweeping line. In component form,
\begin{equation}
    (s_i)_l^a = \delta_{l,l_a^i},
    \label{Eq:circtemplate}
\end{equation}
where $l_a^i$ is the epoch when sensor $a$ is crossed by the line moving with $(\bs{v}_{\perp})_i$, see Eq.~(\ref{Eq:epochsata}).

We are interested in properties of $\mb{\Sigma}$ as these determine SNR-max CDF, PDF and threshold values. The bank covariance matrix elements, Eq.~(\ref{Eq:tempcovelements}), simplify to
\begin{equation}
    \Sigma_{ij} = \frac{1}{\Ns} \sum_{a=1}^{\Ns} \delta_{l_{a}^{i},l_{a}^{j}}.
    \label{Eq:circcorrmatrix}
\end{equation}
The sum counts the number of sensors that are affected in the same epoch in both templates. Apparently, $\Sigma_{ij} \ge 0$, as a result of the signals being always positive. In addition,  $\Sigma_{ij} \le 1$, with $\Sigma_{ij}$ reaching the value of 1 only when both templates are identical. 
 
We fix $v_{\perp}$ and discuss selections of incident directions $\hat{u}$ of sweeping lines. The $C_{\Ns}$ symmetry of our network suggests a symmetric choice of $\hat{u}_i : \hat{u}_i || \hat{n}_a$, i.e. sweeps towards sensors.
An example for an $\Nt = 5$ template bank is shown in Fig.~\ref{Fig:tempplanarnet} (right panel). Because of the rotational symmetry, all templates in the bank can be easily generated by cyclic permutation of sensor labels. For the Fig.~\ref{Fig:tempplanarnet} bank, template 2 can be  generated by relabeling sensors in the template 1, Fig.~\ref{Fig:tempplanarnetslice} : $1 \rightarrow 3, 2 \rightarrow 4, ..., 9 \rightarrow 1, 10 \rightarrow 2$ and then swapping the rows of the resulting template, so that the sensors are normally ordered.

We remind the reader that $\mb{\Sigma}$ is a symmetric matrix, with diagonal elements $\Sigma_{ii} = 1$. Based on the rotational symmetry arguments for the constructed bank, we expect all matrix elements $\Sigma_{i,i \pm 1}$ to be equal; the same applies to groups of matrix elements $\Sigma_{i,i \pm 2}$, $\Sigma_{i,i \pm 3}$, etc. Here $i \pm n$ is understood as mod($i \pm n, \Nt$). In general, $\Sigma_{i,i \pm 1} \ne \Sigma_{i,i \pm 2} \ne \Sigma_{i,i \pm 3} \ne \ldots$.
Such matrices belong to a class of symmetric Toeplitz matrices that offer several computational advantages. This property is a reflection of the underlying symmetry and would not hold for randomly-selected ${\hat{u}_i}$. 

Our explicit computations for the $\Nt=5$ bank of Fig.~\ref{Fig:tempplanarnet} with templates of Fig.~\ref{Fig:tempplanarnetslice} show that all off-diagonal matrix elements $\Sigma_{i,j \ne i} = 1/5$, i.e. all pairwise template overlaps have two simultaneously perturbed sensors. The CDF and PDF of such template bank covariance matrices with $\Sigma_{i,j \ne i} =r$ can be treated analytically, see Sec.~\ref{Sec:SqueezedApprox}. We refer to such banks as {\em perfectly-squeezed}. The CDF $C_{\Nt=5}(Z|\mb{\Sigma}_{r=1/5})$ for such a bank is given by Eq.~(\ref{Eq:sqCDFSigma0}).

Now we expand our bank of Fig.~\ref{Fig:tempplanarnet} to include all $\Ns=10$ possible incident directions towards sensors. The discussed $\Nt=5$ bank in Fig.~\ref{Fig:tempplanarnet} is just a subset of this $\Nt=10$ bank. We find that all off-diagonal matrix elements of the bank covariance matrix between the Fig.~\ref{Fig:tempplanarnet} bank and newly added templates vanish.
The bank breaks into two mutually orthogonal sub-banks.
Thus $\mb{\Sigma}$ assumes a block diagonal form with $\Sigma_{i,j \ne i} = 1/5$ for each block. Thereby, the joint probability distribution~(\ref{Eq:jointsnrpdf}) factorizes and  
the resulting SNR-max CDF is the product 
\begin{equation*}
    C_{\Nt=10}(Z|\mb{\Sigma})  =  \\ 
      C_{\Nt=5}(Z|\mb{\Sigma}_{r=1/5})C_{\Nt=5}(Z|\mb{\Sigma}_{r=1/5}). 
\end{equation*}
We may continue the process of building the symmetry-adapted bank by augmenting the $\Nt=10$ bank with incidence directions $\hat{u}_i$ that subdivide the arks between the adjacent $\Nt=10$ bank directions into equal parts, until we reach the network angular resolution $\Delta_\theta$, introduced in Sec.~\ref{Sec:ToyTemplateBankGeneration}. This procedure relies on the intrinsic $C_{\Ns}$ symmetry of our toy network.  The presented ideas of symmetry-adapted bank construction can be extended to 3D network with icosahedral point group symmetry $I_h$.   
The GPS constellation, discussed in the following section, unfortunately, lacks point symmetries and a different approach is required. 

\subsection{GPS.DM search}
\label{Sec:snrcorr:GPSDM}

Generation of GPS.DM templates for planar transients was discussed in Sec.~\ref{Sec:signaltempgen:GPSDM}. Compared to our toy network, GPS constellation has a reference clock, it has a 3D geometry, lacks point symmetries, and positions of satellites evolve in time.

We start with one of the un-optimized GPS.DM  template banks used in the GPS.DM search.  The bank has $\Nt=1024$ templates and was generated for the GPS constellation orbital configuration for Feb. 14, 2005  08:20  UTC time.
(We specify time because the generated templates depend on the orbital positions of the satellites, which vary throughout the day. Thus the template bank  needs to be re-generated every window.) The network had $\Ns=21$ sensors and relative reference-sensor sensitivity parameter~(\ref{Eq:xi-parameter}) $\xi=0.6$. Identical white frequency noise was assumed for all the clocks. Templates were generated by importance sampling using Monte-Carlo technique described in Sec.~\ref{sec:signaltempgen}, i.e. the template event parameters were randomly drawn from the parameter space weighed by priors. This bank generation did not take into account network angular and normal speed resolutions.

\begin{figure}[h!]
	\begin{center}
		\includegraphics[width=0.42\textwidth]{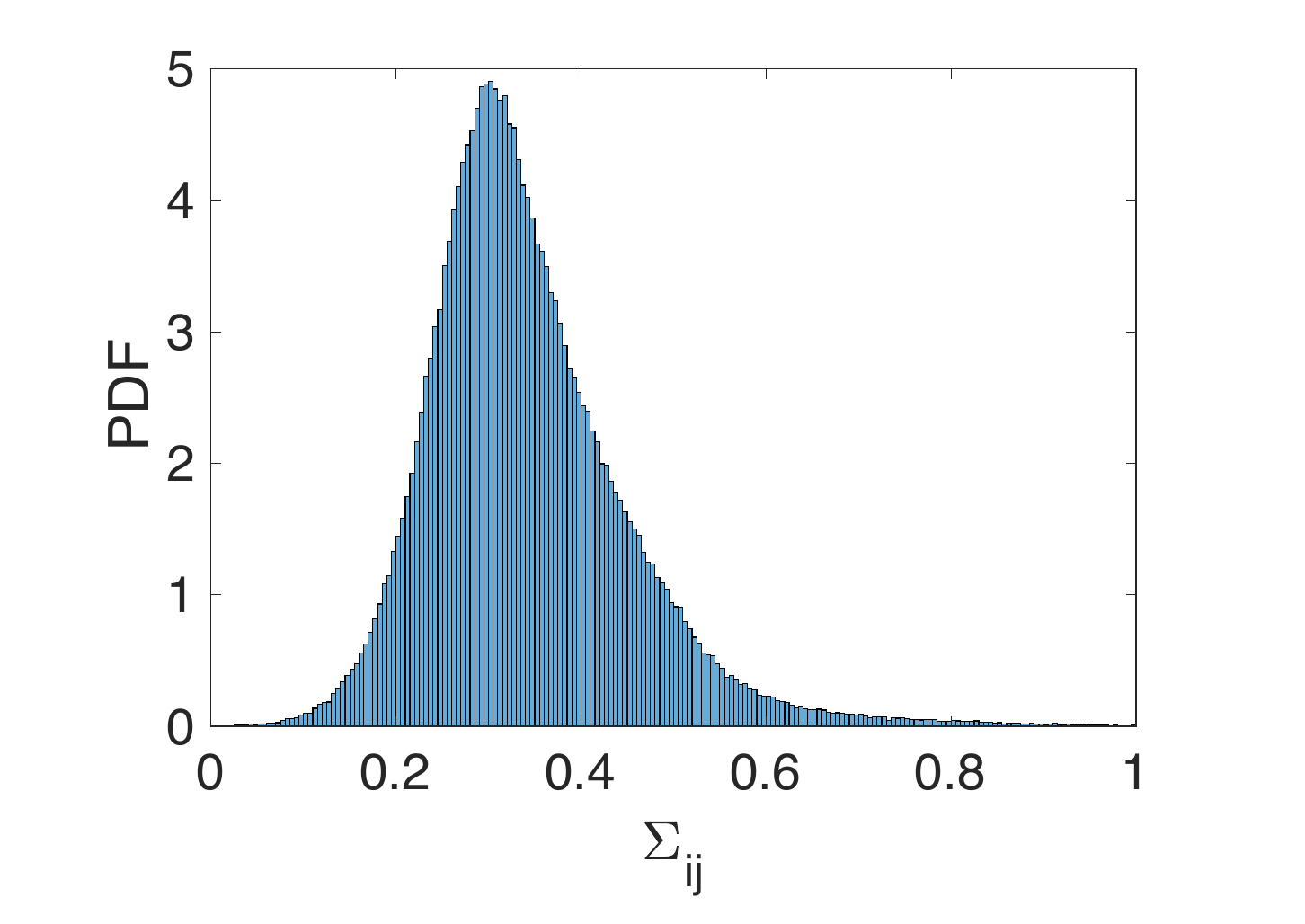}
	\end{center}
	\caption{Distribution of cross-template correlation coefficients $\Sigma_{i\neq j}$ of the template bank covariance matrix $\mb{\Sigma}$ for one of the GPS.DM template banks. Template bank contains $\Nt=1024$ templates. }
	\label{Fig:tempcovelements}
\end{figure}

Fig.~\ref{Fig:tempcovelements} shows our computed  distribution of the bank covariance matrix elements, $\Sigma_{ij}$ for $i \ne j$. We observe that 
$\Sigma_{ij}$'s are centered about the most probable value of $ \Sigma_{ij} \approx 0.33$. All values are positive and populate a range $0.0 \le \Sigma_{ij}\le 1.0$. In this particular template bank, there are 7 templates that are orthogonal to all other templates, while some of the templates in the bank are nearly identical ( $\Sigma_{ij} \approx 1$), and 7 templates are identical, with $\Sigma_{ij} = 1$. 
Clearly, this is not an optimal choice for a template bank, as such aligned (identical) templates would result in wasted computational time. A simple solution to this problem would be to check for  $\Sigma_{ij} \approx 1$ and if so, to draw another template from the Monte-Carlo generator.

To quantify the distribution of $\Sigma_{ij}$  matrix elements and to develop an analytical approach, in Appendix~\ref{App:SNRcovelements} we derive 
several properties of the GPS.DM template bank covariance matrix for
an idealized network of white noise sensors with network noise covariance matrix $\mb{E}$ of Eq.~(\ref{Eq:whitenetEcov}). 
The derivation is similar to that of Sec.~\ref{Sec:snrcorr:ToyProblem} with an addition of the reference sensor perturbation, so that the template elements are given by Eq.~(\ref{Eq:signaltemplate}).    
The derived template covariance matrix depends on a set of three parameters $K_1$, $K_2$, and $R_1$: 
\begin{align} \label{Eq:degennparams}
    K_1^{s,s'} & \equiv \sum_a^{\Ns} \delta_{l_a,l_a^{'}}, \\ \nonumber
    K_2^{s,s'} & \equiv \sum_{a}^{\Ns}\sum_{b\ne a}^{\Ns} \delta_{l_a,l_b^{'}}, \\ 
    R_{1}^{s,s'} & \equiv \Ns - \sum_{a} \left(  \delta_{l_R,l_a} +  \delta_{l_R,l_a^{'}} \right) \, , \nonumber
\end{align}
where $l_a$ are epochs when  sensor $a$ was perturbed in template $\mb{s}$ and primed quantities refer to template $\mathbf{s}'$.
Here $K_1$ is the number of sensors with a perturbation epoch identical to both templates, $K_2$ is the cross-sensor degeneracy between the templates, and $R_1$ quantifies the reference sensor and network sensor degeneracy. 
These expressions apply to both same-template ($\mb{s} \equiv \mb{s}'$) and the cross-template cases. 

With these definitions, the cross-template correlation coefficients between two templates $\mb{s}$ and $\mb{s}'$ read
\begin{align}
    & \Sigma_{ss'}=  \\
    & \frac{a K_1^{s,s'} +  R_{1}^{s,s'} + b K_2^{s,s'} }
    {\sqrt{\left(a K_1^{s,s} +  R_{1}^{s,s} + b K_2^{s,s}\right) \left(a K_1^{s',s'} + R_{1}^{s',s'} + b K_2^{s',s'} \right)}}, \nonumber
    \label{Eq:generalSigmassp}
\end{align}
where template-independent parameters are 
\begin{align}
a &= 1 +(1-1/\Ns)\xi  \,, \\ 
b &= - \xi/\Ns \,.  
\end{align}
Here $\xi$ is the relative reference-sensor sensitivity parameter~(\ref{Eq:xi-parameter}).  

With these definitions, in Appendix~\ref{App:SNRcovelements} we determine ``averaged'' cross-template correlation value $\overline{\Sigma}_{i\neq j}$ for our GPS network data.  Here ``averaged'' means that the degeneracy parameters (\ref{Eq:degennparams}) are equated to the typical values using the most probable epoch spread $\overline{\Delta l}$, Eq.~(\ref{Eq:TemplateSpread}) and the uniform  occupancy  approximation of Sec.~\ref{Sec:signaltempgen:GPSDM}. The average degeneracy parameters, computed in Appendix~\ref{App:SNRcovelements}, are 
$\bar{K}_1^{s,s}  = \bar{K}_1^{s',s'} =  \Ns$, $\bar{K}_1^{s,s'}  = {\Ns}/\overline{\Delta l}$, $\bar{K}_2^{s,s} = \bar{K}_2^{s',s'} = \bar{K}_2^{s,s'} = (\Ns - 1){\Ns}/\overline{\Delta l}$, and $\bar{R}_1^{s,s}  = \bar{R}_1^{s',s'} = \bar{R}_1^{s,s'} =  \left(1 - {2}/{\overline{\Delta l}}\right)\Ns$. These values result in the ``averaged'' cross-template correlation coefficient
\begin{equation}
    \overline{\Sigma}_{i\neq j} \approx    
    \frac{1}{2+(1-1/\Ns)\xi}\, . \label{Eq:SigmaMeanApprox}
\end{equation}

To access the accuracy of approximation~(\ref{Eq:SigmaMeanApprox}), we consider
the computed distribution of cross-template correlation coefficients $\Sigma_{ij}$ of Fig.~\ref{Fig:tempcovelements}. This histogram was computed for the network of $\Ns=21$ Rb clocks  with sensitivity parameter $\xi = 0.6$. Then Eq.~(\ref{Eq:SigmaMeanApprox}) results in $\overline{\Sigma}_{i\neq j} = 0.388$. It is in $\sim 5\%$ agreement with the most probable value inferred from the histogram, $\overline{\Sigma}_{i\neq j} = 0.398$. Over decades of the GPS constellation operation, satellites are replaced as they reach their end of life. The most typical value of $\xi \approx 0.3$, and then for $\Ns=30$ the Eq.~(\ref{Eq:SigmaMeanApprox}) gives $\overline{\Sigma}_{i\neq j} = 0.44$.

\begin{figure}[h!]
	\begin{center}
		\includegraphics[width=0.42\textwidth]{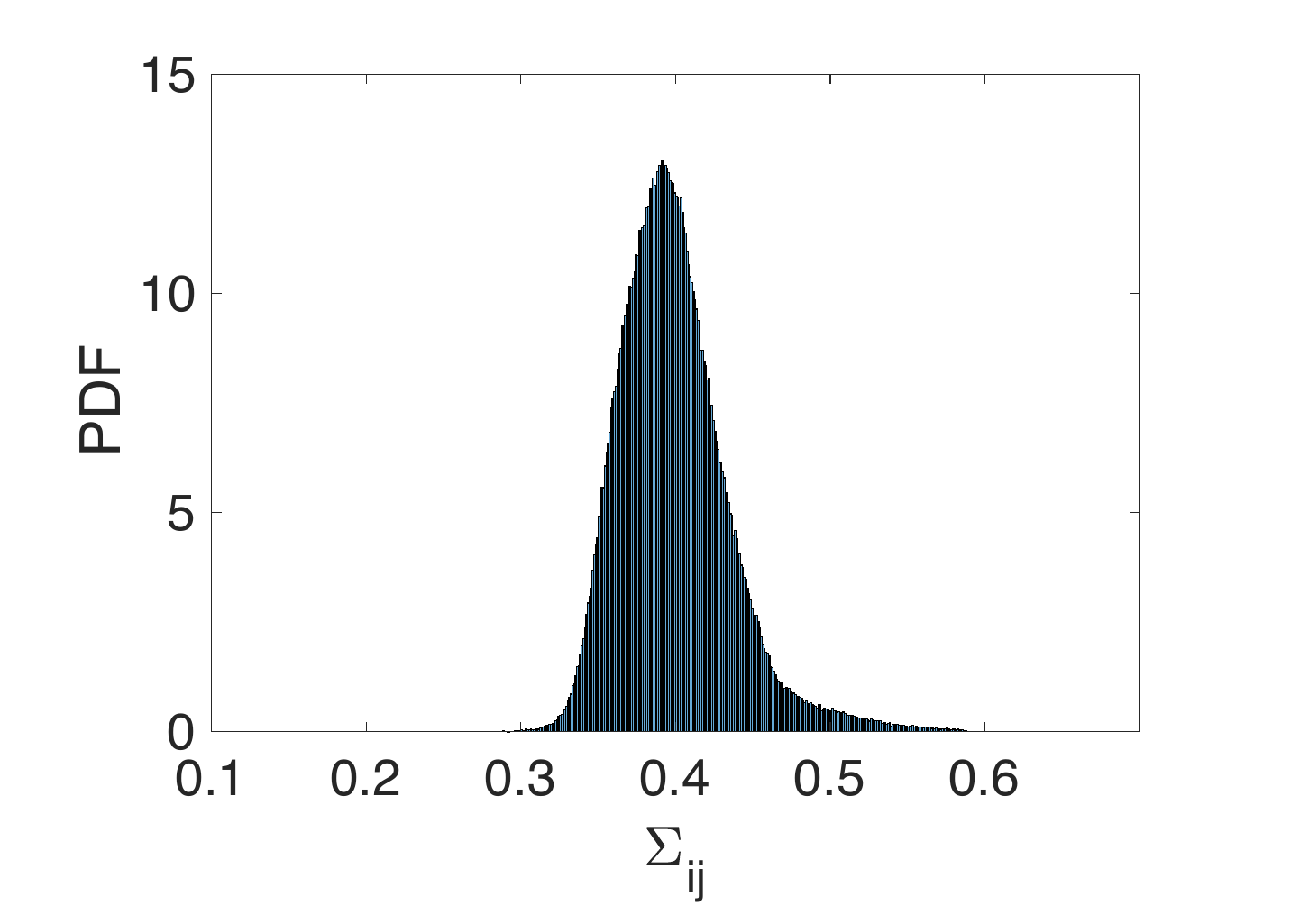}
	\end{center}
	\caption{(Color online) Distribution of off-diagonal matrix elements $\Sigma_{ij}$ of the template bank covariance matrix  for a squeezed template bank with the most probable covariance matrix element $r = 0.39$ and maximum spread in matrix elements $\delta\Sigma_{ij}= \pm 0.2$.  Template bank contains $\Nt=1024$ templates. All diagonal matrix elements $\Sigma_{ii}=1$ and are not accounted for here.}
	\label{Fig:squeeztempbank}
\end{figure}

We discussed template bank optimization strategies in Sec.~\ref{Sec:StrategiesBankChoice}. 
One of the refinements of these strategies is to ``squeeze'' the distribution of off-diagonal template covariance matrix elements around the average value of $\Sigma_{i \ne j}$, $r$, with some small spread $\delta\mb{\Sigma}$, so that $\mb{\Sigma}=\mb{\Sigma}_r + \delta\mb{\Sigma}$, where off-diagonal matrix elements of $\mb{\Sigma}_r$ are set to $r$ and $\left(\Sigma_r\right)_{ii}=1$. 
One ``squeezing'' technique that can be used with success (although with large computational cost) is by performing Monte-Carlo and eliminating templates whose correlation element is above or below a specified spread in correlation.
In Sec.~\ref{Sec:SqueezedApprox} we provide analytical results for the SNR-max CDF and PDF of such $\mb{\Sigma}_r$; the bank ``squeezing'' improves the accuracy of approximating $\mb{\Sigma}$ with $\mb{\Sigma}_r$.
Fig.~\ref{Fig:squeeztempbank} shows an example of ``squeezed'' bank distribution of template covariance matrix elements (not including diagonal 1's) for the most probable covariance matrix element $r = 0.39$ and spread of  $\delta{\Sigma}_{ij}= \pm 0.2$ of the perturbation matrix.


\section{SNR-max distribution for a template bank}
\label{Sec:MAXSNR}
To reiterate, at this point we have discussed the properties of templates and  the template bank covariance matrix $\mb{\Sigma}$. Now we return to the discussion of the SNR-max statistic and 
derive its distributions. The starting point is the joint probability distribution~(\ref{Eq:jointsnrpdf}) for SNRs in the template bank.  
The joint PDF is fully characterized by the template bank covariance matrix $\mb{\Sigma}$,
which is a symmetric and positive semi-definite matrix with diagonal matrix elements $\Sigma_{ii}=1$. Off-diagonal matrix elements of $\mb{\Sigma}$ are given by Eq.~(\ref{Eq:tempcovelements}).
In this section, we determine  an analytic form for the  CDF and PDF of the SNR-max statistic~(\ref{Eq:SNRmaxDefinition}), 
 $  z = \max( \vert \rho _{1}  \vert  ,.., \vert \rho _{\Nt} \vert  ) $.
Apparently, $z\ge 0$.

We start the derivation by computing the CDF of the SNR-max statistic. 
CDF $C_{\Nt}\left(  Z|\boldsymbol{\Sigma}\right)$ 
is the probability that the SNR-max statistic $z$  takes a value less than or equal to $Z$. The condition $z\le Z $ implies that all the individual template SNRs lie in the interval $-Z  \le \rho _{k} \le Z$. 
Thereby, we can  express the CDF as an integral of the joint probability distribution~(\ref{Eq:jointsnrpdf}),
\begin{align} \label{Eq:CDF}
&C_{\Nt}\left(  Z|\boldsymbol{\Sigma}\right) = \\ &\frac{1}{\sqrt{\det(2\pi \boldsymbol{\Sigma})}}\int_{-Z}^{+Z}d\rho
_{1}...\int_{-Z}^{+Z}d\rho_{\Nt}\exp\left(  -\frac{1}{2}\boldsymbol{\rho
}^{T}\boldsymbol{\Sigma}^{-1}\boldsymbol{\rho}\right) \, . \nonumber%
\end{align}
This $\Nt$-dimensional integral can be visualized as an integral of a multivariate normal distribution over an $ \Nt $-dimensional hyper-cube centered at $0$ and having a side of $2Z$. 

To date there is no closed form for the CDF of a multivariate normal distribution, so the $\Nt$-dimensional integral requires numerical evaluation (typically using Monte-Carlo techniques~\cite{PETERLEPAGE1978192,doi:10.1063/1.4822899}). 
For dynamically evolving networks, such as 
the GPS constellation where satellites are moving, this approach is
computationally prohibitive, as templates change with the network geometry, and have to be generated anew for every window.
Every newly generated template bank would have a different template bank covariance matrix $\boldsymbol{\Sigma}$, new CDF, and a new threshold for false-positives. Partially, this problem can be mitigated by observing that there is a 12-hour periodicity in the satellite orbits, i.e. network geometry repeats every 12 hours. Then the computations can be reduced by storing template banks and threshold values for the 12 hour period. 
Even with this shortcut, one  still has to deal with a collection of 1440 template banks for individual $\Delta_t = 30 \, \mathrm{s}$ epochs.

The cumulative distribution can be expressed using the underlying probability density $p_\Nt\left(  \rho|\boldsymbol{\Sigma}\right)$,
$
C_{\Nt}\left(  Z|\boldsymbol{\Sigma}\right)  =\int_{0}^{Z}p_\Nt\left(
z|\mb{\Sigma}\right)  dz
$,
or
\begin{equation}
p_\Nt\left(  z|\mb{\Sigma}\right)  =\left( \frac{d}{dZ}C_\Nt\left(  Z|\mb{\Sigma}\right)\right)_{Z \rightarrow z}.
\label{Eq:pdffromcdf}
\end{equation}
We carry out the differentiation of the SNR-max CDF~(\ref{Eq:maxPDFcorr}) in Appendix~\ref{App:SNRMAXderivation}. The resulting SNR-max PDF reads
\begin{align}\label{Eq:maxPDFcorr} 
p_{\Nt}\left(  z|\boldsymbol{\Sigma}\right)  = 
&\frac{2}{\sqrt{\det\left(
2\pi\boldsymbol{\Sigma}\right)  }}\sum_{k=1}^{\Nt}\left(
{\displaystyle\prod\limits_{n=1,n\neq k}^{\Nt}}
\int_{-z}^{+z}d\rho_{n}\right) \times \\ \nonumber
&\times \left[  \exp\left(  -\frac{1}{2}%
\boldsymbol{\rho}^{T}\boldsymbol{\Sigma}^{-1}\boldsymbol{\rho}\right)
\right]  _{\rho_{k}=z}. \nonumber
\end{align}
In general, this SNR-max PDF is sufficient, yet it still involves $\Nt$  evaluations of $(\Nt-1)$-dimensional integrals. Thus, for large banks this PDF
still requires  computationally-expensive numerical evaluation.

These considerations force us to explore approximate techniques for evaluating SNR-max CDF and PDF. In Sec.~\ref{sec:independtemp} we review the SNR-max CDF and PDF for uncorrelated templates or fully-orthogonal bank ($\bs{\Sigma}=\bs{I}$).
In Sec.~\ref{sec:twocorrtemp}, we consider a simple case of two correlated templates. In Sec.~\ref{sec:smallapprox} we derive an approximate SNR-max PDF and CDF for  a nearly-orthogonal template bank, i.e. when off-diagonal $|\Sigma_{ij}| \ll 1$. Lastly, in Sec.~\ref{Sec:SqueezedApprox} we consider the SNR-max PDF and CDF for a ``squeezed'' template bank, when off-diagonal matrix elements $\Sigma_{ij}\approx r$ for all templates.

Finally, for completeness, we mention an apparent simplification when $\bs{\Sigma}$ has a block-diagonal form, a case encountered in the toy problem, Sec.~\ref{Sec:snrcorr:ToyProblem}. This happens when the template bank can be decomposed into mutually orthogonal sub-banks of dimensions $\Nt'$ and $\Nt''$, $\Nt'+\Nt''=\Nt$.
If $\bs{\Sigma}$ contains two blocks $\bs{\Sigma}'$ and $\bs{\Sigma}''$, spanning SNR sub-spaces $\bs{\rho}'$ and $\bs{\rho}''$, the joint PDF~(\ref{Eq:jointsnrpdf}) factorizes $f\left(  \boldsymbol{\rho}|\mb{\Sigma}\right) = f\left(  \boldsymbol{\rho}'|\mb{\Sigma}'\right) f\left(  \boldsymbol{\rho}''|\mb{\Sigma}''\right)$. Then the CDF factorizes as well,
\begin{equation*}
    C_{\Nt}\left(  Z|\boldsymbol{\Sigma}\right) =  C_{\Nt'}\left(  Z|\boldsymbol{\Sigma}'\right)C_{\Nt''}\left(  Z|\boldsymbol{\Sigma}''\right) \, . 
\end{equation*}

\subsection{Fully-orthogonal bank (independent templates)}
\label{sec:independtemp}
For a fully-orthogonal bank, all the off-diagonal matrix elements of the template bank covariance matrix $\boldsymbol{\Sigma}$ vanish and $\boldsymbol{\Sigma}$ reduces to an identity matrix $\bs{I}$. Then the general SNR-max CDF (\ref{Eq:CDF}) is simply 
\begin{align}\label{Eq:CDForthog}
C_{\Nt}\left(Z|\boldsymbol{I}\right)  &= \frac
{1}{\left(  2\pi\right)  ^{\Nt/2}}\left(  \int_{-Z}^{+Z}d\rho\exp\left(
-\frac{1}{2}\rho^{2}\right)  \right)  ^{\Nt} \\ \nonumber
&=\left[  \text{erf}\left(
\frac{Z}{\sqrt{2}}\right)  \right]  ^{\Nt} \,,%
\end{align}
where Gaussian error function \small$\erf{x} = (2/\sqrt{\pi}) \int_0^x du \, e^{-u^2}$\normalsize. 
Further, differentiating CDF~(\ref{Eq:CDForthog}) with respect to $Z$, we arrive at the SNR-max PDF for a bank of $\Nt$ independent templates 
\begin{equation}
 p_{\Nt}\left(z|\boldsymbol{I}\right)  =M\sqrt{\frac{2}{\pi}}e^{-\frac{z^{2}}{2}%
}\left(\text{erf}\left(  \frac{z}{\sqrt{2}}\right)\right)^{\Nt-1}.
\label{Eq:Mtemppdfnocorr}   
\end{equation}

These results subsume the limiting case of the $M=1$ bank. In this case, we are dealing with a single template, so that the SNR-max statistic~(\ref{Eq:SNRmaxDefinition}), $z=|\rho|$, is an absolute value of the SNR for that one template. From the SNR distribution~(\ref{Eq:snrdistExplicit}) we conclude that $z$ is a half-normally distributed variable with probability distribution
\begin{equation}
    p_1(z) = \sqrt{\frac{2}{\pi}}e^{-\frac{z^{2}}{2}} \,.
    \label{Eq:halfnormalpdf}
\end{equation}
Integrating it, we obtain CDF $C_1(Z) = \erf{Z/\sqrt{2}}$. These results are in agreement with Eqs.~(\ref{Eq:CDForthog},\ref{Eq:Mtemppdfnocorr}) for $M=1$.

\subsection{Bank of $M=2$ correlated templates}
\label{sec:twocorrtemp}
With the results for a fully-orthogonal bank established, we start quantifying the role of cross-template correlations with an analytically tractable case of a bank consisting of two templates. 
For the $M=2$ case, the cross-template covariance matrix depends only on one
off-diagonal matrix element $\Sigma_{12}= \Sigma_{21} \equiv r$. Explicitly,

\begin{align}
\boldsymbol{\Sigma}  &  \boldsymbol{=}\left(
\begin{array}
[c]{cc}%
1 & r\\
r & 1
\end{array}
\right), \\ 
\boldsymbol{\Sigma}^{-1}  &  =\frac{1}{1-r^{2}}\left(
\begin{array}
[c]{cc}%
1 & -r\\
-r & 1
\end{array}
\right)
,%
\label{Eq:Cparameters}
\end{align}
and $\det\mb{\Sigma}=1-r^{2}$. Then the joint probability distribution~(\ref{Eq:jointsnrpdf}) reads
\begin{equation*}
f\left( \rho_{1},\rho_{2}| r\right)  =\frac{1}{ 2\pi
\sqrt{1-r^{2}}}\exp\left(  -\frac{\rho_{1}^{2}%
+\rho_{2}^{2}-2r\rho_{1}\rho_{2}}{2(1-r^{2})}\right).
\end{equation*}
Applying Eq.~(\ref{Eq:maxPDFcorr}), we arrive at the SNR-max PDF for the $M=2$ bank
\begin{align}\label{Eq:twocorrtemppdf}
p_{2}\left( z|\,r\right)  &= \sqrt{\frac{2}{\pi}}e^{-\frac{z^{2}}{2}} \times \\ 
&\left(\text{erf}\left(  \frac{1}{\sqrt{2}}\sqrt{\frac{1+r}{1-r}}z\right)
+\text{erf}\left(  \frac{1}{\sqrt{2}}\sqrt{\frac{1-r}{1+r}}z\right)  \right). \nonumber
\end{align}

\begin{figure}[h!]
	\begin{center}
		\includegraphics[width=0.42\textwidth]{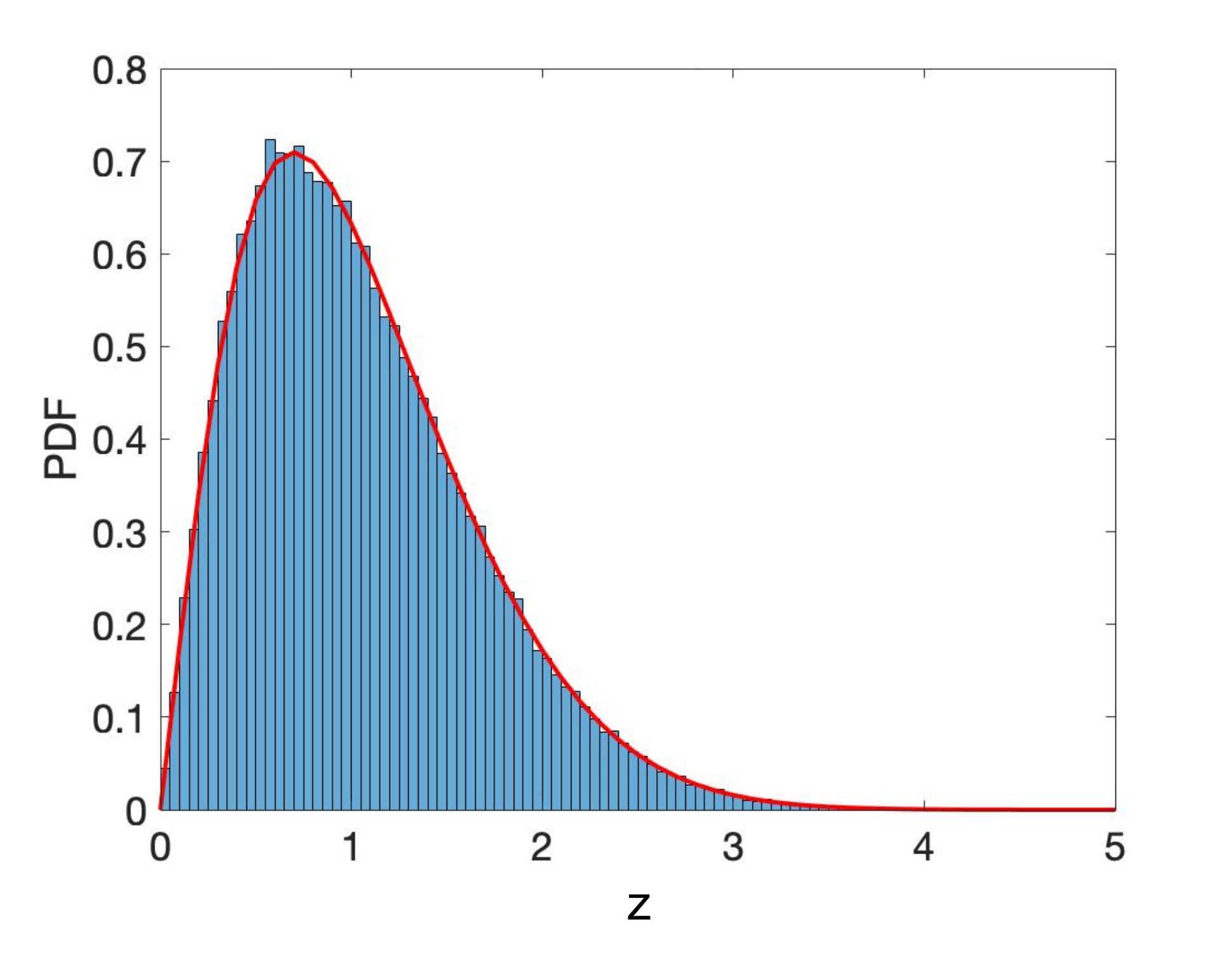}
		\caption{(Color online) Comparison of the Monte-Carlo simulation (blue) with the
		closed-form SNR-max PDF~(\ref{Eq:twocorrtemppdf}) (red curve)
		for a bank of $\Nt=2$ templates with correlation coefficient $r=1/2$.
}
\label{Fig:twodependtemp}
	\end{center}
\end{figure}

To support our analytical result, in  
Fig.~\ref{Fig:twodependtemp} we compare results of a Monte-Carlo simulation
with the derived SNR-max PDF~(\ref{Eq:twocorrtemppdf}), and we find the histogram and the PDF in agreement. 
We carried out these simulations using a single ($\Ns=1$) white noise (Gaussian) data stream with zero mean and $\sigma^{2} = 1$, and with the two templates having a correlation coefficient $r=1/2$.
The positive skew observed in the SNR-max PDF is due to the effect of  templates being correlated. False-positive thresholds determined for correlated template banks are affected by the positive skew observed in this SNR-max PDF,  see Sec.~\ref{sec:thresholds}.


We observe the following properties of the PDF~(\ref{Eq:twocorrtemppdf}).
The distribution $p_{2}\left(  z|r\right) $ is symmetric with respect to
flipping the sign of $r$, i.e. $p_{2}\left(z|-r\right)  = p_{2}\left(
z|r\right)$. 
If there is no correlation between the two SNR variables ($r=0$), Eq.~(\ref{Eq:twocorrtemppdf}) reduces to 
\begin{equation}
p_{2}\left(  z|r=0\right)  =2\sqrt{\frac{2}{\pi}}e^{-\frac{z^{2}}{2}%
}\text{erf}\left(  \frac{z}{\sqrt{2}}\right) \,, 
\label{Eq:twotemppdfnocorr}
\end{equation}
which is identical to the general result~(\ref{Eq:Mtemppdfnocorr}) for the $M=2$ orthogonal template bank. For fully correlated SNRs, $r=\pm1$, the probability distribution tends to the half-normal distribution (\ref{Eq:halfnormalpdf})%
\begin{equation}
p_{2}\left(  z|\, r \rightarrow \pm 1\right) = \sqrt{\frac{2}{\pi}}e^{-\frac{z^{2}}{2}}%
=p_{1}\left(  z\right).
\label{Eq:twotemppdffullcorr}
\end{equation}
In this case, the two templates are identical (or differ by an overall sign) and the
PDF reduces to that of a single template case Eq.~(\ref{Eq:twotemppdffullcorr}). These two cases of $r=0$ and $r=\pm1$  agree with previous literature~\cite{Nadarajah2008}.

\begin{figure}[h!]
	\begin{center}
		\includegraphics[width=0.42\textwidth]{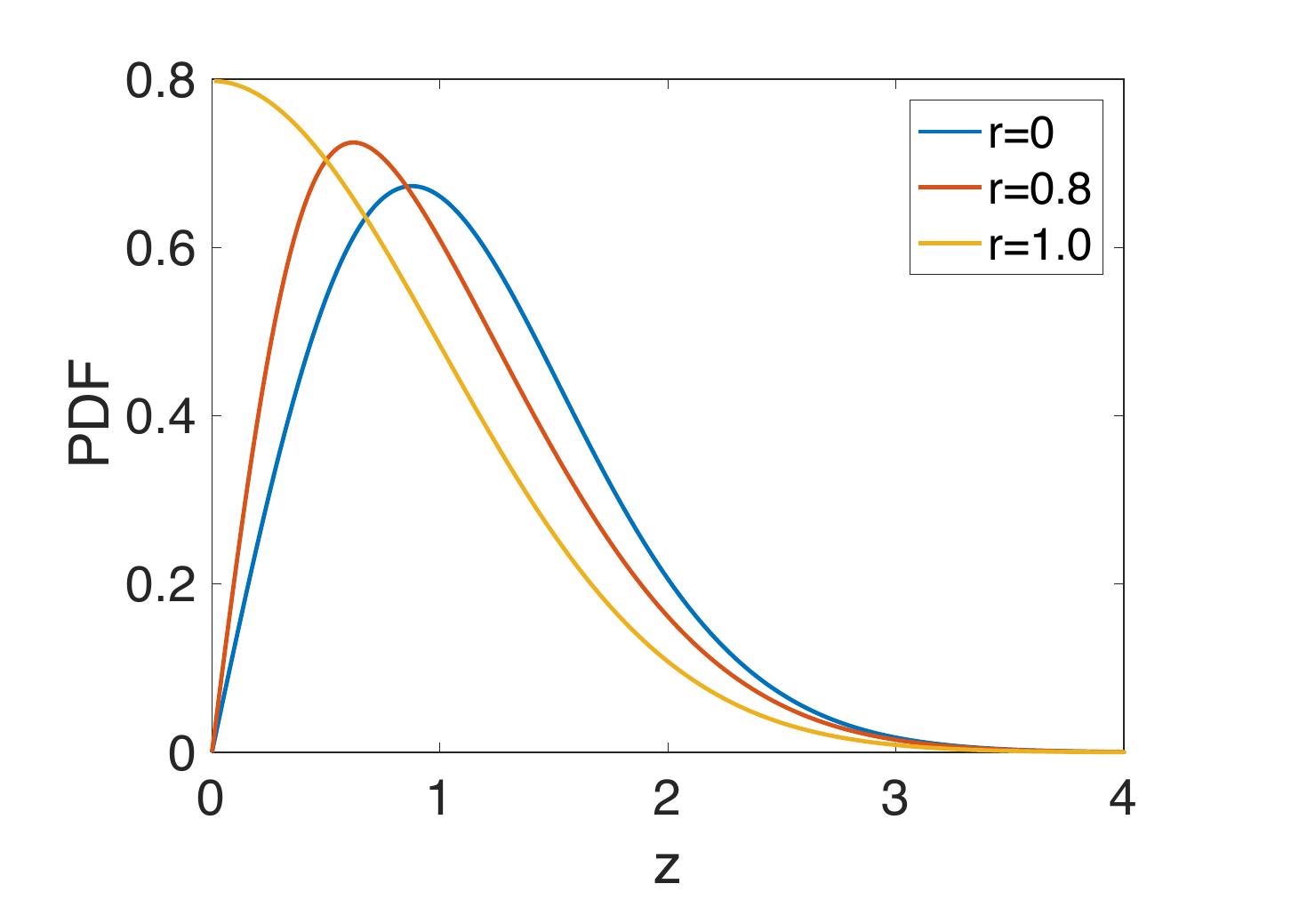}
	\end{center}
	\caption{ (Color online)
	Probability distribution for SNR max statistic for $M=2$ bank with varying cross-template correlation, $0 \le r \le 1$ ( $r=0$ blue curve, $r=0.75$ red curve, and $r=1$ orange curve, respectively). As the magnitude of the correlation coefficient increases from $r=0$ to larger $|r|$, the distribution becomes increasingly positively skewed, and when $r =\pm 1$ (orange curve) we recover the half-normal SNR-max distribution for $M=1$ bank, see Eq.~(\ref{Eq:twotemppdffullcorr}). }
	\label{Fig:SNRmaxtwotemp}
\end{figure}

To further illustrate the effect of correlation on the SNR-max PDF, we vary the correlation coefficient $r$ from 0 to 1, see Fig.~\ref{Fig:SNRmaxtwotemp}. Due to the $r \rightarrow -r$   symmetry of the distribution, positive and negative template correlations  have identical PDFs. Fig.~\ref{Fig:SNRmaxtwotemp} shows that when the correlation $ \vert r \vert>0$, the SNR-max PDF becomes positively skewed (skewed toward zero). When the templates are fully correlated, $ \vert r \vert   = 1$, the distribution $p_2$  reduces to Eq.~(\ref{Eq:twotemppdffullcorr}), i.e. the half-normal distribution $p_1$ characteristic of a single template bank. This is expected qualitatively  as the two random variables $\rho _{1}$ and $\rho _{2}$ become identical, so that $z=\max(|\rho_1|,|\rho_2|)=  |\rho_1| = |\rho_2|$.

For a template bank composed of two templates, $\Nt = 2$, the SNR-max CDF $C_{2}(Z|r)$ can be determined by numerically integrating the two-template SNR-max PDF in Eq.~(\ref{Eq:twocorrtemppdf})
\small\begin{align}
    C_{2}(Z|r) =& \int_{0}^{Z} dz \sqrt{\frac{2}{\pi}}e^{-\frac{z^{2}}{2}} \times \\
    &\left(
\text{erf}\left(  \frac{1}{\sqrt{2}}\sqrt{\frac{1+r}{1-r}}z\right)
+\text{erf}\left(  \frac{1}{\sqrt{2}}\sqrt{\frac{1-r}{1+r}}z\right)  \right). \nonumber
\end{align}
\normalsize
An equivalent representation, 
\begin{align}\label{Eq:sqCDFSigma0:M2}
    &C_2(Z|r) = \frac{1}{4}  \int_{-\infty}^{\infty} du \, 
    \frac{ e^{ - u^{2}/2}}{\sqrt{2 \pi}} \times \\
    &\left[ \erf{\frac{Z+\sqrt{r}u}{\sqrt{2(1-r)}}} - \erf{\frac{-Z+\sqrt{r}u}{\sqrt{2(1-r)}}} \right]^{2} \,, \nonumber
\end{align}
is derived in Sec.~\ref{Sec:SqueezedApprox}.

\begin{figure}[h!]
	\begin{center}
		\includegraphics[width=0.42\textwidth]{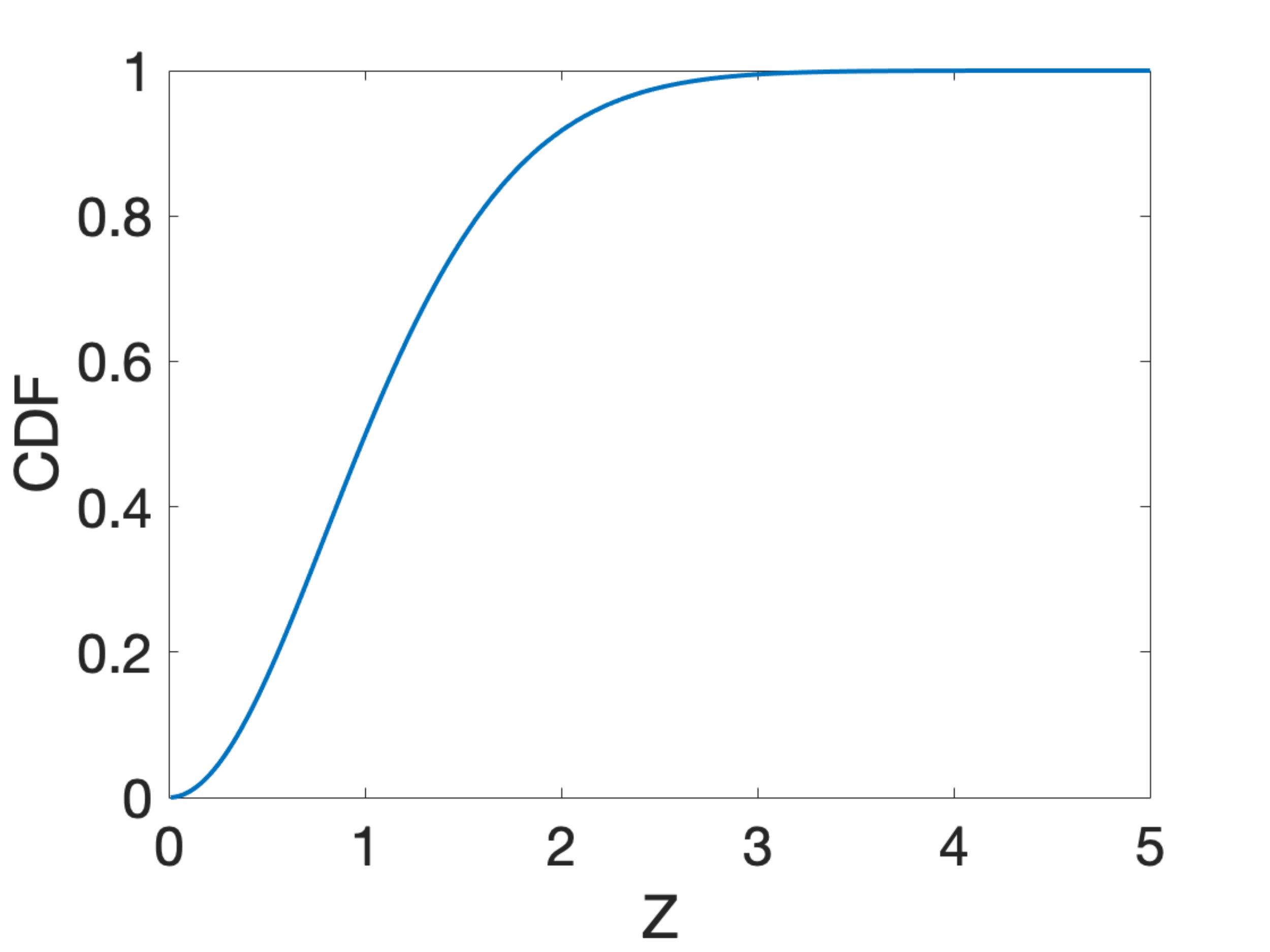}
	\end{center}
	\caption{(Color online) Cumulative distribution for a template bank composed of $\Nt=2$ templates with a correlation coefficient $r=1/2$.}
	\label{Fig:twotempcdf}
\end{figure}

Fig.~\ref{Fig:twotempcdf} shows the CDF $C_{2}(Z|r=1/2)$ resulting from evaluating, Eq.~(\ref{Eq:sqCDFSigma0:M2}) numerically. From this CDF, thresholds for specific false positive rates will be inferred in Sec.~\ref{sec:thresholds}.


\subsection{SNR-max distribution for a nearly-orthogonal bank}
\label{sec:smallapprox}
Suppose that the correlations between various templates are small, i.e. $|\Sigma_{ij}| \ll 1$ for all $i \ne j$, such that the template covariance matrix  $\mb{\Sigma}$ can be represented as a sum of  the identity matrix $\bm{I}$ (fully-orthogonal template bank) and a perturbation matrix $\boldsymbol{A}$.
In this case, the bulk of the SNR-max CDF and PDF is determined by the
distributions for fully-orthogonal bank, 
Eqs.~(\ref{Eq:CDForthog}, \ref{Eq:Mtemppdfnocorr}), with corrections that depend on $\bm{A}$:
\begin{align}
   C_{\Nt}\left(Z|\boldsymbol{\Sigma}\right) \approx & C_{\Nt}\left(  Z|\boldsymbol{\Sigma}=\bm{I}\right) + \delta_I C_{\Nt}\left(  Z|\boldsymbol{\Sigma}\right) \,, \\
p_{\Nt}\left(z|\boldsymbol{\Sigma}\right) \approx & p_{\Nt}\left(  z|\boldsymbol{\Sigma=I}\right) + \delta_I p_{\Nt}\left(  z|\boldsymbol{\Sigma}\right) \,.
\end{align}
Here we use the symbol $\delta_I$ to emphasize that the expansion is about $\boldsymbol{\Sigma}=\bm{I}$. 
In Appendix~\ref{App:SNRsmallcorr}, we demonstrate that these corrections are proportional to the trace of ${\bm{A}^2}$. Below we summarize the derivation and results from Appendix~\ref{App:SNRsmallcorr}.

The general expression for the SNR-max CDF~(\ref{Eq:CDF}) depends on the determinant and the inverse of the template bank covariance matrix.
The determinant can be approximated as
\begin{equation}
\det\left(  \boldsymbol{I}+\boldsymbol{A}\right)  \approx 1
-\frac{1}%
{2}\Tr{\left(  \boldsymbol{A}^{2}\right)} .
\label{Eq:detapprox}
\end{equation}
Here we omitted terms containing $\Tr{\left(  \boldsymbol{A}\right)}=0$ as,  by construction, the diagonal matrix elements of the perturbation matrix $\bm{A}$ vanish. Notice that $\Tr{\left(  \boldsymbol{A}^{2}\right)} = \Tr (\boldsymbol{\Sigma}^{2}) -M$. 
Similarly, the inverse of $ \boldsymbol{\Sigma}$ can  be approximated as 
\begin{equation}
\boldsymbol{\Sigma}^{-1} = \left(  \boldsymbol{I}+\boldsymbol{A}\right)  ^{-1}\approx\boldsymbol{I}%
-\boldsymbol{A}+\bm{A}^{2}.
\label{Eq:ApprxCovMat}
\end{equation}

Based on these approximations and Taylor expansion of the exponential in the SNR-max CDF~(\ref{Eq:CDF}), in Appendix~\ref{App:SNRsmallcorr} we derived the correction to the SNR-max CDF,
\begin{align} \label{Eq:ApproxSmallcdfdiff}
\delta_I C_{\Nt}\left( Z|\boldsymbol{\Sigma}\right) & = \\ \nonumber
&\left( \operatorname{Tr} (\boldsymbol{\Sigma}^{2}) -M\right) \frac{Z^2 e^{-Z^2}}{2\pi}\left[  \text{erf}\left(  \frac
{Z}{\sqrt{2}}\right)  \right]  ^{\Nt-2}\,.%
\end{align}
Similarly, the correction to the SNR-max PDF is found by taking the derivative of the above correction to the CDF, leading to
\begin{align}\label{Eq:ApprxsmallPDF}
&\delta_I p_{\Nt}\left(  z|\boldsymbol{\Sigma}\right) =
\left( \Tr (\boldsymbol{\Sigma}^{2}) -M\right) \frac{z e^{-z^2}}{\pi} \left(\operatorname{erf}\left(\frac{z}{\sqrt{2}}\right)\right)^{\Nt-3} \nonumber \\
& \times \quad\left[\frac{(\Nt-2) z e^{-z^{2} / 2} }{\sqrt{2 \pi}}+\left(1-z^{2}\right) \operatorname{erf}\left(\frac{z}{\sqrt{2}}\right)\right]\,.
\end{align}

The advantage of these approximate formulas is that the multi-dimensional integrations in
the general SNR-max CDF and PDF expressions 
(\ref{Eq:CDF}, \ref{Eq:maxPDFcorr}) have been carried out explicitly. If the bank changes due to time evolution of network geometry, e.g., for orbiting satellites, only $\Tr(\boldsymbol{\Sigma}^{2})$ needs to be reevaluated. 
There is, however, a region of validity of these approximations (see Appendix~\ref{App:SNRsmallcorr}), 
\begin{equation}
    z \ll 1/ \sqrt{\max\left\vert A_{ij}\right\vert} \,,
    \label{Eq:Apprxsmall:ConvergenceRadius}
\end{equation}
which limits practical applicability of these approximations.

As a test of these approximate formulae, we consider a limiting case of two-template bank, see Sec.~\ref{sec:twocorrtemp}, with the closed-form SNR-max PDF~(\ref{Eq:twocorrtemppdf}). In this case, $\Tr (\boldsymbol{\Sigma}^{2}) -M =2 r^2$ and Eq.~(\ref{Eq:ApprxsmallPDF}) becomes
\begin{equation}
    \delta_I p_{2}\left(  z|r \right) =  
2r^2 \left(1-z^{2}\right)  \frac{z e^{-z^2}}{\pi} \,.
\label{Eq:ApprxsmallPDF:M2}
\end{equation}
This result is in agreement with the small-$r$ expansion of Eq.~(\ref{Eq:twocorrtemppdf}).

\subsection{SNR-max distribution for a squeezed template bank}
\label{Sec:SqueezedApprox}

In Sec.~\ref{sec:snrcorr} we encountered a practical search scenario when matrix elements of the bank covariance matrix $\mb{\Sigma}$ are centered about a non-zero value $r$ (see Fig.~\ref{Fig:tempcovelements}). 
Motivated by this scenario, in this section we consider an approximation for the SNR-max CDF when the off-diagonal matrix elements of  $\mb{\Sigma}$ exhibit small deviations  from some  value  $r$. (As follows from the derivation of this section, the optimal choice of $r$ corresponds to the average of off-diagonal matrix elements of $\mb{\Sigma}$.) This is the case of squeezed banks discussed in Sec.~\ref{sec:snrcorr}. Then $\mb{\Sigma}=\mb{\Sigma}_r + \delta \mb{\Sigma}$,
where the perturbation matrix $\delta \mb{\Sigma}$ is ``small'' compared to the highly-structured reference covariance matrix $\mb{\Sigma}_r$. To the leading order in $\delta\mb{\Sigma}$, the SNR-max CDF $C_{\Nt}(Z|\mb{\Sigma}_r + \delta\mb{\Sigma})$ is determined by $C_{\Nt}(Z|\mb{\Sigma}_r)$.
Similar arguments apply to PDF $p_{\Nt}(Z|\mb{\Sigma}_r)$.
Below, we evaluate these leading order 
contributions, leaving the cumbersome derivation of the 
$\delta\mb{\Sigma}$-correction for future work.

Explicitly, we assume that matrix elements of the reference  matrix $\mb{\Sigma}_{r}$ are given by 
\begin{equation}
  \left(\mb{\Sigma}_{r} \right)_{ij} = \delta_{ij} + (1-\delta_{ij})  r \,,  
  \label{Eq:Sigma0MelsExplicit}
\end{equation} 
i.e., all off-diagonal matrix elements of $\mb{\Sigma}_{r}$  are identical and equal to $r$. For example, $\mb{\Sigma}_r$ for $\Nt=3$ reads
\begin{equation}
    \mb{\Sigma}_r = \left(
    \begin{array}
    [c]{ccc}%
    1 & r & r\\
    r & 1 & r\\
    r & r & 1
    \end{array}
    \right) \, .
    \label{Eq:ESigma0example}
\end{equation}
In Sec.~\ref{sec:twocorrtemp} we considered a special case of such matrices for $\Nt=2$. 
It is the highly-structured nature of $\mb{\Sigma}_r$ that allows us to derive a closed-form expression for the SNR-max CDF $C_{\Nt}(Z|\mb{\Sigma}_r)$ below.

We derived the inverse of $\mb{\Sigma}_r$ entering the multivariate normal distribution. 
The inverse retains the general structure of $\mb{\Sigma}_r$ with matrix elements
\begin{equation}
    \left(\mb{\Sigma}_r^{-1} \right)_{ij} = X\delta_{ij} + Y \,,
    \label{Eq:Sigmaijinverse}
\end{equation}
where
\begin{align}
    X & = 1/(1-r) \,,
    \label{Eq:celement} \\
    Y & = - \frac{r}{1+(\Nt-1)r} X \,. \label{Eq:belement}
\end{align}
The derived inverse $\mb{\Sigma}_r^{-1}$ reproduces that for the $M=2$ case, Eq.~(\ref{Eq:Cparameters}).
With the matrix inverse~(\ref{Eq:Sigmaijinverse}), the joint PDF~(\ref{Eq:jointsnrpdf}) reads
\small\begin{align}\label{Eq:jointpdf}
    &f\left(  \boldsymbol{\rho}|\mb{\Sigma}_r\right) = 
    \frac{1}{\sqrt{\det(2\pi \mb{\Sigma}_r)}} \times \\ 
    &\exp{\left\{-\frac{1}{2(1-r)}
    \left(\sum_{i=1}^\Nt \rho^2_i - \frac{r}{1+(\Nt-1)r}\left( \sum_{i=1}^\Nt  \rho_i \right)^2 \right)
    \right\} }  \,, \nonumber
\end{align}
\normalsize
with $\det(\mb{\Sigma}_r) = \left(1-r \right)^{\Nt -1} \left(1 + (\Nt-1)r \right)$.
This joint PDF is equivalent to
\begin{align}\label{Eq:magicpdf}
&f\left(  \boldsymbol{\rho}|\mb{\Sigma}_r\right)  =\frac{1}{(2 \pi(1-r))^{\Nt / 2}} \times \\ &\int_{-\infty}^{+\infty} d u \quad \frac{e^{-u^{2} / 2}}{\sqrt{2 \pi}} \prod_{i=1}^{\Nt} \exp \left(-\frac{\left(\rho_{i}-\sqrt{r} u\right)^{2}}{2(1-r)}\right) \,, \nonumber
\end{align}
which can be verified by direct integration. Remarkably, this integral representation has the advantage of decoupling the SNRs $\rho_i$.

The properties of multi-variate normal distribution for the class of covariance matrices represented by Eq.~(\ref{Eq:Sigma0MelsExplicit}) have been considered in the literature (see Ref.~\cite{Tong1990-multivariate-book} and references therein). We warn the reader, however, that the literature contains misprints (for example, Eq.~(\ref{Eq:jointpdf}) has misprints in Ref.~\cite{Tong1990-multivariate-book}), uses different notation, and proceeds through the less transparent (at least to our taste) ``exchangeable-normal-variables'' proofs.

The CDF $C_{\Nt}(Z|\mb{\Sigma}_r)$ is still an $\Nt$-dimensional integral~(\ref{Eq:CDF}) of the multivariate normal distribution~(\ref{Eq:jointpdf}). With the alternative representation~(\ref{Eq:magicpdf}) of the joint PDF, this integral can be reduced to a one-dimensional integral,
\begin{align}\label{Eq:sqCDFSigma0}
    &C_\Nt(Z|\mb{\Sigma}_r) = \frac{1}{2^{\Nt}}\int_{-\infty}^{\infty} du \, 
    \frac{ e^{ - u^{2}/2}}{\sqrt{2 \pi}}  \times\\
    & \left[ \erf{\frac{Z+\sqrt{r}u}{\sqrt{2(1-r)}}} - \erf{\frac{-Z+\sqrt{r}u}{\sqrt{2(1-r)}}} \right]^{\Nt}. \nonumber
\end{align}
This CDF is equivalent to the result of Ref.~\cite{Tong1990-multivariate-book} expressed in a different notation. For $r=0$, we recover the CDF for independent templates, Eq.~(\ref{Eq:CDForthog}).

Differentiating this CDF with respect to $Z$, we  determine the SNR-max PDF 
\begin{align}
    &p_{\Nt}(z|\mb{\Sigma}_r)  = 
  \frac{1}{2^{\Nt}} \frac{\Nt}{\pi} \frac{1}{\sqrt{1-r^2}} \int_{-\infty}^{\infty} du \, e^{ - u^{2}/2} \times  \\ \nonumber& \left[ \erf{\frac{z+\sqrt{r}u}{\sqrt{2(1-r)}}} - \erf{\frac{-z+\sqrt{r}u}{\sqrt{2(1-r)}}} \right]^{\Nt - 1}  \\  
   & \times \left[  
    \exp\left( - \frac{(z+\sqrt{r}u)^2 }{2(1-r)} \right) + 
    \exp\left( - \frac{(z-\sqrt{r}u)^2 }{2(1-r)} \right)
    \right] \, . \nonumber
    \label{Eq:sqPDFSigma02}
\end{align}

Given the bank covariance matrix $\bs{\Sigma}$, what is the optimal choice of $r$? The answer to this question is not obvious - there are many possibilities like taking $r$ as the most probable value, or the median,  or the average of the $\Sigma_{i\neq j}$ matrix elements distribution. Below we show that 
the optimal choice of $r$ is the mean value of off-diagonal matrix elements of 
the bank covariance matrix $\bs{\Sigma}$. Indeed,  consider the small-$\delta\mb{\Sigma}$  expansion of the determinant normalizing the joint probability distribution, 
\begin{align*} 
    \det(\mb{\Sigma}_r + \delta\mb{\Sigma}) \approx 
    & \left(1+\Tr{(\mb{\Sigma}_r^{-1} \delta\mb{\Sigma})} \right) \det\mb{\Sigma}_r \\ \nonumber
    = & \left(1+Y\sum_{ij} (\delta\mb{\Sigma})_{ij} \right)\det\mb{\Sigma}_r   \, ,
\end{align*}
where $Y$ is the off-diagonal matrix element of $\mb{\Sigma}^{-1}_r$ given in Eq.~(\ref{Eq:belement}). To suppress the $\delta\mb{\Sigma}$ correction, we have to require that $r$ is chosen in such a way that  $\sum_{ij} (\delta\mb{\Sigma})_{ij} = 0$. This condition leads to the optimal choice
\begin{equation}
    r = \frac{1}{\Nt (\Nt -1)} \sum_{i>j} \Sigma_{ij} \,,
    \label{Eq:rOptimalValue}
\end{equation}
corresponding to the mean value of off-diagonal matrix elements of the bank covariance matrix.

Threshold values for specific values of false positive rates  for   squeezed template banks are discussed in Sec.~\ref{sec:thresholds}.  

\section{Threshold values for the SNR-max statistic}
\label{sec:thresholds}
Practical  applications require certain level of confidence in the detection statistic. Typically, confidence levels such as $90\%, 95\%$, and $99\%$ are suitable for many applications requiring less sensitivity far into the tails of a distribution. Threshold values are determined by false positive rates, $q$, and for the enumerated confidence levels, $q$ values are $0.1, 0.05$, and $0.01$, respectively. The meaning of a false positive rate is the probability of falsely accepting the hypothesis in question, when the data should have supported the null hypothesis. In the context of searches for transients, a false positive happens when the SNR-max statistic $z$ exceeds a certain threshold $Z^\ast$, but this in fact, is due to statistical randomness mimicking the transient and not due to the sought exotic physics. 
Suppose for a desired false positive rate, $q$, we want to determine the threshold value $Z^{\ast}$ associated with that false positive rate. We may simply relate the false positive rate to the threshold value by using the CDF
\begin{equation}
1-q=C_{\Nt}\left(  Z^{\ast}|\boldsymbol{\Sigma}\right).
\label{Eq:fpratefromcdf}
\end{equation}
As $C_{\Nt}\left(  Z |\boldsymbol{\Sigma}\right)$ is a monotonically increasing
function of $Z$, $Z^{\ast}$ increases for decreasing rate $q$.

Even without explicit computations, we qualitatively expect that for a fixed false-positive rate $q$,
\begin{enumerate}
    \item[(i)]{ Threshold value $Z^\ast$ increases with the increasing size $\Nt$ of template bank.}
    \item[(ii)]{ Threshold value $Z^\ast$ decreases with the growing  cross-template correlations.}
\end{enumerate}
Indeed, the random data have more chance to match one of the templates in a larger template bank leading to a higher threshold value, supporting statement (i).  As to the statement (ii), suppose we keep the bank size $\Nt$ the same, but increase one of the cross-template correlation coefficients $\Sigma_{ij}$ all the way to 1. Then the two templates $i$ and $j$ become identical, 
effectively reducing the bank to $\Nt-1$ templates. Then we can apply the statement (i) resulting in  $Z^\ast$ being lowered. In particular, this means that for a fixed rate of false positives, a bank of correlated templates 
has a lower SNR-max threshold than a fully-orthogonal bank (uncorrelated or independent templates.)

Evaluation of threshold values for a desired false positive rates for a large bank with arbitrary bank covariance matrix $\boldsymbol{\Sigma}$ requires Monte-Carlo integration of an $\Nt$-dimensional multivariate distribution, see Sec.~{\ref{Sec:MAXSNR}}. While Monte-Carlo integration can yield  the desired threshold value for any $\boldsymbol{\Sigma}$ and dimensionality $\Nt$, the computational cost can be prohibitive when $\Nt$ is large and when sampling far into the tail of the distribution, as needed for small false positive rates. Therefore, in this section we present several analytical results that avoid or greatly reduce numerical costs.
	
For example, if the $\Nt$ templates in the bank are fully orthogonal, i.e. $\bs{\Sigma}=\bs{I}$,  the analytic CDF is given by  Eq.~(\ref{Eq:CDForthog}) and thresholds are
	\begin{equation}
	Z^{*} = \sqrt{2}\, \mathrm{erf}^{-1}\left((1-q)^{1/\Nt}\right)\,.
	\label{Eq:threshnocorr}
	\end{equation}
This formula illustrates our qualitative statement (i). For $q \ll 1$, $Z^{*}$
is given by the conventional threshold formula for a Gaussian distribution,
$ Z^{*} = \sqrt{2}\, \mathrm{erf}^{-1}\left(1-q_\mr{eff}\right)$ with
$q_\mr{eff} = q/\Nt$. As $\Nt$ is increased, $q_\mr{eff}$ decreases, resulting in larger values of $ Z^{*}$.

\subsection{Threshold for two-template bank}
\label{Sec:twotempthresh}

The discussion for the case of the two correlated template SNR-max CDF in Sec.~\ref{sec:twocorrtemp} suggests that for a fixed value of $q$, the threshold value $Z^{\ast}$ moves to lower
values the more correlated the templates are,%
\begin{equation}
Z^{\ast}\left(  r=1\right)  <Z^{\ast}\left(  r\right)  <Z^{\ast}\left(
r=0\right).
\label{Eq:threshqualitative}
\end{equation}

\begin{figure}[h!]
	\begin{center}
		\includegraphics[width=0.42\textwidth]{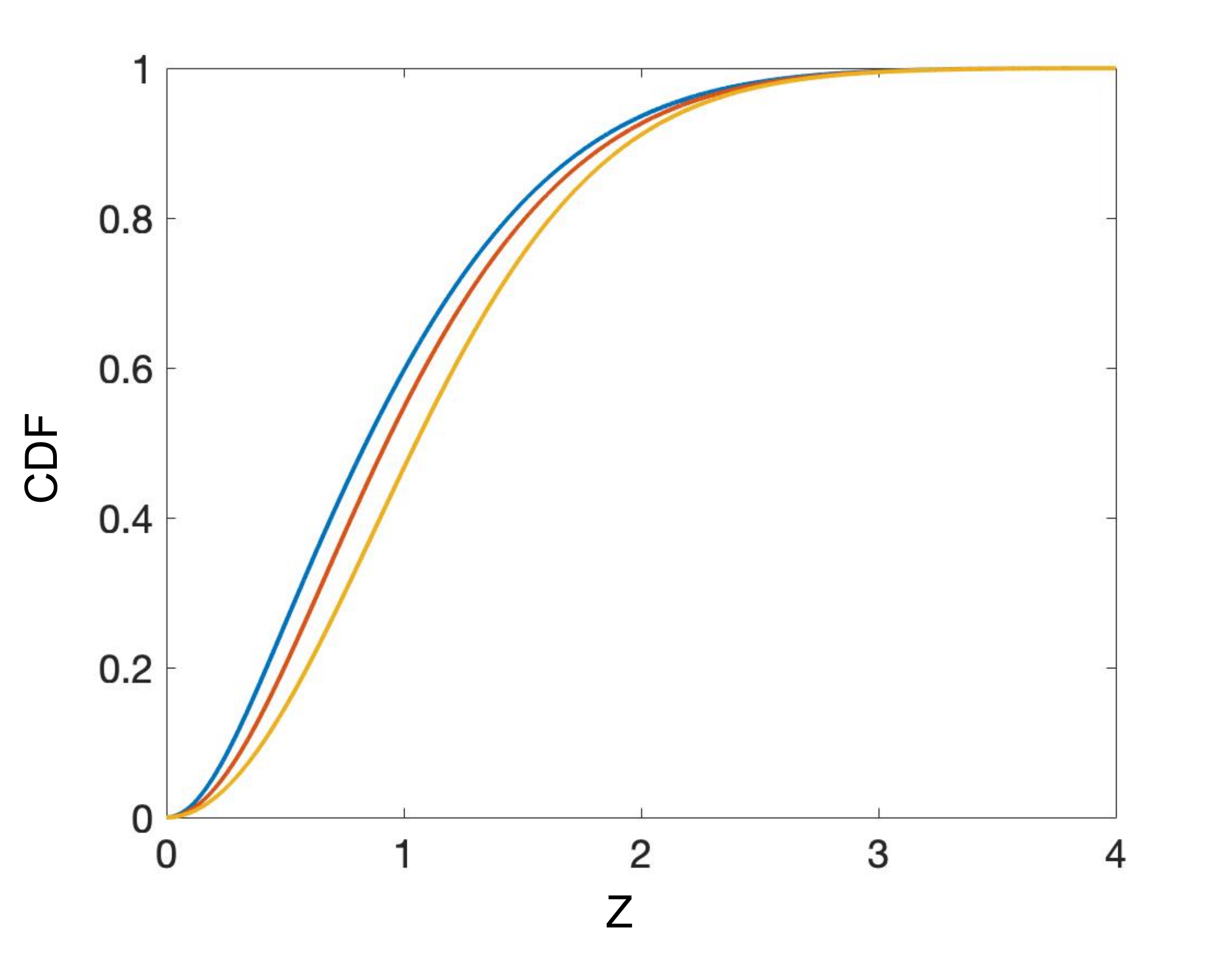}
	\end{center}
	\caption{(Color online) SNR-max CDF for a bank of two templates with varying correlation coefficient: $r=0$ (yellow curve), $r=0.75$ (red curve), and $r=0.9$ (blue curve). The inflection points correspond to the maxima of PDFs.} 
	\label{Fig:SNRmaxCDF}
\end{figure}

Fig.~\ref{Fig:SNRmaxCDF} shows the effect of correlations  on the SNR-max CDF for the case of two templates. Here we vary the correlation coefficient in the range $0 \le r \le 0.9$. It is evident from comparing the $r=0.75$ (red) and the $r=0.9$ (blue) curves that as the correlation increases, the CDF approaches its limiting value of 1 quicker. 
Then the thresholds $Z^{*}$ for specific false positive rates $q$ decrease with increasing cross-template correlation $r$, which is in agreement with the qualitative statement (ii).


Table \ref{Tab:1} compiles the SNR-max threshold values for two templates with varying correlations $r$: $0$, $0.25$, $0.5$, and $0.9$, and for varying false positive rates $q$: $10^{-2}$, $10^{-4}$, $10^{-6}$, and $10^{-8}$. Examination of the table shows that the highest threshold values occur for the case of uncorrelated templates regardless of the false positive rate $q$. For very small false positive rates $q$, the effect of correlation is nearly negligible on the threshold value unless the correlation is very strong, $r\approx1$. Whereas increasing the desired false positive rate, the effect of correlation on the threshold value becomes more pronounced even for mild correlations $r\approx0.25$. This effect of cross-template correlations on thresholds  becomes even more pronounced as we increase the number of templates in the  bank,
see Sec.~\ref{Sec:squeezedbankthresh}.

\begin{table}[h!]
\begin{ruledtabular}
\begin{tabular}{ccccc}
$q$&
$r=0$&
$r=0.25$&
$r=0.5$&
$r=0.9$\\
\colrule

$10^{-8}$ & 5.848 & 5.848 & 5.848 & 5.833\\
$10^{-6}$ & 5.027 & 5.027 & 5.027 & 5.003\\
$10^{-4}$ & 4.056 & 4.056 & 4.054 & 4.014\\
$10^{-2}$ & 2.807 & 2.805 & 2.795 & 2.716\\
\end{tabular}
\end{ruledtabular}
\caption{Threshold values for a template bank consisting of $\Nt=2$ templates.
Threshold values are tabulated for varying cross-template correlation coefficients, $0\leq r \leq 0.9$ and  false positive rates, $10^{-8} \leq q \leq 10^{-2}$. }
 \label{Tab:1}
\end{table}


\subsection{Threshold for a nearly-orthogonal template bank}
\label{Sec:nearlyorthogthresh}

Suppose we can represent the bank covariance matrix as a sum of  the reference matrix  $\mb{\Sigma}_0$ and a small correction $\delta \mb{\Sigma}$. This is the case for a nearly-orthogonal template bank of  Sec.~\ref{sec:smallapprox}, where  $\mb{\Sigma}_0 = \mb{I}$, and for a ``squeezed'' bank of Sec.~\ref{sec:smallapprox}, $\mb{\Sigma}_0 = \mb{\Sigma}_r$. Then 
the thresholds for specific false positive rates can be calculated in a perturbative fashion.  The dominant term in calculating the threshold value is determined by the CDF  $C_{\Nt}(Z^*|\mb{\Sigma}_0)$ with a corrective term $\delta C_{\Nt}(Z^*|\mb{\Sigma}_0)$,
\begin{equation}
    C_{\Nt}(Z|\mb{\Sigma}) = C_{\Nt}(Z|\mb{\Sigma}_0) + \delta C_{\Nt}(Z|\mb{\Sigma}_0) \, .
    \label{Eq:smallcdfapprox}
\end{equation}
The threshold is determined from Eq.~(\ref{Eq:fpratefromcdf}), $1-q = C_{\Nt}(Z^*|\mb{\Sigma})$, with $Z^*  =Z_0^* + \delta Z^*$. Here $Z_0^*$ is the reference threshold and $\delta Z^*$ is the correction to the threshold. The reference threshold satisfies the equation $1-q = C_{\Nt}(Z^*_0|\mb{\Sigma}_0)$.
We would like to find the correction  to the threshold $\delta Z^*$, so that
\begin{equation}
    1-q = C_{\Nt}(Z^*_0 + \delta Z^*|\mb{\Sigma}_0) + \delta C_{\Nt}(Z^*_0|\mb{\Sigma}_0) \, .
    \label{Eq:approxthresholds}
\end{equation}
Explicitly,  
\begin{align*} \label{Eq:approxCDFthresholds}
    C_{\Nt}(Z^*_0 + \delta Z^*|\mb{\Sigma}_0) & \approx C_{\Nt}(Z^*_0|\mb{\Sigma}_0) +   \delta Z^* \times \frac{\partial C_{\Nt}}{\partial Z} \\
    & =  C_{\Nt}(Z^*_0|\mb{\Sigma}_0) +   \delta Z^* \times p_{\Nt}(Z^*_0|\mb{\Sigma}_0). \nonumber
\end{align*}
From here, the correction  to the threshold is
\begin{equation}
    \delta Z^* \approx - \frac{\delta C_{\Nt}(Z^*_0|\mb{\Sigma}_0)}{p_{\Nt}(Z^*_0|\mb{\Sigma}_0)} \,.
    \label{Eq:approxvariationCDFterm}
\end{equation}

 Now we apply this formalism to the determination of the threshold for a nearly-orthogonal template bank, i.e. $\bs{\Sigma}_0=\bs{I}$. Then $Z^{*}_0 = \sqrt{2}\, \mathrm{erf}^{-1}\left((1-q)^{1/\Nt}\right)$, Eq.~(\ref{Eq:threshnocorr}). The corrective CDF term and the SNR-max PDF in Eq.~(\ref{Eq:approxvariationCDFterm}) are per Sec.~\ref{sec:smallapprox},
\begin{align*}
    \delta C_{\Nt}(Z_0^*|\bs{I}) & = \Tr{\left(  \boldsymbol{A}%
^{2}\right)}  \frac{Z_0^{\ast2} e^{-Z_0^{\ast 2}}}{2\pi}\left[  \text{erf}\left(  \frac
{Z^\ast_0}{\sqrt{2}}\right)  \right]  ^{\Nt-2}, \\
    p_{\Nt}(Z^*_0|\bs{I}) & = \Nt \sqrt{\frac{2}{\pi}}e^{-\frac{Z_0^{\ast 2}}{2}%
}\text{erf}\left(  \frac{Z^\ast_0}{\sqrt{2}}\right)^{\Nt-1} \,. \nonumber
\label{Eq:approxtermsCDFthreshexample}
\end{align*}
Plugging these into Eq.~(\ref{Eq:approxvariationCDFterm}), we arrive at the correction to the threshold value, 
\begin{equation}
    \delta Z^* = -\frac{\Tr{\left(  \boldsymbol{A}%
^{2}\right)} }{\Nt}\left[\text{erf}\left(  \frac
{Z^\ast_0}{\sqrt{2}}\right)  \right]^{-1} Z_0^{\ast 2} e^{-Z_0^{\ast 2}} \frac{1}{2\sqrt{2\pi}}.
\label{Eq:approxthreshvariationterm}
\end{equation}
Since $\Tr{\left(  \boldsymbol{A}%
^{2}\right)} > 0$, we see that $ \delta Z^\ast < 0$, so that $Z^\ast < Z_0^\ast$.
This is consistent with our qualitative expectation (ii) that cross-template correlations reduce the threshold values for a given rate of false positives. 

We remind the reader that for a nearly-orthogonal bank, $ \Tr{\left(  \boldsymbol{A}^{2}\right)} = \Tr (\boldsymbol{\Sigma}^{2}) -M$. 
The advantage of our approximation~(\ref{Eq:approxthreshvariationterm}) is that if the bank changes due to time evolution of network geometry, e.g., for orbiting satellites, only $\Tr(\boldsymbol{\Sigma}^{2})$ needs to be reevaluated to readjust the threshold for the new bank.

\subsection{Threshold for squeezed template bank}
\label{Sec:squeezedbankthresh}

Squeezed template banks introduced in Sec.~\ref{Sec:SqueezedApprox} have a special structure of the bank covariance matrix,  $\mb{\Sigma} \approx \mb{\Sigma}_r$, in which all off-diagonal matrix elements have the same value of $r$ (mean value of cross-template correlation coefficients in the bank). This is directly relevant to the GPS.DM search for DM bubble walls.
Analytic SNR-max CDF~(\ref{Eq:sqCDFSigma0}) $C_{\Nt}(Z|\mb{\Sigma}_r)$ was developed in Sec.~\ref{Sec:SqueezedApprox}. This representation sharply reduces  computational cost in determining the thresholds for specific false positive rates, as the $\Nt$-dimensional integral is recast into a one-dimensional integral.

\begin{figure}[h!]
	\begin{center}
		\includegraphics[width=0.42\textwidth]{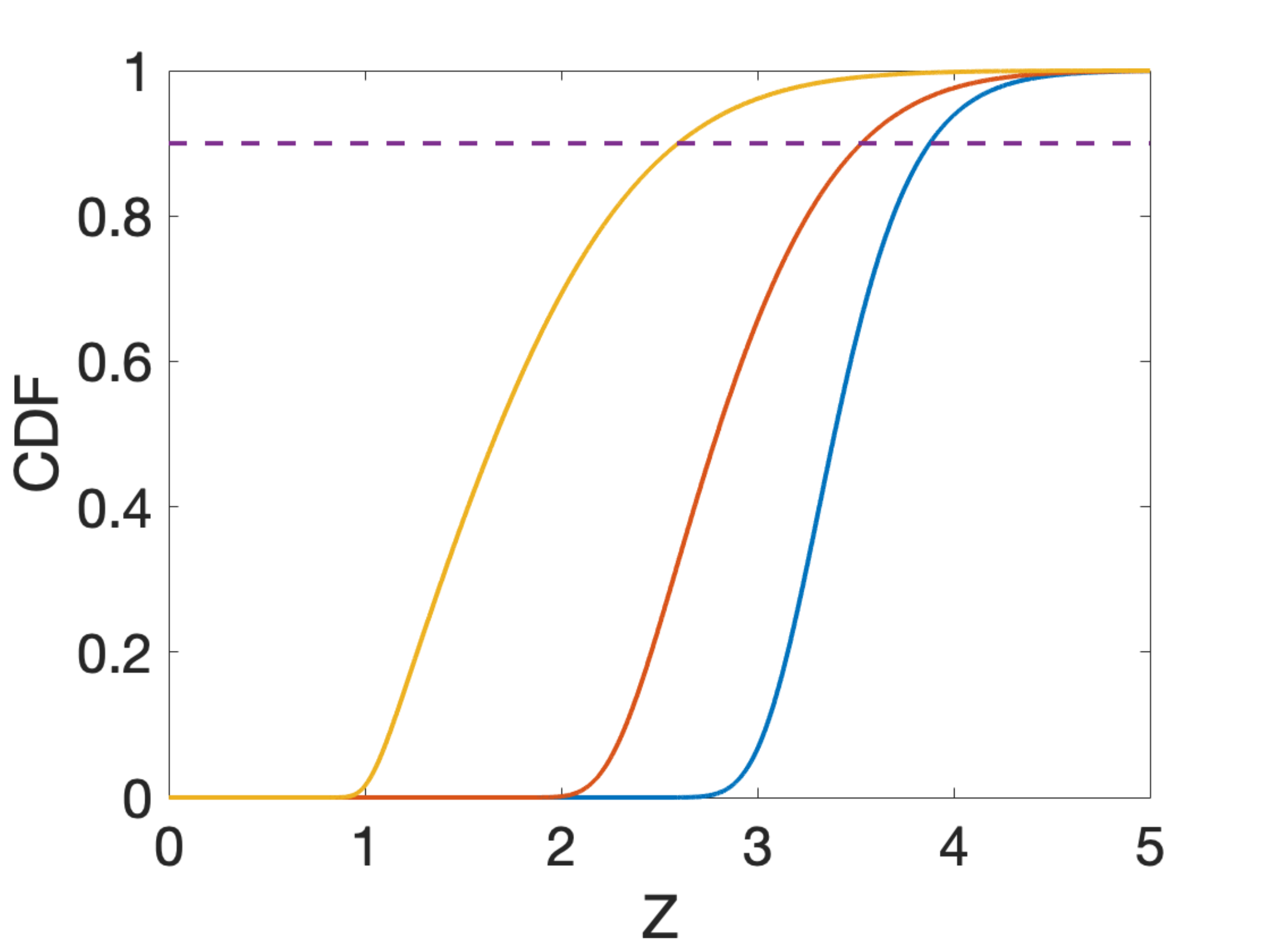}
	\end{center}
	\caption{(Color online) SNR-max CDFs $C_{\Nt}(Z|\mb{\Sigma}_r)$ for a perfectly squeezed  bank of $\Nt=1024$ templates for various values of cross-template correlation coefficient: $r=0$ (blue curve), $r=0.5$ (red curve), and $r=0.9$ (yellow curve). The horizontal purple dashed line is drawn at  
	a constant value of 0.9 corresponding to the rate of false positives $q=0.1$. Intersections of the CDF curves with the horizontal line determine the values of the threshold $Z^\ast$. Increasing cross-template correlations causes a decrease in thresholds values $Z^*$.
	} 
	\label{Fig:SNRmaxfullbankCDF}
\end{figure}

We used the SNR-max CDF~{(\ref{Eq:sqCDFSigma0})} for squeezed banks and Eq.~(\ref{Eq:fpratefromcdf}) to determine SNR-max threshold values for desired rates of false positives. The results are presented in Fig.~\ref{Fig:SNRmaxfullbankCDF} and Table~\ref{Tab:2}.
Fig.~\ref{Fig:SNRmaxfullbankCDF} shows the SNR-max CDFs for a  bank of $\Nt=1024$ templates with  cross-template correlation coefficient varying in the range $0 \le r \le 0.9$.  It is clear that the effect of cross-template correlation strongly affects the underlying CDF: for larger $r$, the CDF  is quickly shifted toward lower values of $Z$. The values of SNR-max threshold are determined by the intersection of the computed CDFs with the horizontal purple dashed line corresponding to a fixed false positive rate.  Increasing cross-template correlations causes decreasing thresholds $Z^*$ in agreement with our qualitative expectation (ii) and discussion for $\Nt=2$ banks. We can understand this shift in the CDF and thresholds, by thinking about how the inflection points (located at the most probable value) vary between the different PDF's associated with the correlation value $r$.

Numerical values for the SNR-max thresholds are compiled in Table~\ref{Tab:2}. Once again we observe that the threshold $Z^*$ decrease with increasing correlation value $r$ for all the tabulated false positive rates. For $r=0.25$ and $r=0.5$ the decrease in the threshold is nearly negligible, while for larger $r=0.9$ the decrease in the threshold becomes much more apparent, similar to the case of $\Nt=2$ template bank. We also see that as the false positive rate decreases, the effect of correlation becomes more drastic on the threshold values.

\begin{table}[h!]
\begin{ruledtabular}
\begin{tabular}{cccccc}
$q$&
$r=0$&
$r=0.25$&
$r=0.5$&
$r=0.75$&
$r=0.9$\\
\colrule

$10^{-8}$ & 6.807 & 6.807 & 6.804 & 6.733 & 6.462\\
$10^{-6}$ & 6.109 & 6.109 & 6.097 & 5.972 & 5.705\\
$10^{-4}$ & 5.327 & 5.325 & 5.283 & 5.076 & 4.745\\
$10^{-2}$ & 4.417 & 4.390 & 4.247 & 3.903 & 3.487\\
\end{tabular}
\end{ruledtabular}
\caption{Compilation of SNR-max statistic threshold values $Z^\ast$ for 
for a perfectly squeezed  bank of $\Nt=1024$ templates. $r$ cross-template correlation coefficient and $q$ is false positive rate.}
 \label{Tab:2}
\end{table}


For completeness, we investigate a more practical case of how the thresholds vary for template banks with template covariance matrix off-diagonal elements not all equal to $r$, but there is a spread in correlation elements values.
As a specific example,  we consider a template bank composed of $\Nt=3$ templates with bank covariance matrix
\begin{equation}
    \mb{\Sigma} = \left(
    \begin{array}
    [c]{ccc}%
    1 & r_{12} & r_{13}\\
    r_{12} & 1 & r_{23}\\
    r_{13} & r_{23} & 1
    \end{array}
    \right) \, .
    \label{Eq:ESigmavaryexample}
\end{equation}
We fix $r_{12} = 0.33$, $r_{13} = 0.23$, and $r_{23} = 0.43$. This bank has a 0.2 spread in cross-template correlation coefficients about mean value of $r=0.33$. In  Table~\ref{Tab:3}, we present results of two threshold computations: 
the Markov-Chain Monte-Carlo (MCMC) method using publicly-available package~\cite{mvNcdf} for integration of the exact CDF~(\ref{Eq:CDF}),
and approximate squeezed-bank CDF~(\ref{Eq:sqCDFSigma0}) with $r=0.33$.
For this bank, the two thresholds agree to three significant figures.

\begin{table}[h!]
\begin{ruledtabular}
\begin{tabular}{ccc}
$q$&
Exact, Eq.~(\ref{Eq:CDF})&
Approx., Eq.~(\ref{Eq:sqCDFSigma0})\\
\colrule
$10^{-8}$ & 5.913 & 5.915\\
$10^{-6}$ & 5.103 & 5.104\\
$10^{-4}$ & 4.149 & 4.150\\
\end{tabular}
\end{ruledtabular}
\caption{Comparison of threshold values from Eq.~(\ref{Eq:CDF}) and Eq.~(\ref{Eq:sqCDFSigma0}) for the case of $\Nt=3$ template bank (see Eq.~(\ref{Eq:ESigmavaryexample}) and text) and varying false positive rates, $10^{-8} \leq q \leq 10^{-4}$.}
 \label{Tab:3}
\end{table}


 As a next numerical experiment, we increase the number of templates to $\Nt=1024$ for a squeezed template bank, with a distribution of cross-template correlations $\Sigma_{ij}$ shown in Fig.~\ref{Fig:squeeztempbank}.
 In this 
 bank, the most probable value is 0.390 with a range of total correlation values being $0.25 \le \Sigma_{ij}\le 0.65$. The median value of the $\Sigma_{ij} $ distribution is 0.395, and an average value $r=0.398$. First, we compute the SNR-max threshold values using the MCMC~\cite{mvNcdf}. MCMC technique evaluates the exact $\Nt$-dimensional integral CDF~(\ref{Eq:CDF}); the results are presented
 in Table~\ref{Tab:4}. Second, we use the approximate CDF $C_{\Nt}(Z|\mb{\Sigma}_r)$ with the most probable 0.390 and an average value of 0.398. The thresholds are evaluated for the false positive rates $q=10^{-4}$, $q=10^{-6}$, and $q= 10^{-8}$.

Table~\ref{Tab:4} demonstrates that choosing the reference cross-template correlation $r$ as the mean value of $\Sigma_{i \ne j}$ (versus most probable value) results in closer agreement with the exact result. This is consistent with our arguments presented in Sec.~\ref{Sec:SqueezedApprox}. The approximation $\mb{\Sigma} \approx \mb{\Sigma}_r$ works better for smaller values of false positives. Even for $q=10^{-8}$, the approximation is in $2\%$ agreement with the MCMC result, demonstrating the utility of our approximation.

\begin{table}[h!]
\begin{ruledtabular}
\begin{tabular}{cccc}
$q$&
Exact&
$r=0.390$&
 $r=0.398$\\ 
\colrule
$10^{-8}$ & 6.67 & 6.8063 & 6.8062\\
$10^{-6}$ & 6.04 & 6.1073 & 6.1070\\
$10^{-4}$ & 5.31 & 5.3122 & 5.3108\\
\end{tabular}
\end{ruledtabular}
\caption{Comparison of threshold values from the exact Eq.~(\ref{Eq:CDF}) (computed with MCMC, column marked ``exact'') and Eq.~(\ref{Eq:sqCDFSigma0}) (both for the most probable (middle column) and average correlation matrix element value (last column)) for the case of $\Nt=1024$ templates (typical template bank vs. perfectly squeezed template bank) and varying false positive rates, $10^{-8} \leq q \leq 10^{-4}$.}
 \label{Tab:4}
\end{table}


\subsection{Threshold for varying number of templates $\Nt$ in template bank}
\label{Sec:varyMthresh}

\begin{figure*}[ht!]
\center
	\begin{center}
		\includegraphics[width=0.9\textwidth]{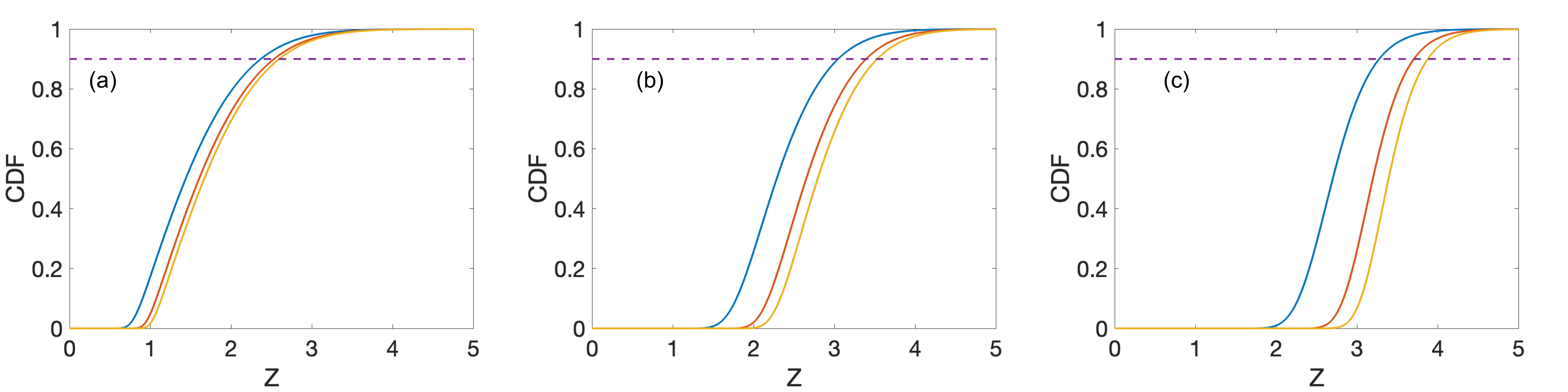}
	\end{center}
	\caption{(Color online) SNR-max CDFs $C_{\Nt}(Z|\mb{\Sigma}_r)$ for a varying number of templates in a  perfectly-squeezed bank: $\Nt = 100$ (blue curve), $\Nt=500$ (red curve), and $\Nt=1000$ (yellow curve).  The  false positive rate is fixed at $q=0.1$ (horizontal purple dotted line is drawn at the $1-q=0.9$ value). The thresholds are determined by the value of $Z$ at the intersection of the CDF curves with the false positive rate horizontal line. 
	Three panels correspond to different values of the cross-template correlation  coefficient $r$: (a) $r=0.9$, (b) $r=0.5$, and (c)  $r=0.1$. }
	\label{Fig:SNRvarybankCDF}
\end{figure*}

SNR-max thresholds for specific false positive rates depend on the number of templates in the template bank. 
{Fig.~\ref{Fig:SNRvarybankCDF} shows the effect of the number of templates $\Nt$ on the thresholds for varying size of  perfectly-squeezed  banks, i.e. with bank covariance matrix (\ref{Eq:Sigma0MelsExplicit})}. In all three panels of that figure, the rate of false positives is fixed (horizontal line at 0.9). In all panels, we plot the same CDFs $C_{\Nt}(Z|\mb{\Sigma}_r)$ for $\Nt = 100$ (blue curve) to $\Nt = 500$ (red curve) to $\Nt=1000$ (yellow curve). The value of the cross-template coefficient $r$ decreases from  $r=0.9$ (panel (a)) to 0.5 in panel (b) and to 0.1 in panel (c). 
For all three panels, we observe that as the number of templates increases from $\Nt = 100$  to $\Nt = 500$ and to $\Nt=1000$ the threshold  (intersection of the horizontal purple dotted line with the CDF curves) shifts to higher $Z$ values. As the correlation coefficient decreases, panels (a) $\rightarrow$ (c),  the differences (spread) in the threshold values increase. These observations further support our qualitative statements (i) and (ii): 
generically, we expect a larger threshold value with increasing number of templates in the template bank, as well as an increasing threshold value with decreasing cross-template correlations.  


\section{Conclusions and the utility of entanglement }
\label{sec:conclusion}

The signal-to-noise ratio (SNR) statistic and template bank searches have seen a variety of applications in modern physics and other fields. Here we focused on the max-SNR detection statistic in the context of searches for transient signals with networks of precision quantum sensors. For large template banks, especially for dynamically evolving networks,
computation of threshold rates is computationally prohibitive. We developed several approximate methods for computing properties of the 
SNR-max statistic that forgo much of the computational costs.
We developed the probability and cumulative distribution for the SNR-max statistic correlated from $\Nt$ templates in a template bank. Generically the cross-template correlations give rise to diminished thresholds for desired false positive rates. 
While our paper was motivated by dark matter and other exotic physics searches, the results of Secs.~\ref{sec:signaltempgen}---\ref{sec:thresholds} have a wide range of applicability. We also anticipate that our detailed expos\`{e} on template bank construction and optimization in the cases of the pedagogical ``toy'' network and the GPS network (Sec.~\ref{sec:signaltempgen}) will be useful in the nascent applications of precision quantum sensors and their networks in fundamental physics.

What is the utility of entanglement in the searches for transients? For individual sensors, the entanglement or spin-squeezing is a useful resource, as it improves the accuracy of a single-shot measurement typical of a search for short transients. Cross-node or geographically-distributed entanglement~\cite{Komar2016} is not useful and is, in fact,  detrimental to the network searches for transients. Indeed, projective measurement on a single node collapses the distributed wave-function, effectively rendering all other nodes deaf to the transient. Then the network loses both velocity and angular resolution.

\section*{Acknowledgments}
We would like to thank M. Pospelov, J. Singh, and T. Tscherbul for discussions.
This work was supported in part by the U.S. National Science Foundation grant PHY-1806672.

\appendix

\section{Derivation of matrix elements of template bank covariance matrix $\bm{\Sigma}$, Eq.~(\ref{Eq:tempcovelements}) }
\label{App:tempcorr}
Matrix elements of the template bank covariance matrix are defined as
\begin{equation}
    \Sigma_{ij} = \langle \rho_i^w \rho_j^w \rangle \,,
\end{equation}
where SNR statistics $\rho_i^w$ and $\rho_j^w$ are given by Eq.~(\ref{Eq:snr}) and the averaging is over signal-free data. From here on we drop the window superscript $w$. Explicitly,
\begin{equation}
\langle \rho_i \rho_j \rangle  =\frac{\langle (\mathbf{d}^{T}\mathbf{E}^{-1}%
\mathbf{s}_i) (\mathbf{d}^{T}\mathbf{E}^{-1} \mathbf{s}_j)\rangle}{
\sqrt{(\mathbf{s}_i^{T}\mathbf{E}^{-1}\mathbf{s}_{i}) (\mathbf{s}_j^{T}\mathbf{E}^{-1}\mathbf{s}_{j})}} \,.
       \label{AppA:Eq:CovExplicit}
\end{equation}
To carry out the averaging, we expand the products in component form
\begin{equation}
    \mb{d}^{T} \mb{E}^{-1} \mb{s}_{i}=\sum_{\alpha, \beta} d_{\alpha} E_{\alpha \beta}^{-1} (s_{\beta})_i, 
\end{equation}
where we used Greek letters for compound indices that encode both the sensor  and the epoch indexes, see Sec.~\ref{sec:SNRprobsetup}.
Then the numerator of Eq.~(\ref{AppA:Eq:CovExplicit}) can be rearranged into
\begin{align}
    &\left\langle\sum_{\alpha, \beta, \gamma, \delta} d_{\alpha} E_{\alpha \beta}^{-1} (s_{\beta})_i d_{\gamma} E_{\gamma \delta}^{-1} (s_{\delta})_j\right\rangle=   \nonumber \\
    &=\sum_{\alpha, \beta, \gamma, \delta}\left\langle d_{\alpha} d_{\gamma}\right\rangle E_{\alpha \beta}^{-1} E_{\gamma \delta}^{-1} (s_{\beta})_i (s_{\delta})_j \,.  \label{AppA:Eq:tmp} 
\end{align}
For signal-free data, the combination $\left\langle d_\alpha d_\gamma\right\rangle$  is the matrix element of the noise covariance matrix,
$
     \left\langle d_\alpha d_\gamma\right\rangle = E_{\alpha\gamma} \,.
$
This simplifies Eq.~(\ref{AppA:Eq:tmp}) to
\begin{align}
&\sum_{\beta, \delta} (s_{\beta})_i (s_{\delta})_j \sum_{\alpha, \gamma} E_{\alpha \gamma}\left(E^{-1}\right)_{\gamma \delta} \left(E^{-1}\right)_{\alpha \beta} = \\ \nonumber
&\sum_{\alpha, \beta} (s_{\alpha})_j\left(E^{-1}\right)_{\alpha \beta} (s_{\beta})_i =\mb{s}_j^{T} \mb{E}^{-1} \mb{s}_{i}.
\label{Eq:Acovnumertorfinal}
\end{align}
Here we  used the identity, 
$ \sum_{\gamma} E_{\alpha \gamma} (E^{-1})_{\gamma \delta}= \bm{I}_{\alpha \delta}= \delta_{\alpha \delta}.
$
With this simplification, the template bank covariance matrix element (\ref{AppA:Eq:CovExplicit})  becomes
\begin{equation}
    \Sigma_{ij} = \frac{ \mb{s}_{j}  ^{T}\mathbf{E}^{-1}\mb{s}_{i}}{\sqrt{ \left(   \mb{s}_{i} ^{T}\mathbf{E}^{-1}\mb{s}_{i} \right)  \left(   \mb{s}_{j}  ^{T}\mathbf{E}^{-1}\mb{s}_{j} \right) }} \, .  
    \label{Eq:Afulltempcovmatrix}
\end{equation}
From here, the variances of individual SNR statistics are $\sigma^2_{\rho_i} = \langle  (\rho_{i})^2 \rangle = \Sigma_{ii}= 1$, in agreement with Ref.~\cite{Panelli2020}. This concludes our proof of Eq.~(\ref{Eq:tempcovelements}).

\section{Template bank covariance matrix element construction}
\label{App:SNRcovelements}

We are interested in the behavior of the template covariance matrix elements $\Sigma_{ij}$ for our generalized signal template construction (Sec.~\ref{sec:signaltempgen}).
To streamline notation, we rename $\mb{s}_i$ and $\mb{s}_j$ to $\mb{s}$ and $\mb{s}'$ respectively.
The template covariance matrix element (\ref{Eq:tempcovelements}) in this notation reads 
\begin{equation}
    \Sigma_{ss'}=  \frac{\mb{s}^{T}\mathbf{E}^{-1}\mb{s'}}{\sqrt{ \left(  \mb{s}^{T}\mathbf{E}^{-1}\mb{s} \right)  \left(\mb{s'}^{T}\mathbf{E}^{-1}\mb{s'} \right) }}  . 
   \label{Eq:Btempcovelements}
\end{equation}
 For concreteness, we focus on a white noise network of identical sensors $a$ referenced to a common reference sensor $R$. In this case, the  inverse covariance matrix is given by Eq.~(\ref{Eq:whitenetEinvcov}), which we reproduce below 
\begin{equation}
    \left(E^{-1}\right)_{lm}^{a b} = A \, \delta_{lm}\delta^{ab} + B \,  (1-\delta^{ab})\delta_{lm},
    \label{Eq:BwhitenetEinvcovsimp}
\end{equation}
where coefficients
\begin{align}
   A &= \frac{1}{\sigma^{2}}\frac{1 +(1-1/\Ns)\xi }{1 + \xi} \,, \\
B &= - \frac{1}{\sigma^{2}}\frac{1}{\Ns}\frac{\xi}{1 + \xi} 
\end{align}
are expressed in terms of $\xi = \Ns \sigma_{R}^{2}/\sigma^2$.
In our index convention, the letters at the beginning of alphabet $a,b,\ldots$ enumerate the sensors (range $\overline{1,\Ns}$) , while letters in the middle of alphabet $l,m,\ldots$ --- the epochs (index range $\overline{1,\Jw}$ in a data window).

Since we consider the signal amplitude $h$ to be the same for the network and reference sensors as described in Sec.~\ref{sec:signaltempgen}, we may simply take out $h$ from the templates. Then the GPS.DM templates (thin DM walls) can be represented using Kronecker symbols,
\begin{equation}
    s_l^a = \delta_{l,l_a} + (-1)\delta_{l,l_R} \,. \label{Eq:BSignal}
\end{equation}
The signal for each sensor $a$ consists of a positive spike at epoch $l_a$, and a negative spike at the reference epoch $l_R$ which is identical for all the templates, $l_R = (\Jw+1)/2$. Then, the individual templates are characterized by the set of network sensor perturbation epochs
\begin{equation}
    \{l\} = {l_1,l_2,...,l_{\Ns}} \,.   \label{Eq:Bsignalsensorepochs}
\end{equation}
for the template $\bm{s}$ and a similar (but not identical) set  $\{l'\}$ for the template $\bm{s}'$.

In general, $l_a$ can be equal to $l_R$, meaning that the sensor $a$ and the reference sensor are affected by the transient in the same epoch; in this case the signal reduces to zero, $s_l^a = 0$. In addition, several sensors can be affected by the transient signal in the same epoch. In this case, we refer to this subset of sensors as  ``degenerate''. When sensor $a$ is degenerate with sensor $b$ for template $\bm{s}$, $l_a = l_b$ in the set  $\{l\}$.

In our derivations below, we will use the fact that $l_a$ and $l_R$ are guaranteed to be within the same template window, so that 
\begin{equation}
    \sum_l \delta_{l,l_a} = \sum_l \delta_{l,l_R} = 1.
    \label{Eq:BSumdeltaepochs}
\end{equation}

First, consider the numerator of Eq.~(\ref{Eq:Btempcovelements}), $\mb{s}^{T}\mathbf{E}^{-1}\mb{s}'$, which can be expanded into
\begin{align} \label{Eq:Bcorrnumerator}
    \mb{s} ^{T}\mathbf{E}^{-1}\mb{s}' &= \\ \nonumber
    & A \sum_{a,b}^{\Ns}\sum_{l,m}^{\Jw} s^a_l \delta^{ab} \delta_{lm}\, s'^{b}_m + B \sum_{a}^{\Ns}\sum_{b\ne a}^{\Ns}\sum_{l,m}^{\Jw} s^a_l \delta_{lm} s'^{b}_m.
\end{align}

The first sum can be represented as
\begin{align*}
    \sum_{l,a} s^a_l s'^{a}_l &= \sum_{l,a} \left(\delta_{l,l_a} - \delta_{l,l_R} \right) \left(\delta_{l,l_a^{'}} - \delta_{l,l_R} \right) = \\ \nonumber
  & \left( \sum_{l,a} \delta_{l,l_a}\delta_{l,l_a^{'}} \right) + \\
  & \left( \sum_{l,a} \delta_{l,l_R} - \sum_{l,a} \delta_{l,l_R}\delta_{l,l_a} - \sum_{l,a} \delta_{l,l_R}\delta_{l,l_a'}   \right) \\
  & \equiv  K_1^{s,s'} + R_{1}^{s,s'}.
    \label{Eq:Bfirsttermsum}
\end{align*}
Here we introduced 
\begin{equation}
    K_1^{s,s'} \equiv \sum_a \delta_{l_a,l_a^{'}},
    \label{Eq:BKqssp}
\end{equation}
which counts the number of sensors with a perturbation epoch identical to both templates. If the two templates are identical, $\mb{s} = \mb{s}'$, then $K_1^{s,s} = \Ns$.
Similarly,
\begin{equation}
    R_{1}^{s,s'} \equiv \Ns - \sum_{a} \left(  \delta_{l_R,l_a} +  \delta_{l_R,l_a^{'}} \right)
    \label{Eq:BR1ssp}
\end{equation}
characterizes the  network-reference sensor  degeneracy.
If both templates have no degeneracies between network and the reference sensors, then $R_{1}^{s,s'} = \Ns$.

The second sum in Eq.~(\ref{Eq:Bcorrnumerator}) simplifies to
\begin{align}
    & \sum_{a}^{\Ns}\sum_{b\ne a}^{\Ns}\sum_{l}^{\Jw} s^a_l s'^{b}_l = \\ \nonumber
    & \sum_{a}^{\Ns}\sum_{b\ne a}^{\Ns}\sum_{l}^{\Jw} \left(\delta_{l,l_a} - \delta_{l,l_R} \right) \left(\delta_{l,l_b'} - \delta_{l,l_R} \right) = \\ \nonumber
    & \sum_{a}^{\Ns}\sum_{b\ne a}^{\Ns} \delta_{l_a,l_b'} + \sum_{a}^{\Ns}\sum_{b\ne a}^{\Ns} (1) 
    - \sum_{a}^{\Ns}\sum_{b\ne a}^{\Ns} \delta_{l_R,l_a}
    - \sum_{a}^{\Ns}\sum_{b\ne a}^{\Ns} \delta_{l_R,l_b'}   \\
   & \equiv K_2^{s,s'} + R_{2}^{s,s'} \, ,\nonumber 
    \label{Eq:Bsecondtermsum}
\end{align}
with 
\begin{align}
K_2^{s,s'} & \equiv \sum_{a}^{\Ns}\sum_{b\ne a}^{\Ns} \delta_{l_a,l_b^{'}} \,,\\ R_{2}^{s,s'} & \equiv \Ns(\Ns -1) -  \sum_{a}^{\Ns}\sum_{b\ne a}^{\Ns} \left( \delta_{l_R,l_a}
    + \delta_{l_R,l_b^{'}}\right)  \\
    &=(\Ns-1)R_1^{ss'} \,. \nonumber
 \end{align}   
$K_2^{s,s'}$ defines the cross-template network degeneracy between network sensors $a,b \ne a$. $R_{2}^{s,s'}$ quantifies the cross-template degeneracy between the reference and the network sensors. If there are no degeneracies between the network sensors $a,b \ne a$ and the reference sensor $R$,  $R_{2}^{s,s'} = \Ns (\Ns -1)$. 
For two identical templates, $K_2^{s,s} \equiv \sum_{a}^{\Ns}\sum_{b\ne a}^{\Ns} \delta_{l_a,l_b}$. This definition of $K_2^{s,s}$ depends on the degeneracy of the network with respect to the template. If, for instance, all the sensors are affected by the transient at different epochs, $l_a \ne l_b$ and $a \ne b$, $K_2^{s,s} = 0$. Thus $K_2^{s,s}$ quantifies the degeneracy of the network with respect to template $\mb{s}$. Further, if all of the sensors are degenerate ($l_a = l_b$ for all $a,b$) then $K_2^{s,s} = \Ns (\Ns - 1)$.

Combining these results, we arrive at the expression for the $\mb{s}^{T}\mathbf{E}^{-1}\mb{s'}$ product in the numerator 
of Eq.~(\ref{Eq:Btempcovelements}) 
\begin{align}
    \mb{s}^{T}\mathbf{E}^{-1}\mb{s'} = A K_1^{s,s'} + C R_{1}^{s,s'} + B K_2^{s,s'} \, ,
    \label{Eq:Bnumerator}
\end{align}
where $C = A + (\Ns -1)B = \frac{1}{\sigma^2}\left(\frac{1}{1+\xi} \right)$. The products in the denominator of Eq.~(\ref{Eq:Btempcovelements}), such as $\mb{s}^{T}\mathbf{E}^{-1}\mb{s}$,  can be obtained from Eq.~(\ref{Eq:Bnumerator}) with substitution $s^{'} \rightarrow s$.

With Eq.~(\ref{Eq:Bnumerator}) we can express the elements of the bank covariance matrix (\ref{Eq:Btempcovelements}) as
\small\begin{align}\label{Eq:BgeneralSigmassp}
    &\Sigma_{ss'}= \\
    &\frac{A K_1^{s,s'} + C R_{1}^{s,s'} + B K_2^{s,s'} }{\sqrt{\left(A K_1^{s,s} + C R_{1}^{s,s} + B K_2^{s,s}\right) (A K_1^{s',s'} + C R_{1}^{s',s'} + B K_2^{s',s'})}}. \nonumber
\end{align}
\normalsize
This expression can be simplified further,
\begin{align}
    &\Sigma_{ss'}= \\
    &\frac{a K_1^{s,s'} +  R_{1}^{s,s'} + b K_2^{s,s'} }
    {\sqrt{\left(a K_1^{s,s} +  R_{1}^{s,s} + b K_2^{s,s}\right) \left(a K_1^{s',s'} + R_{1}^{s',s'} + b K_2^{s',s'} \right)}},\nonumber
    \label{Eq:BgeneralSigmasspsimp}
\end{align}
with 
\begin{align}
a &= 1 +(1-1/\Ns)\xi  \,, \\ 
b &= - \xi/\Ns \,. 
\end{align}
Eq.~(\ref{Eq:BgeneralSigmasspsimp}) is an exact expression for the thin planar transients, such as thin DM walls.

Now we introduce a concept of bank-averaged cross-template correlation coefficient $ \overline{\Sigma}_{ss'}$. This quantity is useful in quantifying
properties of the SNR-max statistic for squeezed banks of Sec.~\ref{Sec:SqueezedApprox}.
We assume that the number of templates is large so that the bank covers the accessible parameter space for the sweep parameters, discussed in Sec.~\ref{sec:signaltempgen}. In addition, we assume that we can average over spatial positions of the nodes, see the uniform-occupancy approximation for a spherically-distributed network of Sec.~\ref{Sec:signaltempgen:GPSDM}. 

With these approximations, we determine the typical values $\bar{K}_1, \bar{K}_2$, and $\bar{R}_1$ of the degeneracy coefficients. The key to computing these quantities is 
the  most probable template spread, Eq.~(\ref{Eq:AverageTemplateSpread})
\begin{equation}
      \overline{\Delta l} = 2R/(v_{\perp}^p\Delta_t)\,,  \label{Eq:App:AverageTemplateSpread}  
\end{equation}
where $v^p_\perp=209 \, \mathrm{km/s}$  the most probable speed of the encounter. For GPS.DM data $\overline{\Delta l}=8$, see Sec.~\ref{Sec:signaltempgen:GPSDM}. Then, a typical template extends over $\overline{\Delta l}$ epochs and perturbs $\Ns/\overline{\Delta l}$ sensors each epoch.
Leading to 
\begin{align}
    \bar{K}_1^{s,s} & = \bar{K}_1^{s',s'} =  \Ns , \nonumber \\ 
    \bar{K}_1^{s,s'} & =\Ns/\overline{\Delta l}, \label{Eq:Bexpectedtdegenvalues} \\ 
    \bar{K}_2^{s,s} & = \bar{K}_2^{s',s'} = \bar{K}_2^{s,s'} = 
    {\Ns(\Ns - 1)}/{\overline{\Delta l}}, \nonumber \\ 
    \bar{R}_1^{s,s} & = \bar{R}_1^{s',s'} = \bar{R}_1^{s,s'} =  \left(1 - {2}/{\overline{\Delta l}}\right)\Ns\,. \nonumber
\end{align}
With these values, the ``bank-averaged'' bank covariance matrix element reads
\begin{align} \label{Eq:Btypicalcorrelementvalue}
    \overline{\Sigma}_{ss'} &= 
    \frac{1}{2+\xi(1-1/\Ns)}\, .
\end{align}

For a network of white noise sensors without a reference sensor, such as the GNOME network of magnetometers~\cite{Pustelny2013}, the noise covariance matrix is determined by Eq.~(\ref{Eq:whitenetEcovnoref}). For this network, $a=\frac{1}{\sigma^2}$, $b=0$, and $R_1^{s,s'} = 0$. Then the bank-averaged cross-template correlation coefficient simplifies to
\begin{align}
    \overline{\Sigma}_{ss'} = 
    {1}/{\overline{\Delta l}} \,. \label{Eq:BwnnorefSigmacentral}
\end{align}

\section{Derivation of SNR-max PDF, Eq.~(\ref{Eq:CDF}) }
\label{App:SNRMAXderivation}
SNR-max CDF is given by an $\Nt$-dimensional integral over a hyper-cube,
\begin{equation}
C_{\Nt}\left(  Z|\boldsymbol{\Sigma}\right) = \int_{-Z}^{+Z}d\rho
_{1}...\int_{-Z}^{+Z}d\rho_{\Nt} f\left(  \boldsymbol{\rho}|\mb{\Sigma}\right) \,, \label{Eq:App:CDF}%
\end{equation}
with joint probability distribution~(\ref{Eq:jointsnrpdf}),
\begin{equation}
f\left(  \boldsymbol{\rho}|\mb{\Sigma}\right)  =\frac{1}{\sqrt{\det(2\pi\boldsymbol{\Sigma}})}\exp\left(  -\frac{1}{2}%
\boldsymbol{\rho}^{T}\boldsymbol{\Sigma}^{-1}\boldsymbol{\rho}\right)\,.
\end{equation}

SNR-max PDF, the focus of our derivation, is a derivative of CDF,
\begin{equation}
p_\Nt\left( Z|\mb{\Sigma}\right)  =\frac{d}{dZ}C_\Nt\left(  Z|\mb{\Sigma}\right) \,.
\label{Eq:Cpdffromcdf}
\end{equation}
To compute this derivative, we apply the Leibniz rule,
\begin{align}\label{Eq:Leibniz}
&\frac{d}{dZ}\int_{a(Z)}^{b(Z)}g(u,Z)du  = \\
&=\int_{a(Z)}^{b(Z)}\frac{\partial g(u,Z)}{\partial Z}du+g(b(Z),Z)\frac{db}{dZ}-g(a(Z),Z)\frac{da}{dZ}. \nonumber
\end{align}
As a preliminary step, we introduce  partial integrals of the joint PDF
\begin{equation}
 I_n \left( \rho_1, \ldots \rho_n; Z\right) \equiv 
 \left( \prod_{k=n+1}^{\Nt} \int_{-Z}^{Z} d\rho_k \right) 
 f\left(  \boldsymbol{\rho}|\mb{\Sigma}\right) 
\end{equation}
with boundary values $I_0(;Z) = C_\Nt\left(Z|\mb{\Sigma}\right)$ and 
$I_\Nt(\bs{\rho};) = f\left(  \boldsymbol{\rho}|\mb{\Sigma}\right)$.
Apparently,
\begin{equation}
    I_{n}\left( \rho_1, \ldots \rho_{n}; Z\right) = \int_{-Z}^{Z} d\rho_{n+1}  I_{n+1}\left( \rho_1, \ldots \rho_{n+1}; Z\right)   \, .
\end{equation}
As we repeatedly apply the Leibniz rule to the differentiating the  CDF~(\ref{Eq:App:CDF}), we encounter the general structure
\begin{align}
&\left( {\prod\limits_{m = 1}^n {\int\limits_{ - Z}^Z {d{\rho _m}} } } \right)\frac{d}{{dZ}}\left( {\prod\limits_{l = n+1}^M {\int\limits_{ - Z}^Z {d{\rho _l}} } } \right)f\left(  \boldsymbol{\rho}|\mb{\Sigma}\right)  = \nonumber\\ 
&=\left( {\prod\limits_{m = 1}^k {\int\limits_{ - Z}^Z {d{\rho _m}} } } \right)\frac{d}{{dZ}}{I_{n}\left( \rho_1, \ldots \rho_{n}; Z\right)}, \nonumber
\end{align}
which can be evaluated as
\[\begin{array}{l}
\frac{d}{{dZ}}{I_n}\left( {{\rho _1}, \cdots ,{\rho _n};Z} \right) = \\ \nonumber
\frac{d}{{dZ}}\int_{ - Z}^Z d {\rho _{n + 1}}{I_{n + 1}}\left( {{\rho _1}, \cdots ,{\rho _{n + 1}};Z} \right) \\
=\int_{ - Z}^Z d {\rho _{n + 1}}\frac{d}{{dZ}}{I_{n + 1}}\left( {{\rho _1}, \cdots ,{\rho _{n + 1}};Z} \right)+\\
+ {\left. {{I_{n + 1}}\left( {{\rho _1}, \cdots ,{\rho _{n + 1}};Z} \right)} \right|_{{\rho _{n + 1}} = Z}} +\\ \nonumber
 +{\left. {{I_{n + 1}}\left( {{\rho _1}, \cdots ,{\rho _{n + 1}};Z} \right)} \right|_{{\rho _{n + 1}} =  - Z}}.
\end{array}
\]
Applying this recursion relation to $I_0(;Z) = C_\Nt\left(Z|\mb{\Sigma}\right)$ we eventually
reach $I_\Nt(\bs{\rho};) = f\left(  \boldsymbol{\rho}|\mb{\Sigma}\right)$ for which the derivative with respect to $Z$ vanishes. The result reads
\begin{align}\label{Eq:CdZCDFfinal}
&p_{\Nt}\left(  Z|\boldsymbol{\Sigma}\right)  = \frac{1}{\sqrt{\det\left(
2\pi\boldsymbol{\Sigma}\right)  }} \times  \\
& \sum_{k=1}^{\Nt}[\left( \int_{-Z}^{+Z}%
{\prod\limits_{n=1,n\neq k}^{\Nt}}
d\rho_{n} \right)\left[  \exp\left(  -\frac{1}{2}\boldsymbol{\rho}^{T}%
\boldsymbol{\Sigma}^{-1}\boldsymbol{\rho}\right)  \right]  _{\rho_{k}%
=+Z} \nonumber \\
&+\left(  \rho_{k}=-Z\right)  ]. \nonumber
\end{align}
The quadratic form $\boldsymbol{\rho}^{T}\boldsymbol{\Sigma}^{-1}%
\boldsymbol{\rho}$ is invariant under $\boldsymbol{\rho}\rightarrow
-\boldsymbol{\rho}$. Thus by changing the variables $\boldsymbol{\rho}\rightarrow
-\boldsymbol{\rho}$ in the second term we can show that it is equal to the first
term. Thereby, the SNR-max PDF reduces to %
\begin{align}\label{Eq:CSNRmaxPDF}
&p_{\Nt}\left( z|\boldsymbol{\Sigma}\right)  = \frac{2}{\sqrt{\det\left(
2\pi\boldsymbol{\Sigma}\right)  }}\times \\ 
&\sum_{k=1}^{\Nt}
{\prod\limits_{n=1,n\neq k}^{\Nt}}
\left( \int_{-z}^{+z}d\rho_{n}\right)  \left[  \exp\left(  -\frac{1}{2}%
\boldsymbol{\rho}^{T}\boldsymbol{\Sigma}^{-1}\boldsymbol{\rho}\right)
\right]  _{\rho_{k}=+z}. \nonumber
\end{align}
This concludes our proof of Eq.~(\ref{Eq:CDF}) of the main text.

\section{Derivation of SNR-max CDF and PDF for a nearly-orthogonal bank, Eqs.~(\ref{Eq:ApproxSmallcdfdiff},\ref{Eq:ApprxsmallPDF})}
\label{App:SNRsmallcorr}
Suppose that the correlations between the various templates were small, i.e. $|\Sigma_{ij}| \ll 1$ for $i\neq j$, such that the template covariance matrix  $\mb{\Sigma}$ can be thought of as the identity matrix $\bm{I}$ plus some perturbation matrix $\boldsymbol{A}$: $\mb{\Sigma} = \bm{ I} + \bm{A}$. 
In this case, the bulk of the SNR-max CDF and PDF is determined by the
distributions for fully-orthogonal bank  
 with corrections that depend on $\bm{A}$:
\begin{align}
   C_{\Nt}\left(Z|\boldsymbol{\Sigma}\right) \approx & C_{\Nt}\left(  Z|\bm{I}\right) + \delta_I C_{\Nt}\left(  Z|\boldsymbol{\Sigma}\right) \,, \\
p_{\Nt}\left(z|\boldsymbol{\Sigma}\right) \approx & p_{\Nt}\left(  z|\boldsymbol{I}\right) + \delta_I p_{\Nt}\left(z|\boldsymbol{\Sigma}\right) \,.
\end{align}
Here we use the symbol $\delta_I$ to emphasize that the expansion is about $\boldsymbol{\Sigma}=\bm{I}$. The leading terms are given by Eqs.~(\ref{Eq:CDForthog}, \ref{Eq:Mtemppdfnocorr}), reproduced below,
\begin{align}
C_{\Nt}\left(Z|\boldsymbol{I}\right) & =\left[  \text{erf}\left(
\frac{Z}{\sqrt{2}}\right)  \right]  ^{\Nt} \,, \\ 
 p_{\Nt}\left(z|\boldsymbol{I}\right) & = M\sqrt{\frac{2}{\pi}}e^{-\frac{z^{2}}{2}%
}\left(\text{erf}\left(  \frac{z}{\sqrt{2}}\right)\right)^{\Nt-1}.    
\end{align}
The goal of this Appendix is to derive the corrections $\delta_I C_{\Nt}\left(  Z|\boldsymbol{\Sigma}\right)$ and $\delta_I p_{\Nt}\left(  z|\boldsymbol{\Sigma}\right)$.

The determinant of the template covariance matrix can be approximated as
\begin{equation}
\det\left(  \boldsymbol{I}+\boldsymbol{A}\right)  \approx1+\Tr{\left(\boldsymbol{A}\right)}%
+\frac{1}{2}\left(  \Tr{\left(  \boldsymbol{A}\right)}  \right)  ^{2}-\frac{1}%
{2}\Tr{\left(  \boldsymbol{A}^{2}\right)}  +...
\label{Eq:Ddetorigapprx}
\end{equation}
Here the elements of the perturbative matrix $\boldsymbol{A}$ are
\begin{equation}
A_{ij}=\left(  1-\delta_{ij}\right)  \Sigma_{ij}.%
\label{Eq:DmatrixAelements}
\end{equation}
Since the perturbation matrix $\boldsymbol{A}$ has zeros along the diagonal,  $\Tr{\left(\boldsymbol{A}\right)}=0$, and the  determinant  simplifies to%
\begin{equation}
\det\left(  \boldsymbol{\Sigma}\right)  =\det\left(  \boldsymbol{I}%
+\boldsymbol{A}\right)  \approx1-\frac{1}{2}\Tr{\left(  \boldsymbol{A}%
^{2}\right)}.
\label{Eq:DdettmpCov}
\end{equation}
Notice that $\Tr{\left(  \boldsymbol{A}^{2}\right)}$ can be related to $\Tr{\left(  \boldsymbol{\Sigma}^{2}\right)}$. Indeed, $\bm{\Sigma}^{2} = \bm{I} + 2 \bm{A} + \bm{A}^2$,
so that 
\begin{equation}
 \Tr{\left(  \boldsymbol{A}^{2}\right)} = \Tr{\left(  \boldsymbol{\Sigma}^{2}\right)} - M \,. \label{Eq:DTrRelation}
\end{equation}
 Also since, by construction, $\boldsymbol{A}$ is a symmetric matrix, $\Tr{\left( \boldsymbol{A}^{2}\right)}  =\sum_{ij}\left(
A_{ij}\right)  ^{2}>0$.

As for the inverse of the template covariance matrix, this can also be approximated as%
\begin{equation}
\boldsymbol{\Sigma}^{-1} = \left(  \boldsymbol{I}+\boldsymbol{A}\right)  ^{-1}\approx\boldsymbol{I}%
-\boldsymbol{A+A}^{2}.
\label{Eq:DinversetmpCov}
\end{equation}
Now that we have our approximate inverse template covariance matrix and determinant, we can simplify the SNR-max CDF~(\ref{Eq:CDF}),%
\begin{align}\label{Eq:DSNRmaxCDF}
&C_{\Nt}\left(  Z|\boldsymbol{\Sigma}\right)  =\frac
{1}{\left(  2\pi\right)  ^{\Nt/2}\sqrt{\det\boldsymbol{\Sigma}}} \\ 
&\int%
_{-Z}^{+Z}d\rho_{1}...\int_{-Z}^{+Z}d\rho_{\Nt}\exp\left(  -\frac{1}%
{2}\boldsymbol{\rho}^{T}\boldsymbol{\Sigma}^{-1}\boldsymbol{\rho}\right) . \nonumber
\end{align}
Using expansion (\ref{Eq:DinversetmpCov}), we represent the exponential in $C_\Nt$ as
\begin{align}\label{Eq:DExpexpansion}
&\exp\left(  -\frac{1}{2}\boldsymbol{\rho}^{T}\boldsymbol{\Sigma}%
^{-1}\boldsymbol{\rho}\right)  \\
&\approx\exp\left(  -\frac{1}{2}\boldsymbol{\rho
}^{T}\boldsymbol{\rho}\right)  \exp\left(  \frac{1}{2}\boldsymbol{\rho}%
^{T}\boldsymbol{A\rho}\right)  \exp\left(  -\frac{1}{2}\boldsymbol{\rho}%
^{T}\boldsymbol{A}^{2}\boldsymbol{\rho}\right). \nonumber
\end{align}
Expanding out the second and the third exponentials up to $O(\bm{A}^2)$ we obtain
\begin{align}\label{Eq:DExpfullexpansion}
&\exp\left(  -\frac{1}{2}\boldsymbol{\rho}^{T}\boldsymbol{\Sigma}%
^{-1}\boldsymbol{\rho}\right)  \\ 
&\approx \exp\left(  -\frac{1}{2}\boldsymbol{\rho}^{T}\boldsymbol{\rho
}\right) \times \nonumber \\
&\left(  1+\frac{1}{2}\boldsymbol{\rho}^{T}\boldsymbol{A\rho+}%
\frac{1}{8}\left(  \boldsymbol{\rho}^{T}\boldsymbol{A\rho}\right)  ^{2}%
-\frac{1}{2}\boldsymbol{\rho}^{T}\boldsymbol{A}^{2}\boldsymbol{\rho}\right). \nonumber
\end{align}
This expansion converges as long as
$|\rho| \ll 1/ \sqrt{\max\left\vert A_{ij}\right\vert}$.

Using these expansions, the correction to the SNR-max CDF reads
\begin{align}
\delta_I C_{\Nt}\left(  Z|\boldsymbol{\Sigma}\right) \approx    &
\frac{1}{4}\Tr{\left(  \boldsymbol{A}^{2}\right)} C_{\Nt}\left(Z|\boldsymbol{I}\right) \label{Eq:DSNRmaxCDFapproxfirst}  \\ 
&  +\frac{1}{2}\sum_{i\neq j}A_{ij} \avZ{ \rho_{i}\rho_{j}}
 \nonumber \\ 
& + \frac{1}{8}\sum_{i\neq j}\sum_{k\neq l}A_{ij}A_{kl}
\avZ{ \left(  \rho_{i}\rho_{j}\right)  \left(
\rho_{k}\rho_{l}\right) }
\nonumber  \\ 
&  -\frac{1}{2} \sum_{ij} \left(\boldsymbol{A}^{2}\right)_{ij} \avZ{\rho_{i}\rho_{j}}
\,.\nonumber
\end{align}
Here we kept all the terms up to~$O(\boldsymbol{A}^{2})$ and
introduced ``averaging'' notation
\begin{equation}
    \avZ{ g( \bs{\rho} )} \equiv \int_{-Z}^{+Z}d\rho_{1}...\int_{-Z}%
^{+Z}d\rho_{\Nt}  g( \bs{\rho})  f\left(  \boldsymbol{\rho
}|\boldsymbol{I}\right) \,,
\end{equation}
where subscript $Z$ emphasizes that the \Nt-dimensional integral is evaluated 
over hyper-cube centered at $\bs{\rho}=0$ and of side $2Z$.  $f\left(\boldsymbol{\rho}|\boldsymbol{I}\right)$ is a joint PDF for fully orthogonal bank,
\begin{equation}
    f\left(\boldsymbol{\rho}|\boldsymbol{I}\right) = \frac{1}{\left(  2\pi\right)^{\Nt/2}} 
\exp\left(  -\frac{1}{2}\boldsymbol{\rho}^{T}\boldsymbol{\rho}\right)  \,.
\label{Eq:DjointPDFOrtho}
\end{equation}
This joint PDF factorizes into the product of exponentials, simplifying evaluation of averages.  The required  averages in Eq.~(\ref{Eq:DSNRmaxCDFapproxfirst}) are 
\begin{align}
    &\avZ{ \rho_{i}\rho_{j}}  = \delta_{ij} \, \avZ{\rho^{2}} 
    C_{\Nt-1}\left(Z|\boldsymbol{I}\right)  \,, \\
    &\left\langle \left(  \rho_{i}\rho_{j}\right)  \left(  \rho_{k}\rho_{l}\right)
\right\rangle_Z ^{i\neq j,k\neq l}  = \nonumber \\
 & = \left( \avZ{\rho^{2}} \right)^2 C_{\Nt-2}\left(Z|\boldsymbol{I}\right)
\left( 
\delta_{ik}\delta_{jl}+\delta_{il}\delta_{jk}
\right)^{i\neq j,k\neq l} \,, \nonumber
\end{align}
with 
\begin{align}
   \left\langle \rho^{2}\right\rangle_Z  & = \frac{1}{\sqrt{2\pi}}\int_{-Z}%
^{+Z}\rho^{2} e^{-\rho^{2}/2}d\rho \\
& =\erf{
\frac{Z}{\sqrt{2}} }  -\sqrt{\frac{2}{\pi}}Z e^{-\frac{Z^{2}}{2}} \nonumber \,.
\end{align}
Then the correction to the SNR-max CDF reduces to 
\begin{align}\label{Eq:DdiffSNRmaxCDFs}
&  \delta_I C_{\Nt}\left(  Z|\boldsymbol{\Sigma}\right) =\frac{1}{4}\Tr{\left(  \boldsymbol{A}^{2}\right)}  C_{\Nt-2}\left(
Z|\boldsymbol{I}\right)  \times \\
&\left[  
C_{2}\left(  Z|\boldsymbol{I}\right)  -
2C_{1}\left(
Z|\boldsymbol{I}\right)  \left\langle \rho^{2}\right\rangle_Z + \left\langle \rho^{2}\right\rangle^{2}_Z\right] \, . \nonumber
\end{align}
The factor in square brackets can be recognized as $\left(C_{1}\left(
Z|\boldsymbol{I}\right) - \left\langle \rho^{2}\right\rangle_Z \right)^2$.
Finally,
\begin{equation}
\delta_I C_{\Nt}\left(Z|\boldsymbol{\Sigma}\right) = \Tr{\left(  \boldsymbol{A}%
^{2}\right)}  \frac{Z^{2} e^{-Z^{2}}}{2\pi}\left[  \text{erf}\left(  \frac
{Z}{\sqrt{2}}\right)  \right]  ^{\Nt-2}\,.
\label{Eq:DSNRmaxCDFapproxdiff}
\end{equation}
With the relation (\ref{Eq:DTrRelation}) this is  Eq.~(\ref{Eq:ApproxSmallcdfdiff}) of the main text.

Finally, differentiating $\delta_I C_{\Nt}\left(Z|\boldsymbol{\Sigma}\right)$ with respect to $Z$ and using the Leibniz rule~(\ref{Eq:Leibniz}), we arrive at the correction to the SNR-max PDF
\begin{align*}
  & \delta_I p_{\Nt}\left(z|\boldsymbol{\Sigma}\right)  =  \operatorname{Tr}\left(\boldsymbol{A}^{2}\right) \frac{z e^{-z^{2}}}{\pi} \left(\operatorname{erf}\left(\frac{z}{\sqrt{2}}\right)\right)^{\Nt-3} \times \\
& \left[\frac{(\Nt-2) z e^{-z^{2} / 2} }{\sqrt{2 \pi}}+ \left(1-z^{2}\right) \operatorname{erf}\left(\frac{z}{\sqrt{2}}\right)\right] \, ,
\label{Eq:DSNRmaxPDFapproxdiff}
\end{align*}
which is Eq.~(\ref{Eq:ApprxsmallPDF}) of the main text.

\end{document}